\documentclass[pdftex,twocolumn,epjc3,numbook,final]{svjour3}

\RequirePackage[T1]{fontenc}

\smartqed  % flush right qed marks, e.g. at end of proof

\RequirePackage{graphicx}
\RequirePackage{mathptmx}      % use Times fonts if available on your TeX system
\RequirePackage[numbers,sort&compress]{natbib}
\RequirePackage[colorlinks,citecolor=blue,urlcolor=blue,linkcolor=blue]{hyperref}

\journalname{Eur. Phys. J. C}

\usepackage{graphicx}% Include figure files
\usepackage{dcolumn}% Align table columns on decimal point
\usepackage{bm}% bold math
\usepackage{color}
\usepackage[dvipsnames]{xcolor}
\usepackage{url}
\usepackage{xspace}
\usepackage{slashed}
\usepackage{multirow,bigstrut}
\usepackage{mathrsfs} % pretty maths
\usepackage{relsize,amsmath}
\usepackage[geometry]{ifsym}
\usepackage{amssymb}% http://ctan.org/pkg/amssymb
\usepackage{pifont}% http://ctan.org/pkg/pifont
\usepackage{physics}
\usepackage{comment}
\usepackage{hyperref}
\RequirePackage[capitalise]{cleveref}

\usepackage{graphicx}
\usepackage{rotating}
\usepackage{ragged2e}
\usepackage{microtype} %for adjustment of interword spacing (to avoid overflow)

\usepackage{fontawesome} % for code link icons
\definecolor{blue-violet}{rgb}{0.33, 0.17, 0.89}

\usepackage[normalem]{ulem}

\usepackage{siunitx}

\newcommand\gmu{(g-2)_{\mu}}
\newcommand{\refref}[1]{Ref.~\cite{#1}}

\begin{document}

\title{Constraining Light Thermal Inelastic Dark Matter with NA64}

%%%%%%%%%%%%%%%%%%%%%%%
% Authors
\author{
Martina~Mongillo\thanksref{addrETH,e1}\and
Asli~Abdullahi\thanksref{addrFNAL,addrDH}\and
Benjamin~Banto~Oberhauser\thanksref{addrETH}\and
Paolo~Crivelli\thanksref{addrETH,e4}\and
Matheus~Hostert\thanksref{addrW, addrSPA_MN,addrWIFTPI_MN}\and
Daniele~Massaro\thanksref{addrBO,addrINFN, addrLU}\and
Laura~Molina~Bueno\thanksref{addrVA}\and
Silvia~Pascoli\thanksref{addrBO,addrINFN,addrCERN}
}

\thankstext{e1}{e-mail: mmongillo@student.ethz.ch}
\thankstext{e4}{e-mail: Paolo.Crivelli@cern.ch}
%\thankstext{e2}{e-mail: asli.abdullahi@durham.ac.uk}
%\thankstext{e3}{e-mail: bantoobb@student.ethz.ch}
%\thankstext{e5}{e-mail: mhostert@pitp.ca}
%\thankstext{e6}{e-mail: daniele.massaro5@unibo.it}
%\thankstext{e7}{e-mail: laura.molina.bueno@cern.ch}
%\thankstext{e8}{e-mail: silvia.pascoli@unibo.it}

\institute{
Institute for Particle Physics and Astrophysics, ETH Z\"urich,
Z\"urich, CH-8093, Switzerland \label{addrETH}
\and
Theoretical Physics Department, Fermi National Accelerator Laboratory, Batavia, IL 60510, USA  \label{addrFNAL}
\and
Institute for Particle Physics Phenomenology, Department of Physics, Durham University, South Road, Durham DH1 3LE, United Kingdom  \label{addrDH}
\and
Perimeter Institute for Theoretical Physics, Waterloo, ON N2J 2W9, Canada  \label{addrW}
\and
School of Physics and Astronomy, University of Minnesota, Minneapolis, MN 55455, USA \label{addrSPA_MN}
\and
William I. Fine Theoretical Physics Institute, School of Physics and Astronomy, University of Minnesota, Minneapolis, MN 55455, USA \label{addrWIFTPI_MN}
\and
Dipartimento di Fisica e Astronomia, Universit\`a di Bologna, via Irnerio 46, 40126 Bologna, Italy \label{addrBO}
\and
INFN, Sezione di Bologna, viale Berti Pichat 6/2, 40127 Bologna, Italy \label{addrINFN}
\and
CERN, Theoretical Physics Department, Geneva, Switzerland \label{addrCERN}
\and
Centre for Cosmology, Particle Physics and Phenomenology (CP3), Universit\'e Catholique de Louvain, B-1348 Louvain-la-Neuve, Belgium \label{addrLU}
\and
Instituto de Fisica Corpuscular (CSIC/UV),
Carrer del Catedrátic José Beltrán Martinez, 2, 46980 Paterna, Valencia, Spain \label{addrVA}
}

\date{Received: date / Accepted: date}

\maketitle

%%%%%%%%%%%%%%%%%%%%%%%%%%%%%%%%%%%%%
\begin{abstract}
%%%%%%%%%%%%%%%%%%%%%%%%%%%%%%%%%%%%%
A vector portal between the Standard Model and the dark sector is a predictive and compelling framework for thermal dark matter.
Through co-annihilations, models of inelastic dark matter (iDM) and inelastic Dirac dark matter (i2DM) can reproduce the observed relic density in the MeV to GeV mass range without violating cosmological limits.
In these scenarios, the vector mediator behaves like a semi-visible particle, evading traditional bounds on visible or invisible resonances, and uncovering new parameter space to explain the muon $(g-2)$ anomaly.
By means of a more inclusive signal definition at the NA64 experiment, we place new constraints on iDM and i2DM using a missing energy technique.
With a recast-based analysis, we contextualize the NA64 exclusion limits in parameter space and estimate the reach of the newly collected and expected future NA64 data.
Our results motivate the development of an optimized search program for semi-visible particles, in which fixed-target experiments like NA64 provide a powerful probe in the sub-GeV mass range.

\end{abstract}

%%%%%%%%%%%%%%%%%%%%%%%%%%%%%%%%%%%%%
\section{Introduction}
%%%%%%%%%%%%%%%%%%%%%%%%%%%%%%%%%%%%%

Dark matter (DM) represents one of the main challenges faced by modern particle physics and the most striking evidence for the incompleteness of the Standard Model (SM). 
One paradigm in the broad landscape of proposed solutions~\cite{Feng:2010gw} is that of dark sectors (DS).
This framework considers a new sector of particles below the electroweak scale that are not charged under the SM gauge symmetries but could participate in dynamics of its own. 
Because of the weak interactions between SM and the DS, this elusive sector can address some of the outstanding issues of the SM. 
In particular, it can tackle the DM problem without violating experimental observations through the existence of new light particles.
The interactions between these DS states and the SM can then only proceed via gravity or feeble portal interactions~\cite{Battaglieri:2017aum, Alexander:2016aln}. 
The latter are renormalizable terms involving SM and DS fields and are classified as the vector (in the presence of kinetic mixing with a dark photon), scalar (for mixing between a dark scalar and the Higgs), or neutrino (for a Yukawa interaction with the SM lepton doublet and heavy neutral leptons) portals. 
For a recent review, see Ref.~\cite{Agrawal:2021dbo}.

Because of the low scale, DS models provide a fertile ground for phenomenology, motivating the development of a program in high-intensity and low-energy experiments~
\cite{ArkaniHamed:2008qn,Pospelov:2008jd,Hooper:2012cw,Pospelov:2007mp,Pospelov:2008zw,Essig:2013lka}. 
Apart from the direct search for DM, the DS hypothesis drives a complementary effort to search for the mediator particles of the portal interactions. 

Among the most popular and studied models of DS is that of a vector portal to DM, where light dark matter (LDM) particles interact with the SM via a kinetically-mixed \textit{dark photon} $A'$~\cite{Holdom:1985ag,Okun:1982xi}, the mediator of a new dark gauge symmetry $U(1)_D$.
As the mediator particle, the dark photon is responsible for keeping DM in thermal equilibrium with the SM in the early Universe.
In addition, due to its coupling to the SM sector, dark photons could be produced in lab-based experiments and leave a measurable imprint in precision observables like the anomalous magnetic moment of the muon.

Analogously to the SM, it is possible that the DS contains more than one generation of particles charged under the $U(1)_D$.
In the presence of mixing, the DS spectrum would allow for the decay of the heavier states, $\chi_{2,3,\dots}$, into the lightest and most stable particle, $\chi_1$, identified with DM.
If such decays involve SM particles, as in $\chi_2 \to \chi_1 e^+e^-$, the heavy partners can decay into final states that are neither fully visible SM states, nor entirely invisible DS states. 
Therefore, if the mediator $A'$ decays predominantly into the excited states $\chi_{2,3,\dots}$, its semi-visible branching ratios would be much larger than the visible, e.g., $A' \to e^+e^-$, and invisible decays, e.g., $A' \to \chi_1\chi_1$.
This semi-visible dark photon can evade several constraints from missing energy and visible resonances, thus providing a new target for experimental searches~\cite{Mohlabeng:2019vrz,Tsai:2019buq,Duerr:2019dmv,Abdullahi:2020nyr}.
A thorough exploration of the physics of semi-visible dark photons can be found in Ref.~\cite{future}.

In this article, we critically assess the potential of the NA64 experiment~\cite{NA64:2017vtt} in testing this class of dark sectors in the MeV to sub-GeV mass range, where NA64 is sensitive to on-shell dark photon production. 
The NA64 experiment at CERN pioneered the missing energy technique for the study of DS physics in fixed-target experiments~\cite{Izaguirre:2014bca}.
By using an active beam-dump to search for missing energy, it has a distinct advantage over direct search methods: the signal detection relies on identifying solely the \textit{production} of the dark mediator and not on its subsequent decays or scatterings.
Besides improving the experimental sensitivity to small couplings, this strategy makes the $A' \rightarrow DS$ search mostly insensitive to the dark photon decay channels and provides the leading limits in a large region of parameter space of suggested DS models, reaching the thermal DM relic density targets.

In this study, we focus on LDM models where the dark photon is semi-visible. We consider two scenarios: inelastic dark matter (iDM) and inelastic Dirac dark matter (i2DM).
The next Section describes and motivates these frameworks, while \cref{sec3} introduces the NA64 experiment and explains the recast method adopted to derive the experimental coverage and sensitivity. 
\Cref{sec4} gathers the relevant existing constraints on semi-visible dark photons. 
The results of our analysis are illustrated and discussed in \cref{sec5} and \cref{sec6} for different parametrizations; the first part builds on the previous investigation of a $(g-2)_\mu$ driven model point, while the second focuses more generally on the thermal DM targets. 
Lastly, the conclusions drawn from the obtained results can be found in \cref{sec7}.

%%%%%%%%%%%%%%%%%%%%%%%%%%%%%%%%%%%%%
\section{Semi-visible dark photons}
%%%%%%%%%%%%%%%%%%%%%%%%%%%%%%%%%%%%%

The dark photon as a vector portal to dark sectors has received great interest because it can provide a simple realization of freeze-out LDM~\cite{Pospelov:2007mp,ArkaniHamed:2008qn,Pospelov:2008jd}.
Due to its simplicity and predictivity, it serves as a benchmark in studies of both direct and indirect LDM detection~\cite{Knapen:2017xzo,Agrawal:2021dbo,Essig:2022dfa}. 
Several constraints have been discussed in the literature, including astrophysical limits~\cite{Pospelov:2008jd,Dreiner:2013mua,Chang:2018rso}, meson factories~\cite{BNL-E949:2009dza,Mirra:2018zdd,NA62:2020xlg}, neutrino and beam-dump experiments~\cite{Batell:2009di,Batell:2014mga,MiniBooNE:2017nqe}, and $e^+e^-$-colliders~\cite{Batell:2009yf,Essig:2013vha,BaBar:2017tiz}.
The two most studied types of $A'$ models are the invisible dark photon, where $A'$ decays predominantly to invisible final states, such as sub-GeV LDM candidates, or the visible dark photon, where $A'$ decays predominantly to all visible SM particles. 

A dark photon can also lead to sizable deviations to the anomalous magnetic moment of the electron and the muon~\cite{Gninenko:2001hx,Pospelov:2008zw}.
Notably, the latter observable is currently the subject of increased scrutiny 
due to recent measurements~\cite{Muong-2:2006rrc,Muong-2:2021ojo}.
At present, a $4.2\sigma$ discrepancy exists between the combined measurements and the data-driven theoretical predictions based on the dispersive method~\cite{Jegerlehner:2009ry,Miller:2012opa,Aoyama:2020ynm}.
The discrepancy is much less significant between the measurements and lattice calculations~\cite{Borsanyi:2020mff}, but the reason for the disagreement between the two theoretical methods is unknown.
For an invisible $A'$, the parameter space where kinetic mixing can resolve the $(g-2)_\mu$ anomaly has already been excluded by the missing energy searches at the NA64~\cite{Banerjee:2016tad} and BaBar~\cite{BaBar:2017tiz} experiments.
In the alternative case of visible $A'$, experimental searches for decays to SM states ($A' \to e^+e^-$, $\mu^+\mu^-$, $\pi^+\pi^-$) at BaBar~\cite{BaBar:2014zli}, NA48~\cite{NA482:2015wmo}, and KLOE~\cite{Anastasi:2015qla} have also ruled out the $(g-2)_\mu$ region.
However, in DS models with multiple generations, the dark photon can have additional decay modes involving both SM and DM final states, and this feature can significantly weaken the bounds just mentioned.
This possible scenario, characterized by the presence of visible and invisible final states, will be referred to as the \textit{semi-visible} dark photon, following \refref{Mohlabeng:2019vrz}. 
Such a channel could give another chance to kinetically-mixed dark photons as a viable explanation of the observed DM abundance and the $(g-2)_\mu$ discrepancy.
In what follows, we introduce the two semi-visible dark photon models studied in this article.

We start from the usual kinetically-mixed dark photon~\cite{Holdom:1985ag}.
In the physical basis for the gauge bosons, the dark photon Lagrangian is
\begin{equation}
    \mathcal{L}_{\rm DP} = \frac{m_{A'}^2}{2} A'_{\mu} A'^\mu + A'_\mu \left(g_D \mathcal{J}_{\rm DS}^\mu - e\epsilon\mathcal{J}_{\rm EM}^\mu\right),
\end{equation}
where $\epsilon$ parametrizes the kinetic mixing, $g_{D}$ represents the dark coupling constant and $\mathcal{J}_{\rm DS}^\mu$ denotes the dark sector current.
The physical dark photon $A'_\mu$ acquires a coupling to the electromagnetic current $\mathcal{J}_{\rm EM}^\mu$ after the diagonalization of kinetic terms. 
The mass $m_{A'}$ can be generated by the St\"uckelberg mechanism or by the vacuum expectation value of a dark Higgs scalar field, which spontaneously breaks the associated $U(1)_{D}$ symmetry.

%%%%%%%%%%%%%%%%%%%%%%%%%%%%%%%%%%%%%
\subsection{Inelastic Dark Matter}

Semi-visible dark photons appear in models of inelastic dark matter (iDM), where a DM fermion is split into two Majorana states, often referred to as pseudo-Dirac DM. 
The idea was first proposed in Ref. \cite{Tucker-Smith:2001myb,Tucker-Smith:2004mxa} as a mechanism to avoid direct detection searches.
Since the two Majorana states only interact with each other, but not with themselves, DM particles could only be detected through their inelastic scattering on nuclei, requiring larger kinetic energy to overcome the mass gap.
In these models, the relic abundance of dark matter is achieved via the freeze-out mechanism.
Because of the off-diagonal interactions, DM freeze-out is achieved through co-annihilation between the two pseudo-Dirac states.
The heavier component eventually decays, leaving only the lightest particle as a stable DM relic.

In the dark photon realization, $A'$ can couple off-diagonally to the two components of a pseudo-Dirac fermion or complex scalar.
The phenomenology of the fermion and scalar cases is very similar, so in this work, we only consider the former.
The model introduces a fermion $\Psi = \Psi_L + \Psi_R$, where $\Psi_L$ and $\Psi_R$ are chiral spinors.
To ensure anomaly cancellation, $\Psi$ is a vector-like fermion, with $Q_{\Psi_L} = Q_{\Psi_R} = 1$.
The Lagrangian is given by
\begin{align}
    \mathcal{L}_{\rm DS} & = \overline{\Psi_L}i(\slashed{\partial} - i g_D A'^\mu)\Psi_L + \overline{\Psi_R}i(\slashed{\partial} - i g_D A'^\mu)\Psi_R 
    \\ \nonumber
    &\quad - m_D\overline{\Psi}\Psi - \left( \frac{\mu_L}{2} \overline{\Psi^c_L}\Psi_L + \frac{\mu_R}{2} \overline{\Psi_R^c}\Psi_R + \text{h.c.}\right)
\end{align}
The Dirac mass $m_D$ is gauge invariant and can take any value, while the Majorana mass terms $\mu_L$ and $\mu_R$ are generated from the breaking of the $U(1)_D$ symmetry. 
In the minimal model, the Majorana masses can arise from the vacuum expectation value of a scalar with $Q_\Phi = 1$.

\begin{figure}[h]
\centering
\includegraphics[width=0.8\columnwidth]{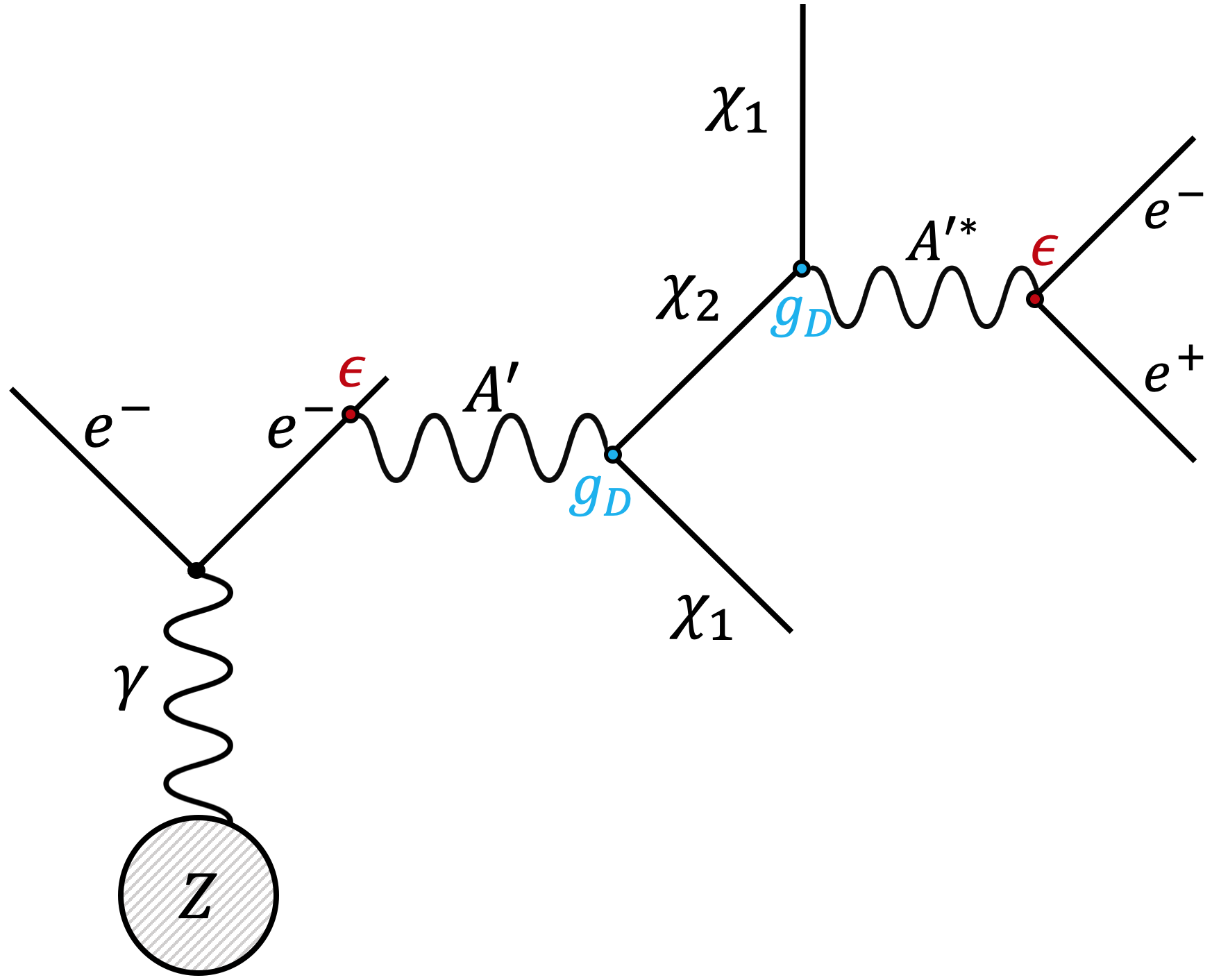}
\caption{Production of $A'$ via a dark bremsstrahlung reaction ${e^-Z\rightarrow e^-ZA'}$ and subsequent semi-visible decay $A'\rightarrow\chi_2\chi_1$;$\chi_2 \rightarrow \chi_1 A'^* \rightarrow \chi_1 e^+ e^-$ \label{fig:decay}.}
\end{figure}

In the iDM regime, in the limit where $\Delta \mu = \mu_L - \mu_R \simeq 0$, the rotation from the interaction basis $\Psi_{L,R}$ to the mass basis $\chi_{1,2}$ is maximal, $\tan 2\theta=m_D/\Delta \mu \to \infty$.
Therefore, the two mass eigenstates can only interact with the dark photon via off-diagonal terms, 
\begin{align}
    \mathcal{J}_{\rm DS}^\mu & = \overline{\chi_2} i\gamma^\mu \chi_1 + \mathcal{O}\left( \Delta \mu/m_D \right).
\end{align}
In this limit, the splitting of the pseudo-Dirac particle is given by $\mu = \mu_L + \mu_R$.
The free parameters needed to describe the phenomenology of this model are therefore five: $m_{A'}$, $\epsilon$, $m_{\chi_1}$, $\alpha_D = g_D^2/4\pi$ and the mass splitting $\Delta=m_{\chi_2}-m_{\chi_1}$. 
We focus on the regime where the mediator is heavier than the dark fermions, $m_{A'}>m_{\chi_1}+m_{\chi_2}$, and consider sizeable mass splitting of at least $\Delta = 0.01 m_{\chi_1}$ and at most $\Delta = 0.4 m_{\chi_1}$.  
In this regime, the production of an on-shell dark photon is followed by its prompt  decay to $A' \rightarrow\chi_1\chi_2$. 
As illustrated in~\cref{fig:decay}, due to the mass splitting between the two states, the heavier partner then decays by emitting an off-shell dark photon, that, via kinetic mixing, undergoes lepton pair production, $\chi_2 \rightarrow \chi_1 A'^* \rightarrow \chi_1 l^+ l^-$. 
According to this model, the lightest, stable state $\chi_1$ is responsible for the observed DM relic abundance.

%%%%%%%%%%%%%%%%%%%%%%%%%%%%%%%%%%%%%
\subsection{Inelastic Dirac Dark Matter}

Another recently proposed candidate which broadens the inelastic secluded matter category is the so-called inelastic Dirac dark matter (i2DM)~\cite{Filimonova:2022pkj}. 
In this model, the spectrum is composed of two Dirac fermions, a light sterile fermion and a heavier dark one.
Due to the small mixing, the dark photon has hierarchical couplings to the dark sector states, coupling more strongly to the heavier partner.
In terms of its cosmological history, the DM freeze-out is determined by a combination of its co-annihilation with the heavy fermion, $\chi_1 \chi_2 \rightarrow A^{'*} \rightarrow f^+ f^-$, as well as by co-scattering, $\chi_2 \chi_2 \rightarrow A^{'*} \rightarrow \chi_1 \chi_2$.
As long as the $\chi_1$ and $\chi_2$ sectors are in chemical equilibrium, the self-annihilation of the heavy partner, $\chi_{2} \chi_{2} \to f^+ f^-$ also contributes~\cite{Filimonova:2022pkj}. 
Both $\chi_2$ self-annihilations and $\chi_1 \chi_2$ co-annihilations are exponentially suppressed for large mass splittings, but can dominate over the mixing-suppressed $\chi_1$ self-annihilation.

This scenario, instead of predicting a DS consisting of two almost degenerate \textit{Majorana} fermions, forming a pseudo-Dirac pair, envisages two \textit{Dirac} fields, $\Psi = \Psi_L+\Psi_R$ and $\eta = \eta_L + \eta_R$, one being charged and the other neutral under the hidden $U(1)_{D}$ gauge symmetry: $Q_{\Psi} = +1$ and $Q_\eta = 0$. 
In summary,
\begin{align}
    \mathcal{L}_{\rm DS} & = \overline{\Psi}i(\slashed{\partial} - i g_D A'^\mu)\Psi + \overline{\eta}i\slashed{\partial}\eta
    \\ \nonumber
    &\quad - m_\eta \overline{\eta}\eta - m_\Psi \overline{\Psi}\Psi  - \left(\mu\, \overline{\Psi}\eta + \mu^\prime\, \overline{\Psi^c}\eta + \text{ h.c.}\right),
\end{align}
where $\mu, \mu^\prime$ could be generated via spontaneous breaking of the $U(1)_D$ symmetry by a dark scalar with charge $Q_{\Phi} = +1$.
Assuming the Majorana masses of $\Psi$ and $\eta$ to be negligibly small and, for simplicity, taking $\mu = \mu^\prime$, the spectrum is composed of two exact Dirac particles~\footnote{In principle, $\eta$ may also have Yukawa couplings with SM neutrinos. A dark parity $Z_2$ symmetry or lepton number conservation, where $L(\eta) = 0$, could be imposed to forbid such terms, and would explain the stability of DM.}.
In the physical basis, the dark current is given by
\begin{equation}\label{eq2.5}
    \mathcal{J}_{\rm DS}^\mu = s_\theta^2 \overline{\chi_1} \gamma^\mu \chi_1 + c_\theta^2 \overline{\chi_2} \gamma^\mu\chi_2 + \left(s_\theta c_\theta \overline{\chi_2} \gamma^\mu \chi_1  + \text{ h.c.} \right),
\end{equation}
where $\theta \simeq \mu/(m_\Psi - m_\eta)$ controls the interaction strength of the dark photon to the fermions.
Here, we assume $\chi_1$ is mostly in the direction of $\eta$ so that $\theta$ is small and the lightest particle interacts less strongly with $A'$.

The mixing angle now determines the branching ratios of the three allowed decay channels $A'\rightarrow \chi_i \chi_j$, with $i, j \in \{1,2\}$, and
the phenomenology is then determined by six parameters: $m_{A'}$, $\epsilon$,  $m_{\chi_1}$, $\alpha_D = g_D^2/4\pi$, ${\Delta=m_{\chi_2}-m_{\chi_1}}$ and $\theta$. 

%%%%%%%%%%%%%%%%%%%%%%%%%%%%%%%%%%%%%%%%%%%%%%%%%%
\section{The NA64 experiment}\label{sec3}
%%%%%%%%%%%%%%%%%%%%%%%%%%%%%%%%%%%%%%%%%%%%%%%%%%
NA64 is a fixed-target experiment located in the North Area at the CERN Super Proton Synchrotron (SPS) accelerator, using electron, positron, muon and hadron beams.
The NA64 search plan is focused on sub-GeV LDM, visible~\cite{NA64:2019auh} and invisible~\cite{Banerjee:2019pds} dark photons, and other new light particles which may belong to a dark sector. 

The here presented study is based on the $e^-$ invisible setup configuration, where the $A'$ search is conducted through a missing energy approach. The idea behind this technique is to intercept all the SM secondaries produced from nuclear collisions in the dump, to be able to label as signal an event with an imbalance between initial and final total energy. A description of the setup design and the analysis carried out on the two semi-visible models described in the previous section is provided in the following.

%%%%%%%%%%%%%%%%%%%%%%%%%%%%%%%%%%%%%
\subsection{The experimental setup and technique}
\begin{figure}[b]
\centering
\includegraphics[clip, trim=0 6cm 0cm 4cm, width=0.95\columnwidth]{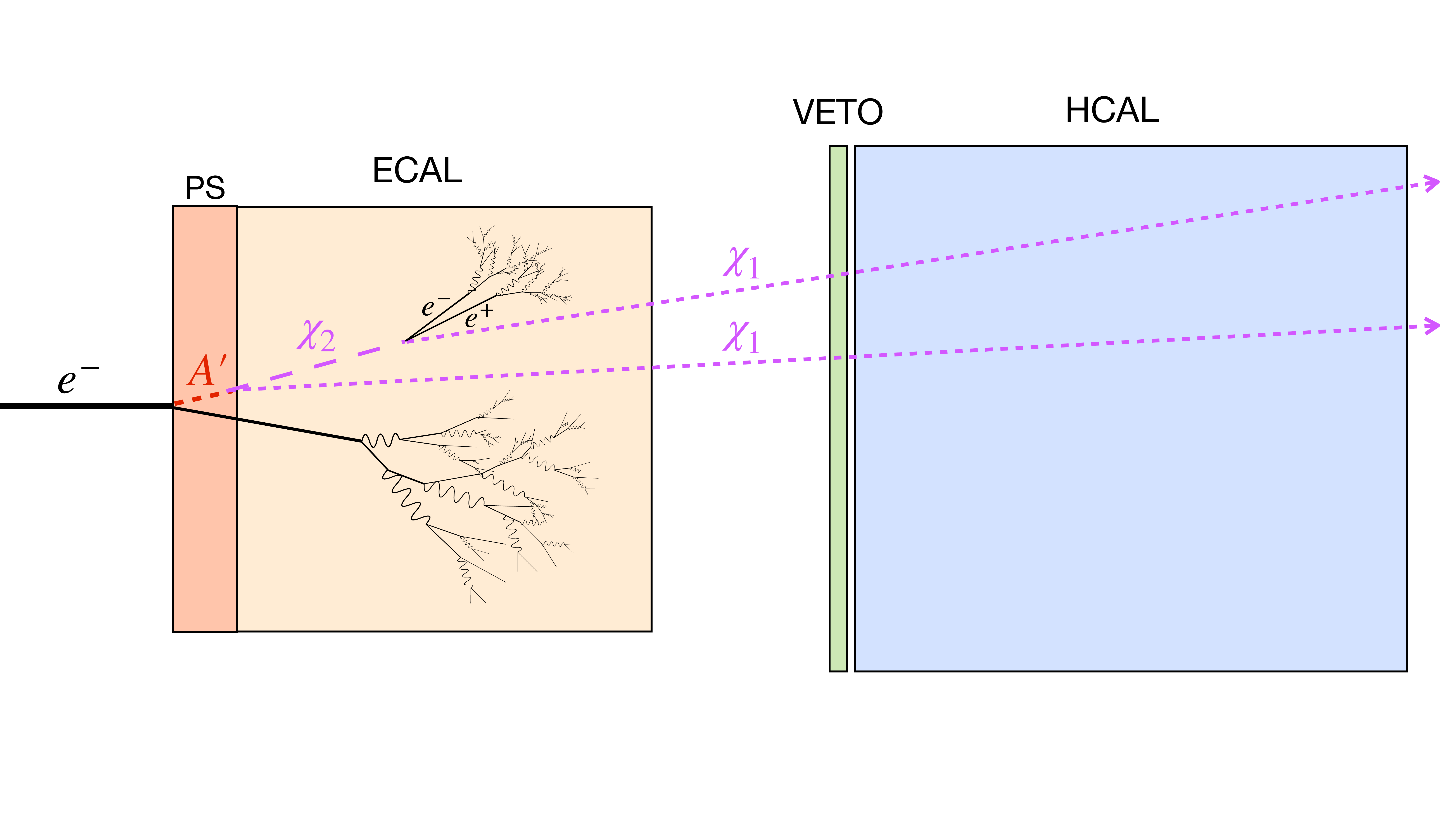}
\includegraphics[clip, trim=0 6cm 0cm 6cm, width=0.95\columnwidth]{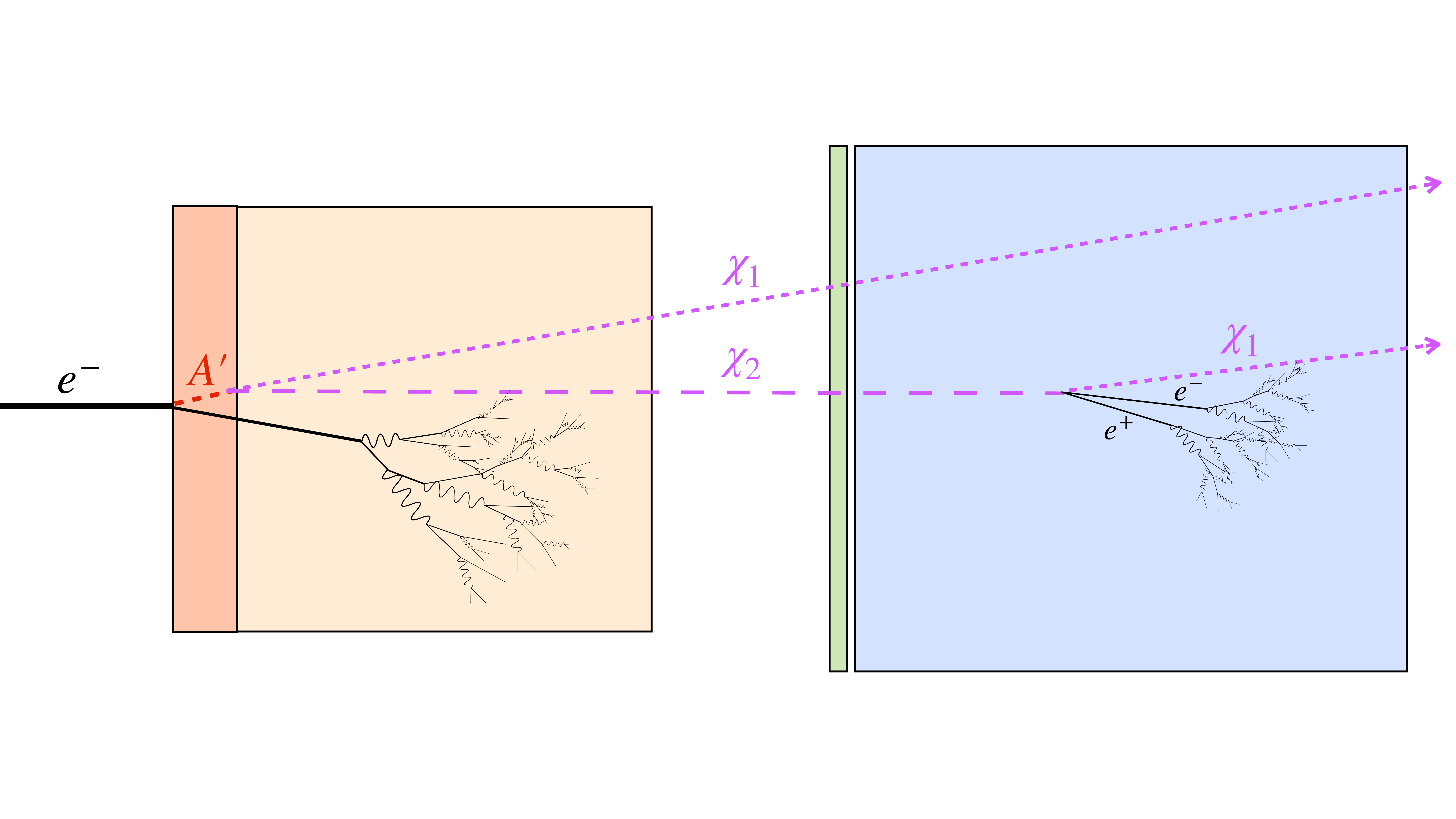}
\caption{Sketch of a semi-visible event in the NA64 setup. Note that only the detectors relevant for the discussion are depicted. The first drawing represents a prompt $\chi_2$ decay in the target, while in the second, $\chi_2$ has a longer lifetime and produces a leptonic pair in the HCAL.
\label{fig:sketchNA64}}
\end{figure}

The NA64 experiment employs a 100 GeV electron beam from the H4 beamline, a high momentum resolution secondary line fed by the SPS accelerator, where the beam purity was measured to be at the level $\pi^{-}/e^{-}\lessapprox10^{-2}$ \cite{Gninenko:2013rka}. To ensure the necessary precision in the determination of the incoming particles and suppress background from hadronic or low-energy $e^{-}$ contamination, the upstream part of the detector serves as a tagging system (see e.g. \cite{Banerjee:2019pds,Depero:2017mrr} for more details). 
The core of the experiment on which the $A'$ production relies is the electromagnetic calorimeter (ECAL). 
Due to the mentioned kinetic mixing, the electrons impinging on this dense target can undergo dark bremsstrahlung processes $e^- Z \rightarrow e^- Z A'$~\cite{Bjorken:2009mm} (Z being an active target nucleus), where a dark photon is radiated (see ~\cref{fig:decay}). The experiment sensitivity scales only with the $A'$ bremsstrahlung rate ($\propto\epsilon^2$), since the $A'$ observable signature is the missing energy carried away by the dark particles escaping the setup. 
This is a key figure that differentiates the active target method from beam-dump facilities, where the hypothetical existence of DM is inferred by seeking DM recoil reactions in a far calorimeter ($\propto\epsilon^2\alpha_D$), or, when investigating semi-visible models, by measuring the SM products from the heavier state down-scattering ($\propto\epsilon^2\alpha_D$). 
The total signal yield for these traditional approaches is then governed by $\epsilon^4\alpha_D$ and the experimental sensitivity is limited by the size of the detector. 
Besides this general advantage, in the NA64 context the radiated $A'$ is strongly boosted and it is likely to take over a significant fraction of the incident beam energy~\cite{Gninenko:2017yus,Liu:2017htz}. In semi-visible decays producing soft SM final states, this allows signal detection also in the case of short-lived DS particles leading to a visible energy deposit in the target, as depicted in the first sketch of \cref{fig:sketchNA64}.

Regarding the detector design, the ECAL is composed of 36 modules arranged in a $6\times6$ matrix. 
The transversal segmentation allows to exploits the shape of the electromagnetic shower to further reduce any possible hadronic background. 
Each module of the calorimeter is made of 150 alternating layers of lead absorber and scintillator material. 
In addition to the transversal, longitudinal segmentation is also present to strengthen the primary electron identification: the ECAL is subdivided into two segments, the first of which is called pre-shower (PS) and represents the first 16 layers of the entire detector.
Finally, to measure particles overcoming the $\sim40$ radiation lengths of the main target and to ensure the complete hermeticity needed to detect missing energy, a VETO counter, and three large hadronic calorimeters (HCAL) modules are installed after the ECAL. 
Each HCAL module is arranged in a $3\times3$ matrix with approximately $160$~cm of total length and consists of a sandwich of 48 stacked iron absorber-scintillator plates.

%%%%%%%%%%%%%%%%%%%%%%%%%%%%%%%%%%%%%
\subsection{The semi-visible recast analysis}\label{sec3b}

During the period 2016-2018, the NA64 experiment accumulated in the invisible mode a total of $2.84 \times 10^{11}$ electrons on target (EOT) and hereafter the resulting exclusion limits~\cite{NA64:2019auh} are reinterpreted in the context of the two inelastic models through a recast-based analysis. With this procedure, the already claimed absence of $A'$ observations in the original search can be translated to constraints in the parameter space of the iDM and i2DM models without a novel unblinding of the collected data.

As previously stated, the detection strategy in the invisible regime defines a signal event through missing energy and therefore the signal region is identified as ($E_{ECAL}<50$~GeV, $E_{HCAL}<1$~GeV). In addition, other selection criteria are applied to minimize possible background sources, and the candidate $A'$ event is identified by a list of requirements determined and optimized on studies based on simulations and data~\cite{NA64:2019auh}. 
To extend the sensitivity of invisible searches to test these hybrid-decay theories, a very accurate Monte Carlo (MC) signal simulation based on the Geant4 software~\cite{GEANT4:2002zbu} was employed to obtain the detector responses to semi-visible $A'$ signatures. 
The DM events generation relied on the DMG4 package~\cite{Bondi:2021nfp}, where both the iDM and i2DM decay models have been implemented in version 2.2. The analysis was conducted by correcting the signal efficiency provided by the MC simulation with the signal selection performance determined in the invisible search (see \refref{Andreev:2021fzd} for a comprehensive explanation). 
Moreover, due to the presence of the $e^+e^-$ pairs in the $A'$ decay chain that characterizes this peculiar mode, all the cuts effective after the primary $e^-$ hits the ECAL were applied to the specific signal simulation and accounted for in the signal expectation. To be selected, an event must exhibit a PS energy deposit and an ECAL shower profile consistent with one predicted for an incoming $e^-$, the energy recorded in the cells outside the central $3\times3$ matrix of the ECAL must constitute a small fraction of the total ECAL energy, no activity is expected in the VETO, and, finally, the candidate event is rejected if the energy deposition in the first HCAL periphery cells is higher than the one in the central cell. 

By imposing this correction and selection workflow on grid simulations, a parameter-dependent signal expectation $N_{A'}$ is obtained and the 90\% confidence level (C.L.) exclusion limits can be set. Given that the calculated boundaries have proven to be virtually unaltered when taking into consideration the expected number of backgrounds estimated in the invisible search $(0.53\pm0.17)$, the limits were calculated with a background-free hypothesis, excluding a specific parameter point with the simple condition $N_{A'/EOT} \times EOT \geq 2.3$.

This newly developed recast provides a complementary and enlarged study with respect to the first one described in~\refref{NA64:2021acr}, which focused on a specific iDM benchmark realization and exploited a distinct signature by looking at long-lived $\chi_2$ only, decaying either inside the last two HCAL modules or beyond the setup.

%%%%%%%%%%%%%%%%%%%%%%%%
\section{Other constraints on semi-visible dark photons}\label{sec4}
%%%%%%%%%%%%%%%%%%%%%%%%

The strongest existing constraints on semi-visible models come from both visible and invisible dark photon searches. However, as previously mentioned, the hybrid nature of the channel allows to relax the main exclusion limits.
Among these, the reinterpretation of the bounds set by the BaBar $A'$ invisible search~\cite{BaBar:2017tiz} is especially relevant as it represents the strongest probe in the large kinetic mixing region of the parameter space studied.
In BaBar, dark photons can be produced in $e^+ e^-$ collisions along with initial state radiation, $e^+ e^- \to \gamma A'$.
The production is followed by the $A'$ decay into a pair of dark fermions $A' \to \chi_{\{1,2\}} \chi_{\{1,2\}}$, where $\chi_2$ can further decay, producing $e^+ e^-$ pairs and depositing visible energy in the detector.
This semi-visible chain would resemble an invisible event in case of either long-lived $\chi_2$'s or collected energy below the energy cuts performed by the analysis.
Ref.~\cite{future} simulated the BaBar experiment and provided the recasted limits in the parameter space of iDM and i2DM models.
We make use of their results in our plots.

In addition to BaBar, other $A'$ experimental searches and several bounds of different origin set limits on the parameter space of the $U(1)_D$ models. 
We list all relevant constraints below:
\begin{itemize}
    \item  The CHARM and NuCal searches for long-lived particles~\cite{CHARM:1983ayi,Blumlein:1990ay,Blumlein:1991xh} have been recasted as constraints on inelastic dark matter using the decays in flight of $\chi_2$ into lepton-antilepton pairs in a downstream decay volume~\cite{Blumlein:2011mv,Gninenko:2012eq,deNiverville:2018hrc}.
    Both are proton beam-dump facilities and are designed with similar fiducial decay lengths, which is what determines the reach of these experiments. 
    We make use of the recast analysis of \refref{Tsai:2019buq}.
    For the benchmark that were not considered in \refref{Tsai:2019buq}, we apply a rescaling procedure to take into account the different lifetimes and production modes of $\chi_2$.
    \item The Liquid Scintillator Neutrino Detector (LSND) \cite{LSND:2001akn} is another high-intensity proton beam-dump placing tight constraints on MeV-scale DM~\cite{deNiverville:2011it}. 
    If the unstable $\chi_2$ is sufficiently long-lived, LSND can detect semi-visible $A'$ processes through $\chi_2$ down-scattering as well as via its decay signals, as  demonstrated in \refref{Izaguirre:2017bqb}.
    \item The SLAC E137 experiment~\cite{Bjorken:1988as} is instead an electron beam-dump that can measure the electromagnetic shower initiated either by leptonic pairs produced in $A'$ decays~\cite{Bjorken:2009mm} or via DM scatterings~\cite{Batell:2014mga}. 
    The recasted bound for iDM is computed in Ref.~\cite{Mohlabeng:2019vrz}.
    \item Other constraints can be obtained by direct production in meson decays, including limits from NA62 searches of $\pi^0\to \gamma A'$~\cite{NA62:2019meo} and E949 searches for $K\to \pi A'$, with $A'$ invisible~\cite{BNL-E949:2009dza,NA62:2020xlg}.
    \item The dark photon contribution to deep-inelastic-scattering (DIS) of electrons on nuclei has an impact on the measurement of parton distribution functions.
    Ref.~\cite{Kribs:2020vyk,Carrazza:2019sec,Thomas:2021lub} have derived the corresponding limits based on HERA data~\cite{H1:2015ubc}.
    \item Electro-weak precision observables (EWPO) set bounds on large kinetic mixing~\cite{Hook:2010tw,Curtin:2014cca}.
    \item The dark photon contribution to the electron $(g-2)_e$ can also be used to derive constraints on kinetic mixing~\cite{Pospelov:2008zw}. 
    \item Late annihilation of DM at the time of recombination can affect the cosmic microwave background (CMB) by injecting additional charged particles in the SM plasma. 
    Constraints on this late annihilation have been derived in the literature~\cite{Slatyer:2015jla}.
    However, they do not apply to the iDM model \cite{Izaguirre:2015zva,Izaguirre:2017bqb} because the only available DM annihilation channel involves co-annihilation with the heavier partner $\chi_2$.
    Since this one is unstable, its number density would drop, suppressing any indirect detection contribution at late time.
    These constraints do apply to the i2DM model but are relaxed due to $\sin^4{\theta}$ suppression in $\chi_1$ self-annihilations.
\end{itemize}
Note that some of these bounds will not be shown in the presented plots for the sake of clarity, as they cover regions already highly constrained by other $A'$ experimental searches shown in our figures. 
For an in-depth discussion on the re-interpretation of the constraints for semi-visible dark photons, see~\cite{future}.

%%%%%%%%%%%%%%%%%%%%%%%%%%%%%%%%%%%%%
\section{Bridging light thermal DM and the muon \texorpdfstring{$\gmu$}{(g-2)\textmu} puzzles}\label{sec5}
%%%%%%%%%%%%%%%%%%%%%%%%%%%%%%%%%%%%%

Besides the potential explanation of the DM nature, the unresolved muon dispute boosted the interest in a new boson mediating inelastic scatterings within the dark sector. With the increase of the mass splitting between the hypothetical DS states, the weakening of the invisible decaying dark photon bounds is enhanced by the larger phase space of the leptonic pairs.
By combining this feature with a thorough choice of the remaining free parameters of the models, an interesting parameter space region simultaneously resolving the
observed DM yield as a freeze-out relic, and the $(g-2)_\mu$ discrepancy arises.\\

%iDM results
%%%%%%%%%%%%%%%%%%%%%%%%%%%%%%%%%%%%%

\begin{figure}[b]
\centering
\includegraphics[trim=0.2cm 0cm 0.6cm 0cm, width=\columnwidth]{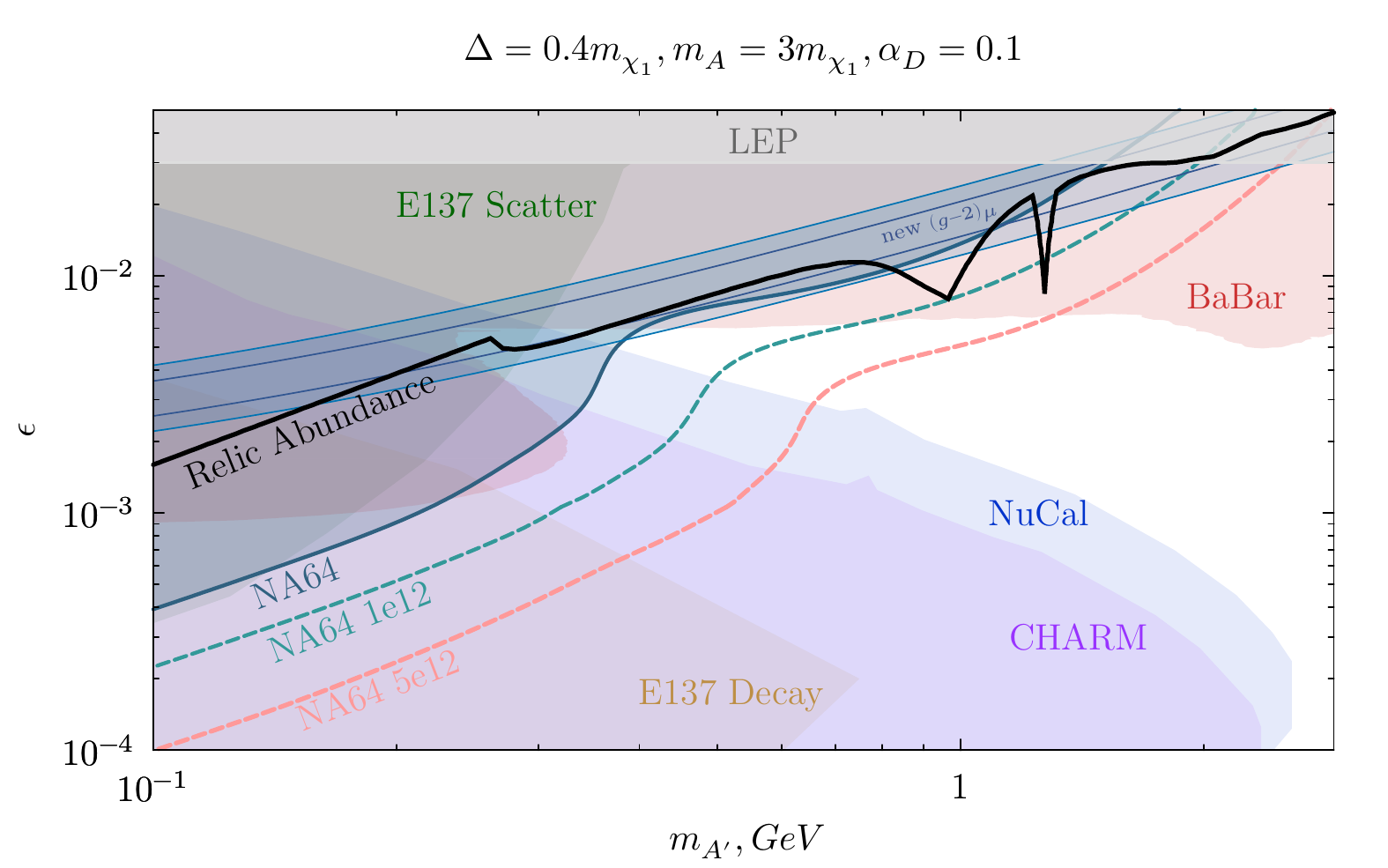}
\caption{\label{fig:iDMmuon_ma_vs_eps}Existing constraints for the $A'$ mediated iDM model in the ($m_{A'},\epsilon$) plane, fixing$\Delta/m_{\chi_1}=0.4$, $m_{A'}/m_{\chi_{1}}=3$ and $\alpha_D=0.1$. The NA64 90\% C.L. exclusions evaluated from the new recast are shown as a blue-shaded area and the projections for two different future statistics are drawn with dashed curves: the total EOT accumulated considering the new 2022 data ($10^{12}$ EOT) and the $5\times 10^{12}$ EOT expected before LS3. The BaBar exclusion limits~\cite{future} are depicted in red and the restrictions from E137~\cite{Mohlabeng:2019vrz}, NuCal, and CHARM~\cite{Tsai:2019buq} are also included. The model-independent area probed at LEP was estimated in Ref.~\cite{Hook:2010tw} to be $\epsilon \gtrsim0.03$ and is shown in grey. The blue band indicates the region of parameter space favored for the $(g-2)_\mu$ explanation (the internal blue curves stand for the 2$\sigma$ deviation considering the latest measurement performed at Fermilab~\cite{Muong-2:2021ojo} while the external lines correspond to the $2\sigma$ band from the BNL E821 result~\cite{Muong-2:2006rrc}) and the thick black line represents the combination of parameters compatible with a DM thermal relic scenario.}
\end{figure}

\emph{iDM --- } For the minimal iDM model, this was realized in \refref{Mohlabeng:2019vrz}, where the constraints in ($m_{A'},\epsilon$) plane for an appealing benchmark solution are presented, together with an analysis tuned to explain the $(g-2)_\mu$ anomaly, studying the limits' behavior for different dark coupling constants $\alpha_D$ and mass splittings~$\Delta$. Following the same idea, the outcome of the performed recast analysis on the NA64 bounds described in \cref{sec3b} is summarized in~\cref{fig:iDMmuon_ma_vs_eps} in the plane spanning the mediator mass and kinetic mixing strength, by imposing $\Delta/m_{\chi_1}=0.4$ and the commonly selected $\alpha_D=0.1$ and $m_{A'}/m_{\chi_{1}}=3$.

Remarkably, the NA64 refined limits exclude the 2$\sigma$ deviation band calculated by including the latest $(g-2)_\mu$ measurement performed at Fermilab up to $m_{A'}\sim1.14$ GeV. The comparison of this mass point with the $\sim0.39$ GeV value obtained in the pioneering NA64 semi-visible recast~\cite{NA64:2021acr} suggests that this alternative analysis method allows probing a significantly larger portion of parameter space corresponding to a short-lived $\chi_2$. The $\chi_2$ decay width scales in fact as~$\propto\epsilon^2m_{A'}$ and for the iDM case it is given by \cite{Mohlabeng:2019vrz,Izaguirre:2015zva}
\begin{equation} \label{width}
    \Gamma_{iDM}(\chi_2\rightarrow\chi_1 e^+e^-)\simeq \frac{4\epsilon^2\alpha\alpha_D\Delta^5}{15\pi m_{A'}^4}.
\end{equation}
In our analysis, we apply a rescaling factor to account for full dependence on the final state masses~\cite{NA64:2021acr}.

The previous NA64 exclusions were attained by seeking semi-visible displaced vertices on macroscopic distances or by looking for long-lived $\chi_2$ states decaying after the setup.
\begin{figure*}[t!]
\centering
\includegraphics[trim=1cm 0cm 0.2cm 0cm,width=\columnwidth]{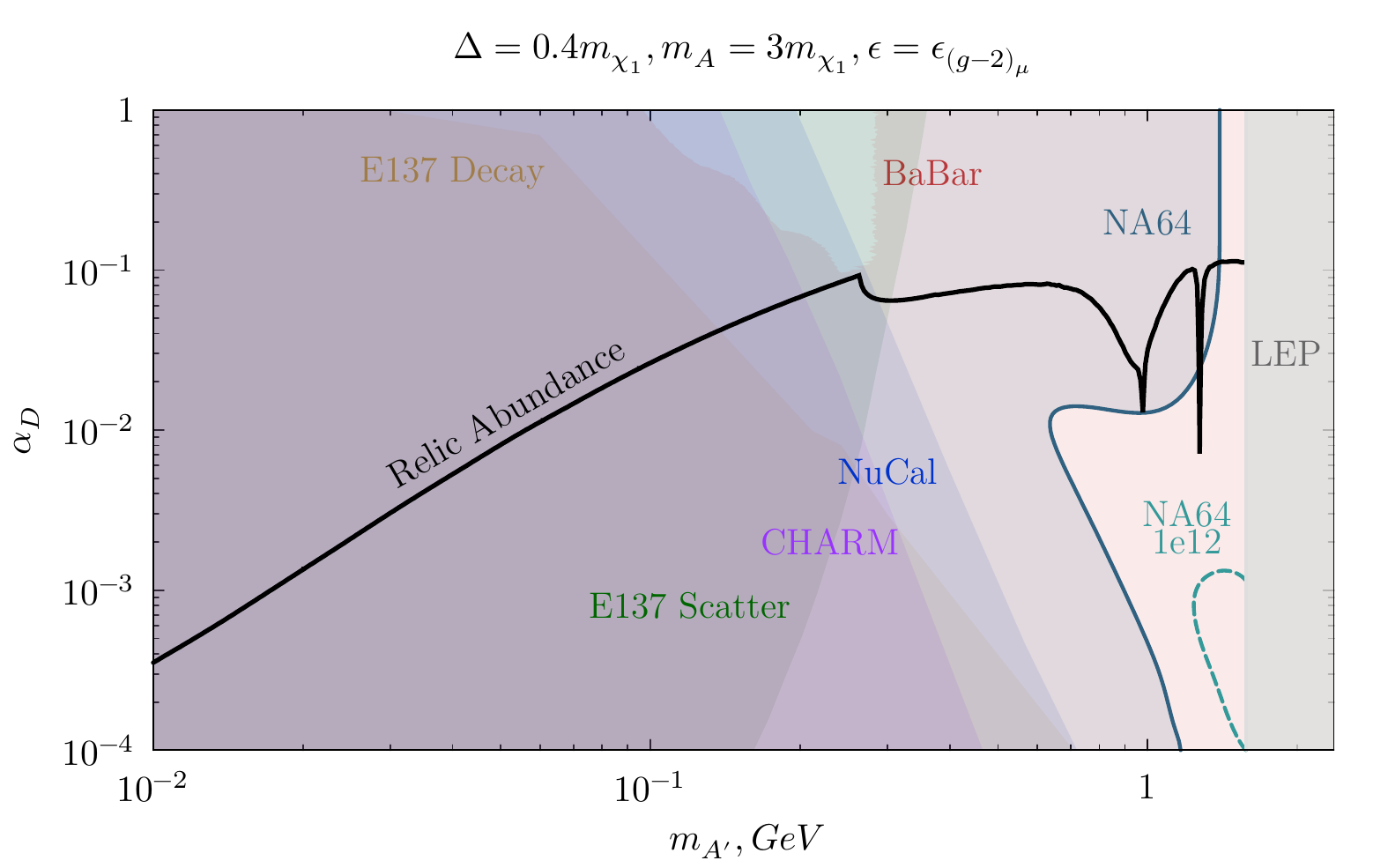}
\includegraphics[trim=0.2cm 0cm 1cm 0cm,width=\columnwidth]{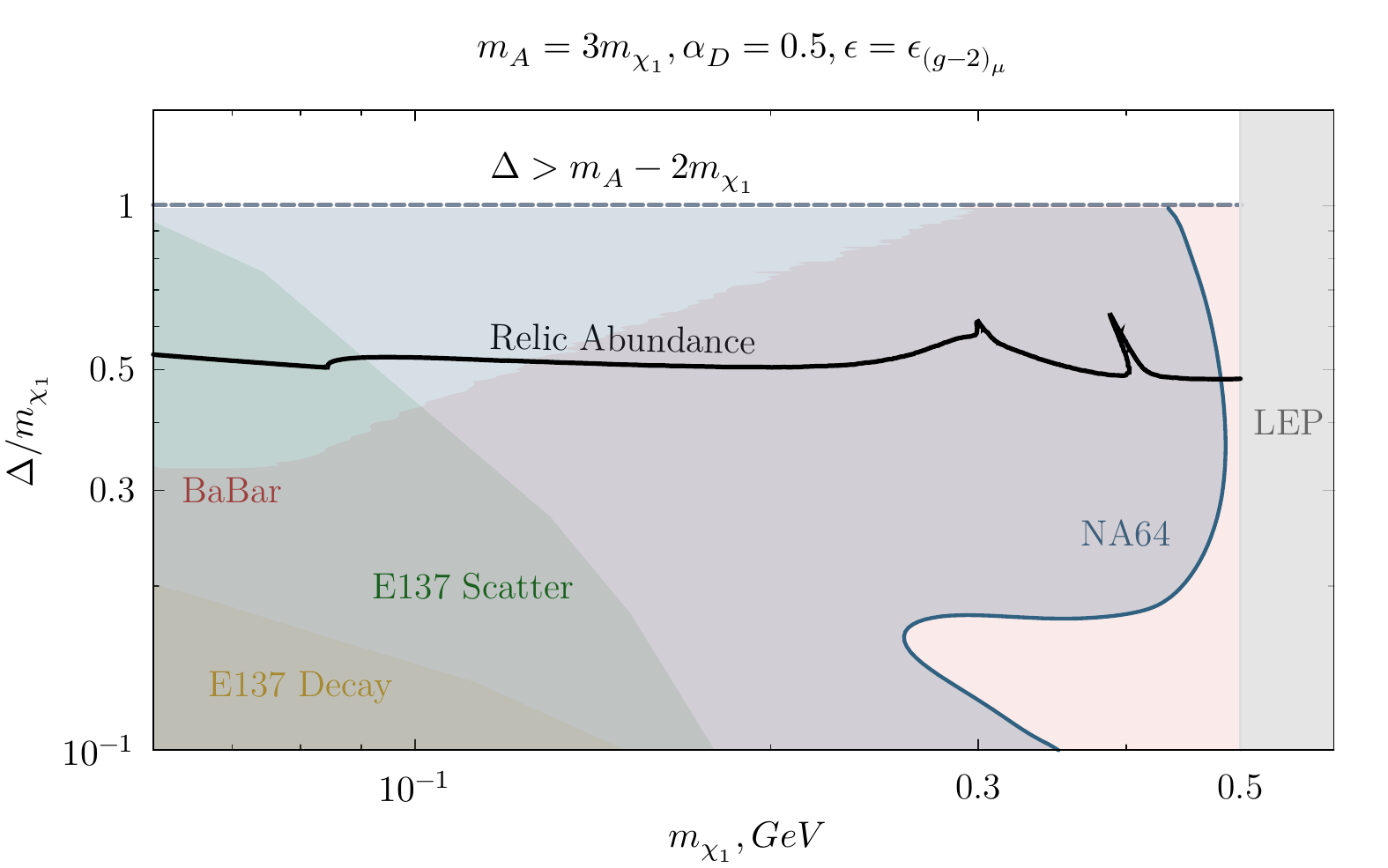}
\caption{\label{fig:iDMmuon_otherplanes} Constraints on the iDM model for $\epsilon=\epsilon_{(g-2)_\mu}$. Left: the plane of dark coupling $\alpha_D$ versus dark photon mass $m_{A'}$. Right: mass splitting $\Delta/m_{\chi_1}$ versus the DM candidate mass $m_{\chi_1}$ parameter space.
In both planes, we represent the relic density as a solid black line. The NA64 90\% C.L. limits are shown with a solid blue boundary and the projected one for the accumulated statistics to date ($10^{12}$ EOT) is displayed as a dashed curve.
}
\end{figure*}
This strategy restricted the NA64 sensitivity to a specific decay length range of the unstable state ($d_{\chi_2}\gtrapprox2.5$~m, considering $A'$ decays close to the production point), preventing detection in the prompt decay region which corresponds to the upper-right part of~\cref{fig:iDMmuon_ma_vs_eps}; an area which is explored here. 
Combined with the revised BaBar boundary, both the internal and external $(g-2)_\mu$ bands for the full mass range are essentially ruled out, practically discrediting this $A'$ model point as a candidate explanation to the $(g-2)_\mu$.

The possible combinations of $m_{A'}$ and $\epsilon$ that would generate the observed relic abundance for a pseudo-Dirac iDM candidate are illustrated with a black line in the figure.
We use our own calculation of the relic density in all of the plots shown. 
For masses below $m_{\chi_1}\sim 0.17$~GeV we find good agreement with and between the literature~\cite{Berlin:2018bsc,Mohlabeng:2019vrz,Izaguirre:2017bqb, Filimonova:2022pkj,Duerr:2019dmv,Tsai:2019buq} to below $20\%$.
The here proposed recasts extend the coverage of this curve to the higher mass region. With the examined statistics, NA64 probes this target up to $m_{A'}\sim0.82$~GeV, point above which the BaBar restrictions apply.

The fluctuation feature present in the shape of the NA64 contour in the mass range ${m_{A'}\sim0.3-0.8}$~GeV  is due to the presence of the $e^+e^-$ pair in the $A'$ decay chain.
The visible particles leaking out of the ECAL or being produced afterward within the setup are detected by the VETO counter or in the HCAL, as shown in the last sketch of~\cref{fig:sketchNA64}. This leads to a rejected event by the requirement of null activity in the modules, a selection criteria imposed in the original invisible analysis to mitigate hadronic background. 
The area above the feature is characterized by fast decaying $\chi_2$, where $e^+e^-$ are contained in the main target (first sketch of~\cref{fig:sketchNA64}). 
In this case, being the lepton pairs mostly soft, the additional visible energy deposited in the ECAL has a reduced impact on the missing energy selection and, consequently, on the coverage. 
The region underneath the feature corresponds instead to invisible signatures, given that for these parameter settings, the SM particles would be produced after the experimental setup, and the long-lived $\chi_2$ mimics a stable state at the experiment's scale. 

The loss in the semi-visible coverage due to medium-living $\chi_2$'s decaying between the final ECAL part and the end of the setup can be restored through a modification of the signal event definition. For instance, this is possible by combining the search for missing energy in the ECAL with the displaced leptons approach, exploiting both the visible and invisible energy distinctive of this channel. 
The NA64 sensitivity to semi-visible $A'$ could also benefit from the study of a secondary dark boson production method, namely the $A'$ resonant annihilation of secondary shower positrons with atomic electrons. This mechanism was successfully employed in the NA64 invisible search and boosted the experimental reach in the $m_{A'}\sim0.2-0.3$ GeV window~\cite{Andreev:2021fzd}. A dedicated semi-visible analysis will be carried out on the larger sample of EOT collected in 2022, with the purpose of providing a wider and more robust investigation of these models through an optimized and complete analysis.

In addition to the current bounds, the NA64 projections for the future statistics, namely the total cumulative number of EOT including the data collected in 2022 (amounting to $\sim10^{12}$ EOT) and the foreseen prospects before the forthcoming LHC shutdown (LS3) starting in 2026 ($5\times 10^{12}$ EOT), are shown with dashed lines. 
By performing the same data analysis without any optimization, with $10^{12}$ EOT only the tip of one spike characterizing the relic line at higher masses (given by $\chi_1\chi_2$ co-annihilation to hadronic final states \cite{Izaguirre:2015zva,Ilten:2018crw}) would remain unprobed, while an increase in the statistics by a factor 5 would allow testing the entire relevant region of parameter space.

\cref{fig:iDMmuon_otherplanes} focuses specifically on the $(g-2)_\mu$ anomaly by enforcing the kinetic mixing strength $\epsilon$ to the central value of the band~\cite{Muong-2:2021ojo}. The left plot depicts the parameter space spanned by varying the dark coupling $\alpha_D$ and the mediator mass $m_{A'}$, whereas the right one shows the fractional mass splitting $\Delta/m_{\chi_1}$ versus the DM candidate mass $m_{\chi_1}$ plane. For both cases, the blue-shaded area showcasing the NA64 exclusion sets strong constraints by covering the whole vertical axis range and the relic target up to high mass values. In this region, the fluctuation in the limits characteristic of the signal definition employed reveals a small open area which is, however, discarded by the BaBar study.

Summing up the new findings on the reviewed parameters restrictions for the iDM model, it can be asserted that this scenario is almost completely excluded as a feasible interpretation of the muon contentious results.
\bigskip

%i2DM results
%%%%%%%%%%%%%%%%%%%%%%%%%%%%%%%%%%%%%
\begin{figure}[b]
\centering
\includegraphics[ trim=0.6cm 0cm 0.6cm 0cm,width=\columnwidth]{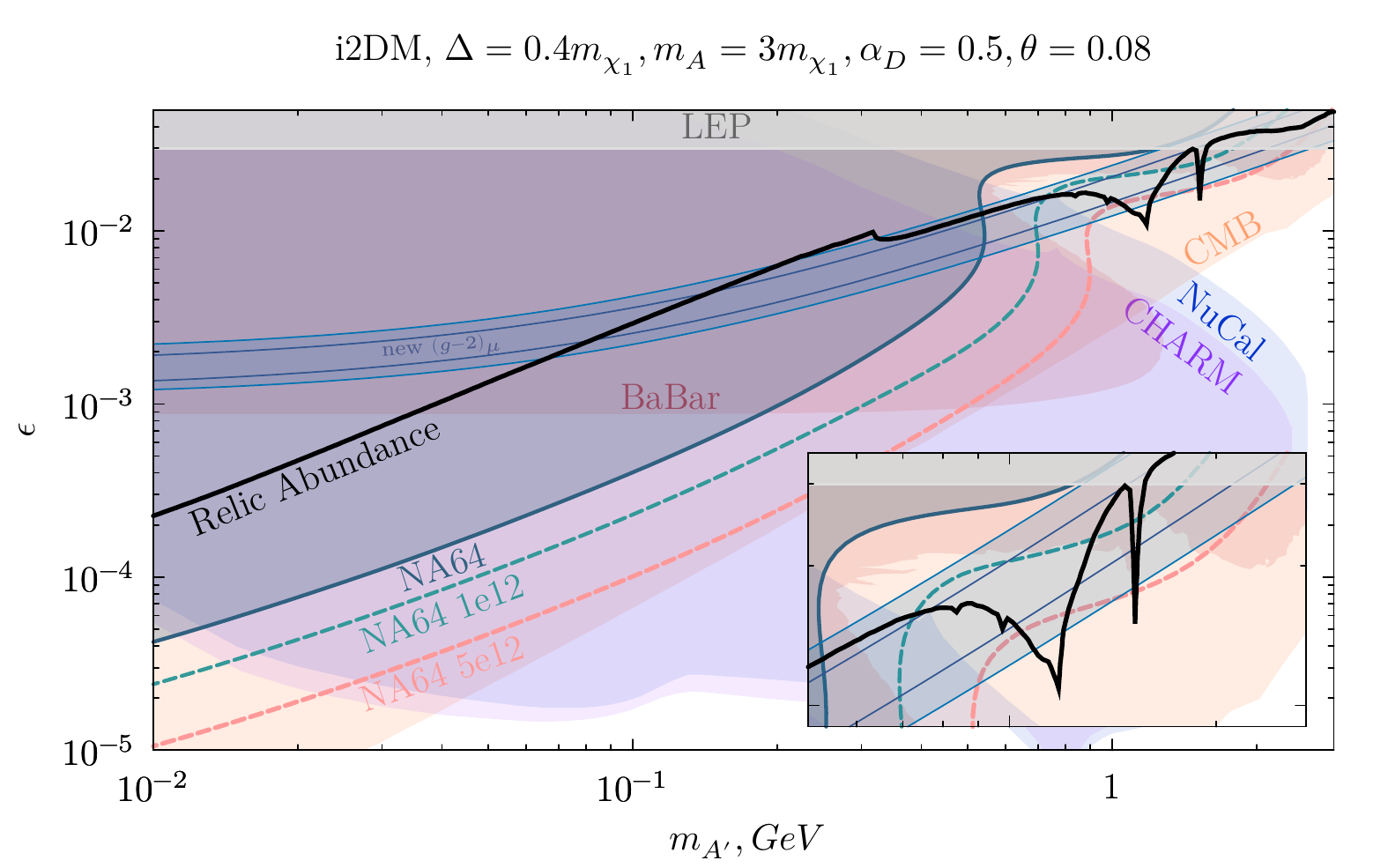}
\includegraphics[trim=0.6cm 0cm 0.6cm 0cm,width=\columnwidth]{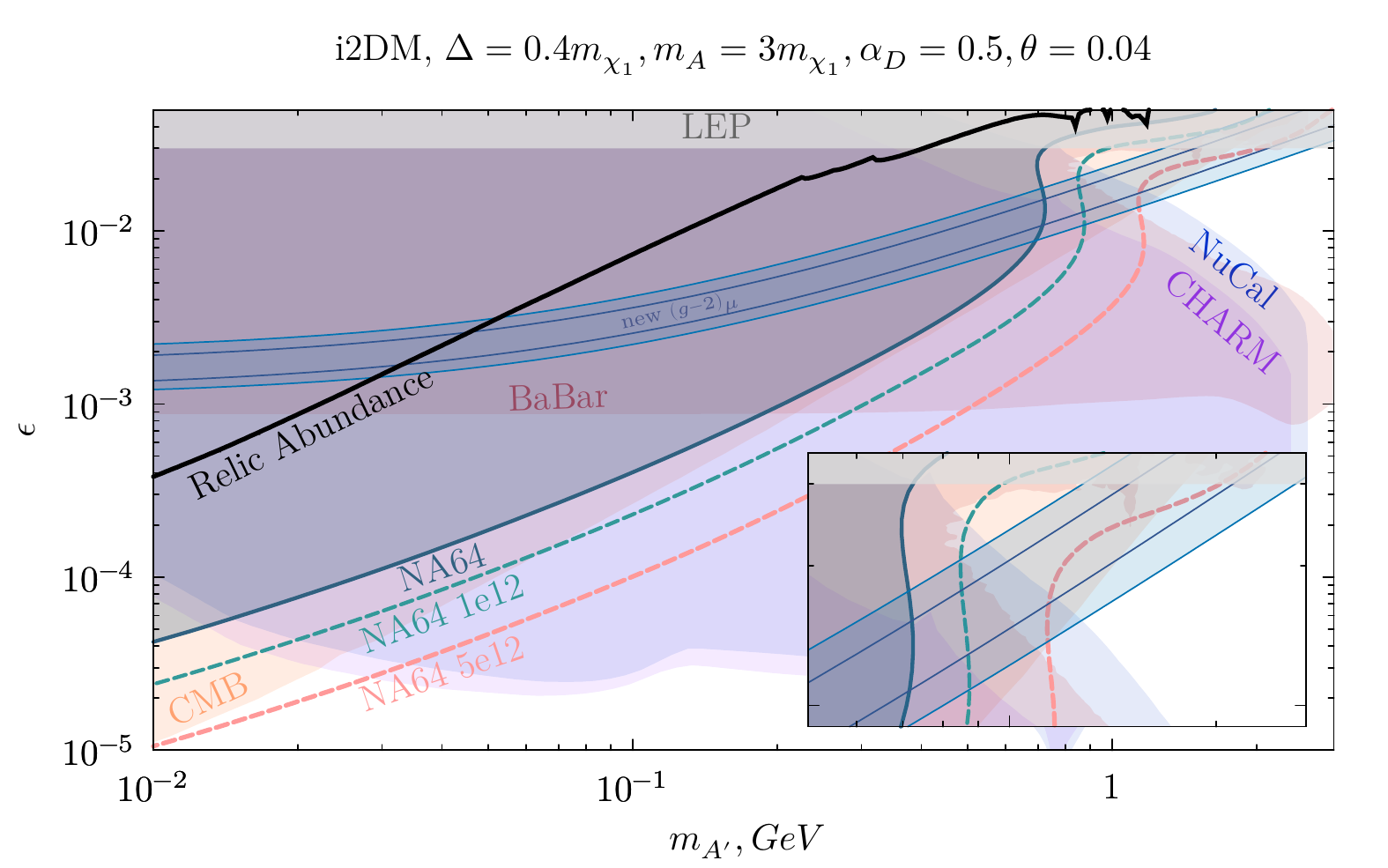}
\caption{\label{fig:i2DMmuon_ma_vs_eps}i2DM parameter space for two specific model points focused on a potential explanation of the $(g-2)_\mu$ anomaly, presenting constraints and projections in the  ($m_{A'},\epsilon$) plane. The NA64 current exclusion limits are shown with the blue colored area, and the sensitivities for two statistics ($10^{12}$ EOT and $5\times 10^{12}$ EOT) are drawn with dashed curves.}
\end{figure}

\emph{i2DM --- } The i2DM framework was suggested as an alternative theory to address light thermal dark matter and recover a plausible explanation of the $(g-2)_\mu$ anomaly.
To enhance the amount of visible energy present in this $A'$ model, the benchmark realizations proposed here require a very small $\theta$ value (see \cref{eq2.5}), so that, given the dark photon branching ratios~\cite{Filimonova:2022pkj}
\begin{align}
    Br(A'\rightarrow\chi_1\chi_1) &\propto \sin^4{\theta},
    \\
    Br(A'\rightarrow\chi_1\chi_2) &\propto \sin^2{\theta}\cos^2{\theta},
    \\
    Br(A'\rightarrow\chi_2\chi_2) &\propto \cos^4{\theta},
\end{align}
the decay is driven by the channel producing two unstable $\chi_2$ states, thus leading to two pairs of SM final states. 

As a feasibility study, we present in~\cref{fig:i2DMmuon_ma_vs_eps} the first derived constraints and projections in the ($m_{A'},\epsilon$) plane. 
As illustrated in the first plot, the value $\theta=8\times 10^{-2}$ allows for a simultaneous explanation of the DM abundance and the $(g-2)_\mu$ discrepancy. Nevertheless, the bounds calculated from CMB data~\cite{Slatyer:2015jla} stretch through the entire region of interest, rendering this solution unviable. 
Smaller mixing angles, such as $\theta=4\times 10^{-2}$ chosen for the second panel, suppress the late self-annihilation of the DM particle and loosen the CMB constraints.
The shift of the relic line to larger kinetic mixing strengths is, however, an inevitable consequence: the co-annihilation processes between $\chi_1$ and $\chi_2$ are suppressed by a lower $\theta$ value, implying a higher $\chi_1\chi_1$ annihilation rate to avoid an overabundant scenario.
This renders the $(g-2)_\mu$ band and the relic line irreconcilable over the window opening in the semi-visible constraints.

The i2DM regime is thereby unable to provide an admissible interpretation of the $(g-2)_\mu$ discrepancy. 
Regardless of such finding, the study of this model demonstrates the versatility of NA64 in the search for rich dark sector models. Several other complex scenarios, involving more generations of unstable DS particles and thus more visible final states, e.g. heavy neutrinos theories \cite{future}, can be studied at NA64 by means of an improved analysis and with the milestone statistics planned to be collected before LS3.

%%%%%%%%%%%%%%%%%%%%%%%%%%%%%%%%%%%%%
\section{Probing the thermal inelastic target}\label{sec6}
%%%%%%%%%%%%%%%%%%%%%%%%%%%%%%%%%%%%%

This section aims to examine parametrically different light dark matter scenarios by illustrating the existing constraints to both thermal co-annihilating iDM and i2DM and by assessing the NA64 discovery potential for these searches.\\

\begin{figure*}[t]
\centering
  \includegraphics[trim=0.6cm 0cm 0.2cm 0cm,width=\columnwidth]{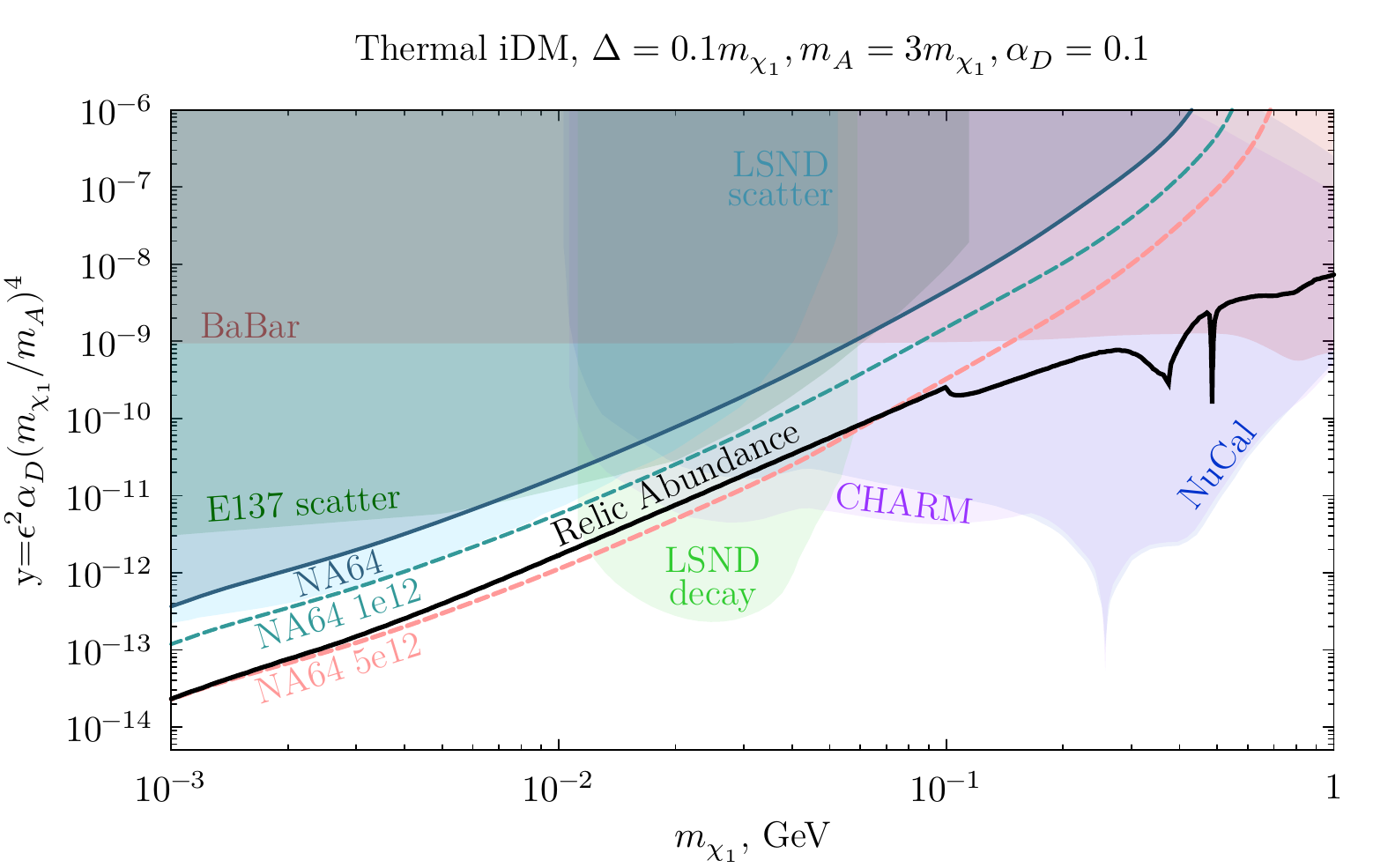}
  \includegraphics[trim=0.2cm 0cm 0.6cm 0cm,width=\columnwidth]{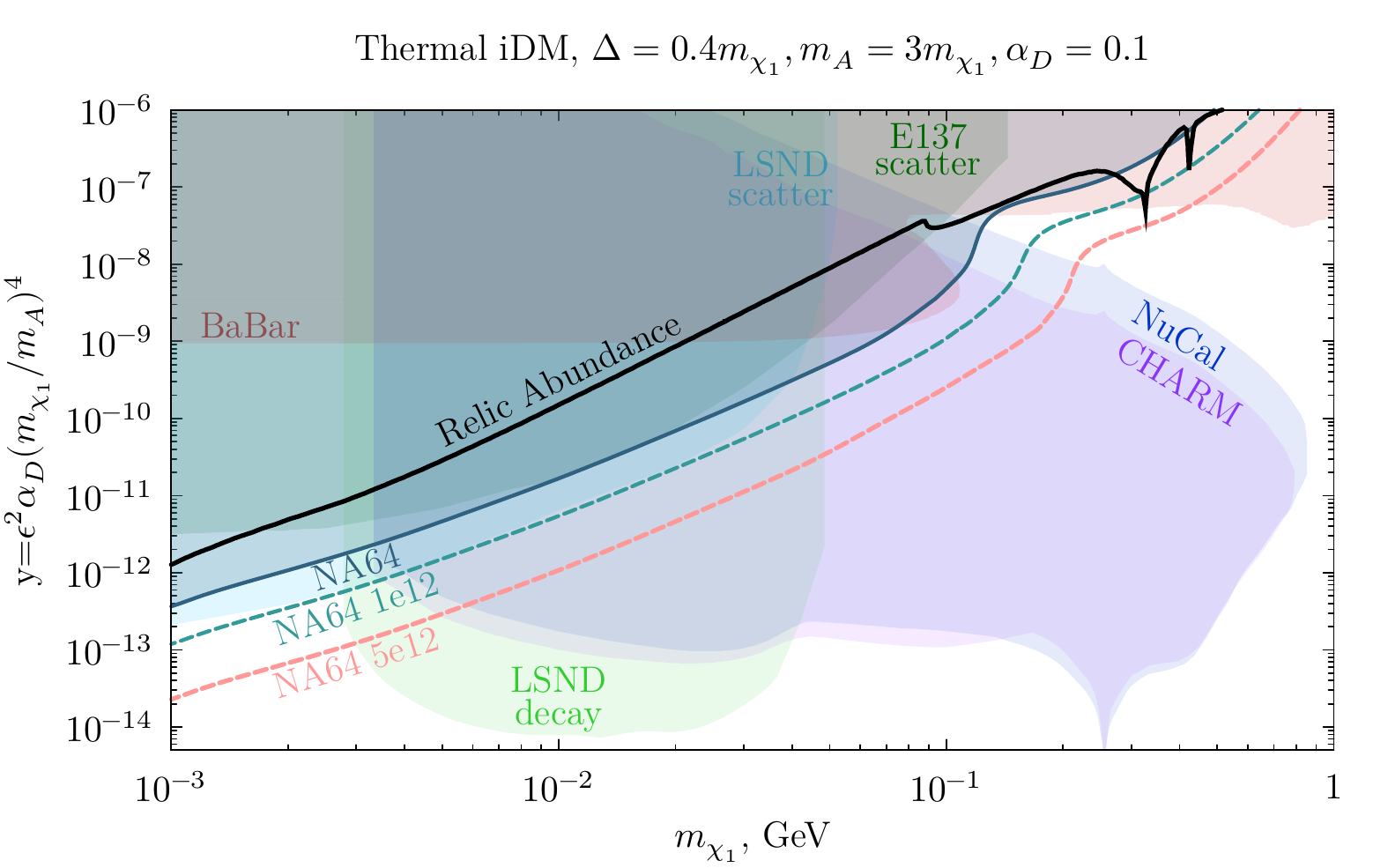}\\
    \includegraphics[trim=0.6cm 0cm 0.2cm 0cm,width=\columnwidth]{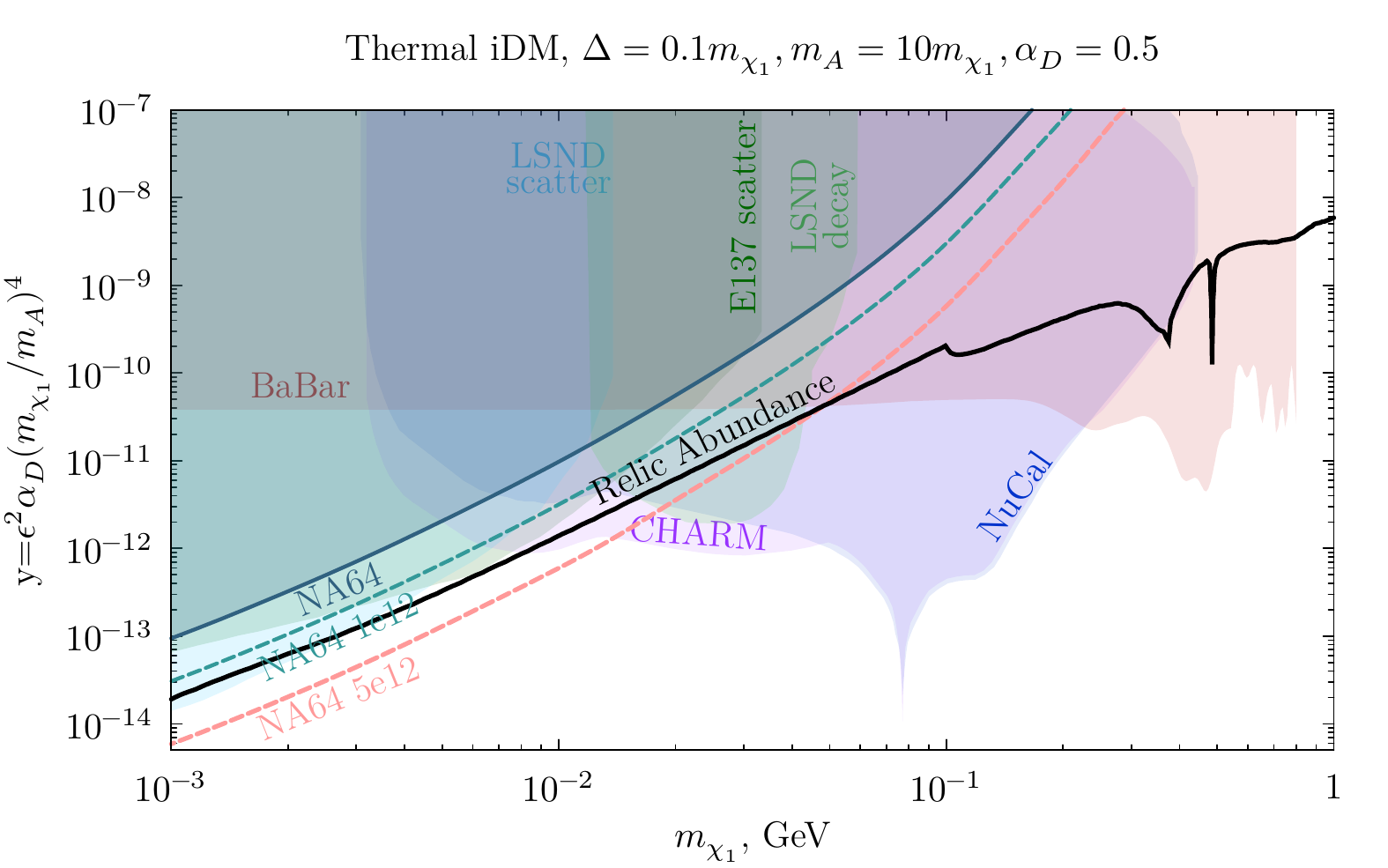}
    \includegraphics[trim=0.2cm 0cm 0.6cm 0cm,width=\columnwidth]{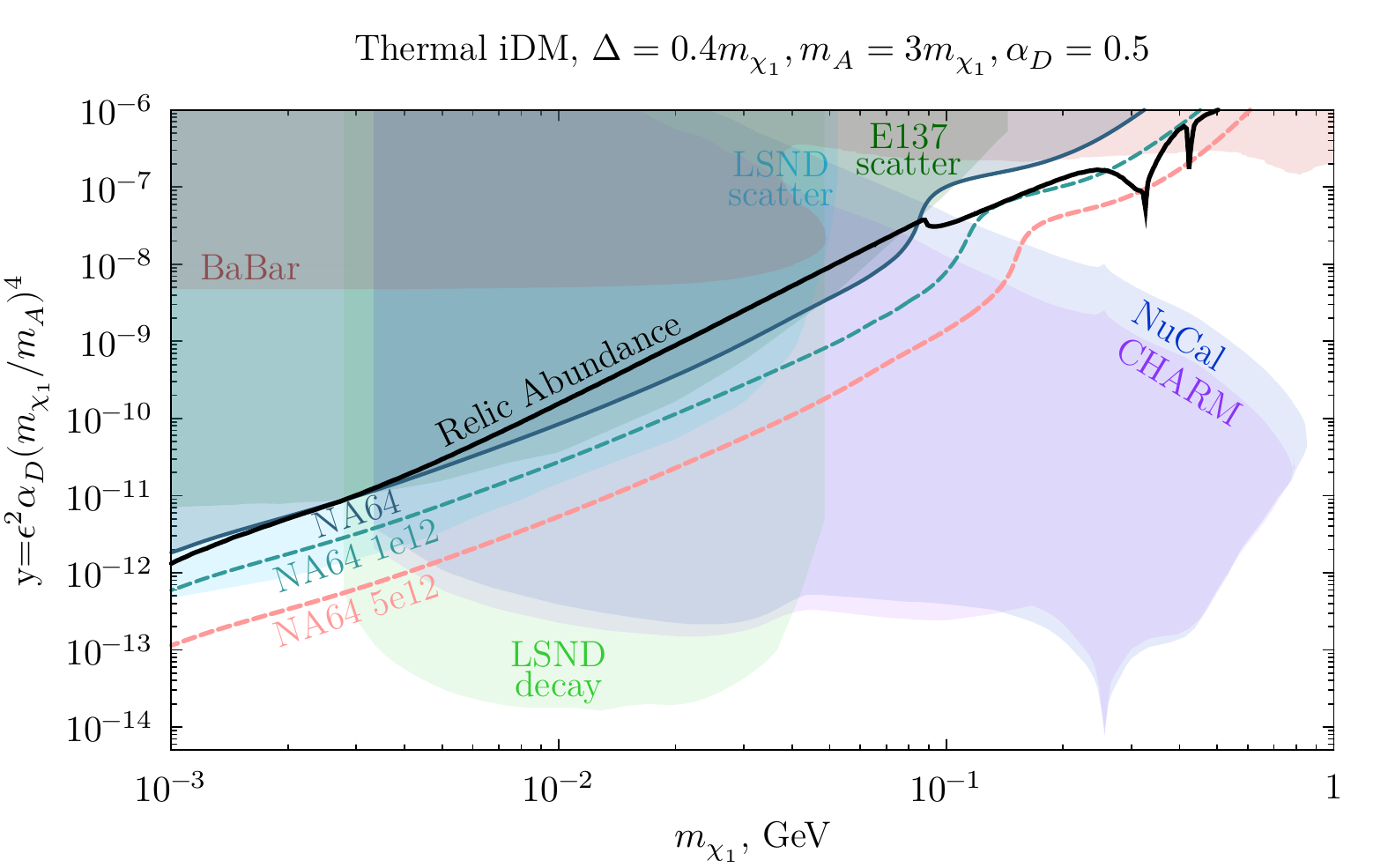}
 \caption{Current exclusions and future NA64 projections for fermionic iDM in the parameter space compatible with the thermal target for different combinations of $\Delta$, $\alpha_D$ and $m_{A'}/m_{\chi_{1}}$. The boundaries from E137 and LSND data are taken from \refref{Izaguirre:2017bqb} (to obtain the beam-dumps limits for the two plots in the bottom row the lines were rescaled). The NA64 and BaBar constraints and the NA64 future sensitivities depicted for two different statistics were instead derived in this work\label{fig:iDM_m1_vs_y}}
\end{figure*}

%iDM results
%%%%%%%%%%%%%%%%%%%%%%%%%%%%%%%%%%%%%
\emph{iDM --- }\cref{fig:iDM_m1_vs_y} quantifies the NA64 exclusion bounds and its sensitivity on iDM for four distinct combinations of $\alpha_D$, $\Delta/m_{\chi_1}$ and $m_{A'}/m_{\chi_{1}}$ values and compares them to the parameters setting the observed relic abundance and to the existing experimental limits. 
The contours are shown in the $m_{\chi_1}-y$ plane, where $y$ is a dimensionless variable defined as $y=\epsilon^2\alpha_D(m_{\chi_1}/m_{A'})^4$ that reduces the large dimensionality of the model parameter space. 
This choice is also dictated by a facilitated comparison with the DM target, given that the tree-level annihilation cross section determining the freeze-out rate is proportional to~$y$~\cite{Izaguirre:2015yja}. 
The relic line in the variable $y$ remains invariant upon different choices of $\epsilon$, $\alpha_D$, and $m_{A'}/m_{\chi_{1}}$. 
For co-annihilation, however, the number density of the co-annihilator is exponentially suppressed for large mass splitting $\Delta$.
The increase of $\Delta$ makes the co-annihilation less efficient, and, therefore, dark matter more over-abundant.
As a result, a larger kinetic mixing is required to achieve the correct abundance, improving the experimental prospects for discovery.
The $\Delta/m_{\chi_1}=0.4$ scenarios are shown on the right panels of \cref{fig:iDM_m1_vs_y}, and prove that the thermal target is already tightly constrained in this case. 
Given that the NA64 sensitivity scales as $\epsilon^2$, when translating the bounds to the $y$ variable, there is an overall linear shift by a factor $\alpha_D$, as can be seen by comparing the two panels on the right column. 
In fact, the $\alpha_D=0.1$,  $m_{A'}/m_{\chi_{1}}=3$, $\Delta/m_{\chi_1}=0.4$ relic curve is essentially excluded, as discussed in~\cref{sec5}, whereas in the $\alpha_D=0.5$ case, an open window above $m_{\chi_1}>110$ MeV emerges, also due to the reduced $\chi_2$ lifetime (see~\cref{width}) that shifts the fluctuation feature in the NA64 boundary to lower $m_{A'}$ and $\epsilon$. 
NA64 has a partial coverage of the unprobed parameter space below $m_{\chi_1}\sim 0.25$ GeV with an optimized analysis on the new 2022 data, and almost full coverage of the relic target for the milestone $5\times 10^{12}$ EOT statistics.
In conclusion, NA64 alone can probe the full mass range of this thermal light iDM parameter space within a reasonable time frame.
%\mh{IN THE MIDDLE OF THIS PARAGRAPH, WE NEED A STATEMENT ABOUT THE NUCAL AND CHARM.}

In the case of a spectrum with reduced splitting, where $\Delta/m_{\chi_1}=0.1$, the overall y-reach of the bounds is severely weakened. 
In the upper-left panel of \cref{fig:iDM_m1_vs_y}, the beam-dump constraints span only the $m_{\chi_1}\gtrsim12$~MeV part of the thermal relic line. NA64 can provide complementary coverage at low masses, approaching the relic density curve in the ${m_{\chi_1}\sim1-100}$~MeV region for thermal co-annihilating iDM by the start of LS3. 

\begin{figure*}[t!]
\centering
    \includegraphics[trim=1cm 0cm 0.2cm 0cm,width=\columnwidth]{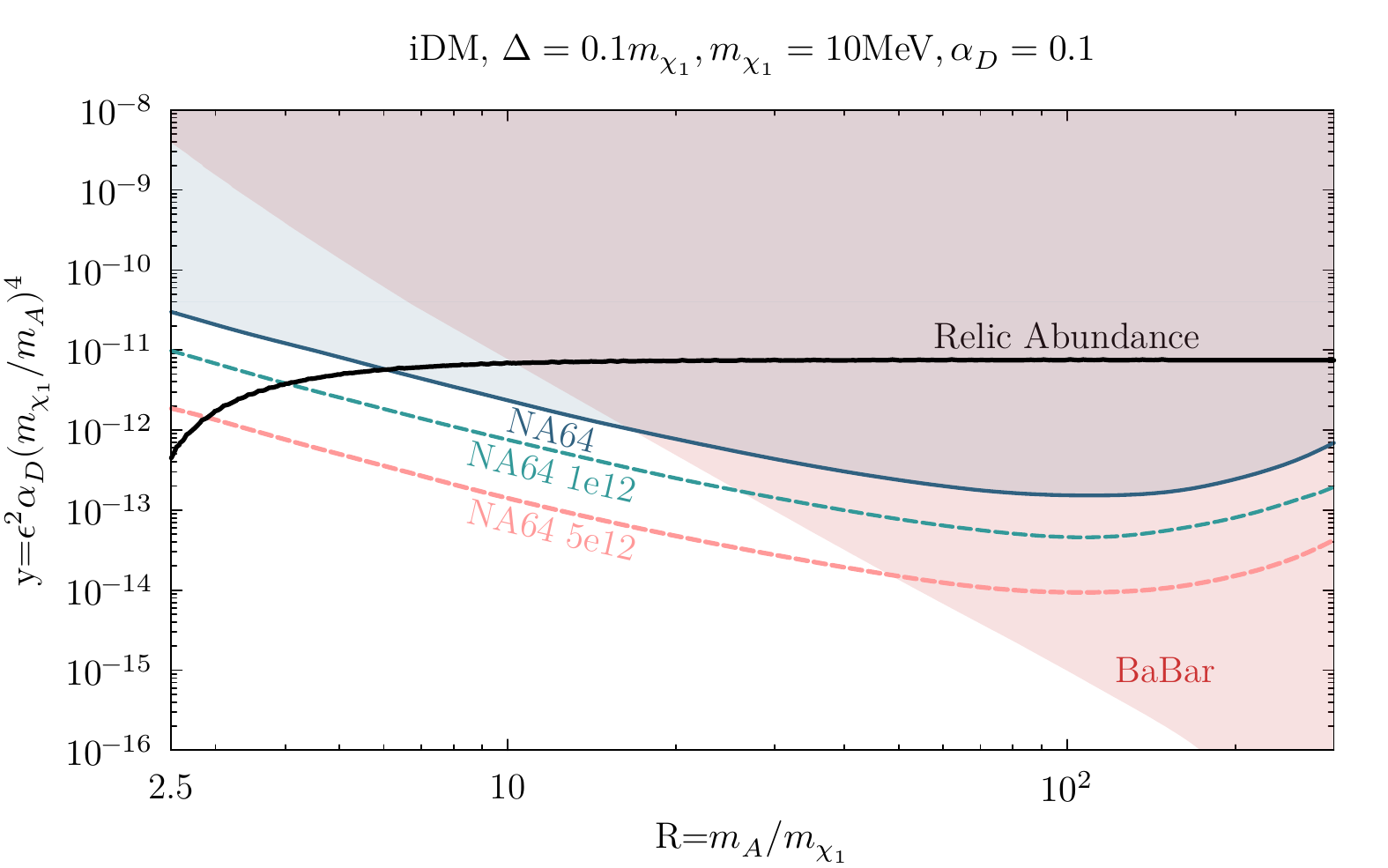}
    \includegraphics[trim=0.2cm 0cm 1cm 0cm,width=\columnwidth]{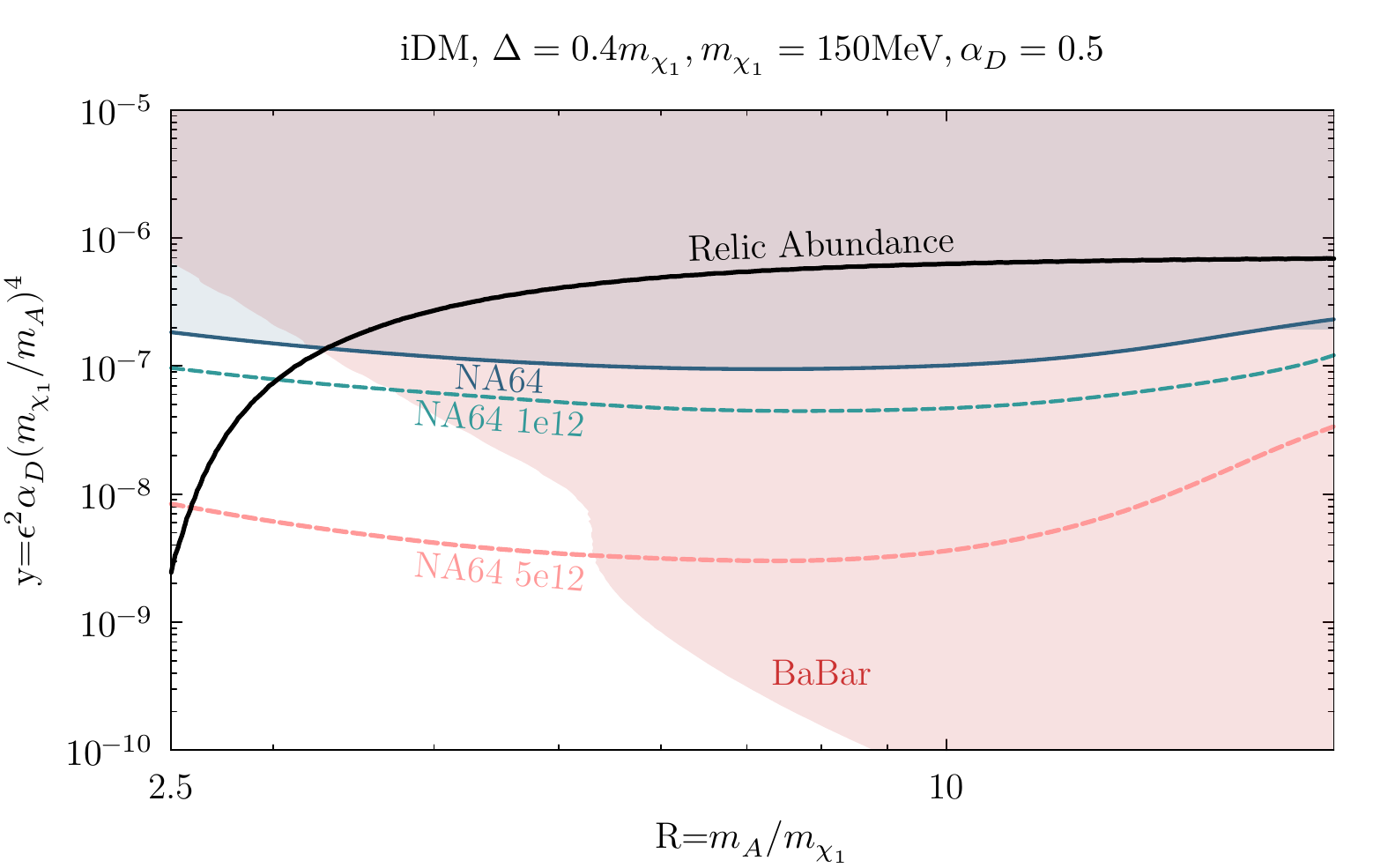}
\caption{NA64 exclusion limits and projections in the R-y plane for two different model points. Note that both the plots' x-ranges start from $R=2.5$ to avoid the resonance region.}
\label{fig:iDM_R_vs_y}
\end{figure*}
Finally, the last degree of freedom that influences the existing limits is the ratio between the mediator and the DM candidate masses. 
The bottom left plot of \cref{fig:iDM_m1_vs_y} illustrates the change in the existing constraints when this ratio is increased w.r.t. the commonly used value $R=m_{A'}/m_{\chi_{1}}=3$ (this parameter slice is usually chosen to yield a conservative evaluation of the experimental sensitivities and avoid the resonance in the DM annihilation at $m_{A'}\sim 2 m_{\chi_1}$~\cite{Feng:2017drg,Berlin:2020uwy}). 
The experimental reach is much improved for lighter DM particles;however, the reach of beam-dump experiments quickly decreases as it approaches the kinematical threshold for  $\chi_2 \to \chi_1 e^+e^-$ decays.
Since NA64 does not rely on the observation of the decay products, its sensitivity is not affected by this fact.
In this context, the NA64 projections for $5\times 10^{12}$ EOT can unequivocally test the relic curve from $m_{\chi_1} =1$~MeV up to the point where the BaBar constraint begins.

As discussed in \refref{Berlin:2020uwy}, for thorough coverage of the LDM parameter space, it is also necessary to consider the interplay between experimental sensitivity and the DM relic density as a function of the $R=m_{A'}/m_{\chi_{1}}$ ratio. 
We show the $R$ dependence of the NA64 bounds in~\cref{fig:iDM_R_vs_y} for two, currently unprobed, benchmark points in the iDM model. 
The left panel displays the constraints for a light $\chi_1$, $m_{\chi_1} = 10$~MeV, as well as a small mass splitting $\Delta=0.1 m_{\chi_1}$ and dark coupling $\alpha_D =0.1$.
In this case, the DM relic target remains unprobed in the range $R \sim 2.5 - 6$ by the current NA64 and BaBar exclusion limits. 
On the right panel, we show a scenario with a heavier DM candidate, $m_{\chi_1} = 150$~MeV, larger splitting $\Delta = 0.4 m_{\chi_1}$, and larger coupling $\alpha_D = 0.5$, where only a small $R$ window of the DM relic curve between $\sim2.5$ and $\sim3.3$ is still untested. 
In both cases, the NA64 projections for $5\times 10^{12}$ EOT extend the experiment sensitivity below the standard $R=3$ point, where DM predominantly undergoes resonant co-annihilations, $\chi_1 \chi_2 \to A'$, and the relic abundance is strongly enhanced.
Below the resonance, for $R \lesssim 2$, the dark photon can only be produced off-shell and the experimental sensitivity is, therefore, very suppressed.

%i2DM results
%%%%%%%%%%%%%%%%%%%%%%%%%%%%%%%%%%%%%
\begin{figure}
\centering
    \includegraphics[trim=1cm 0cm 0.6cm 0cm,width=\columnwidth]{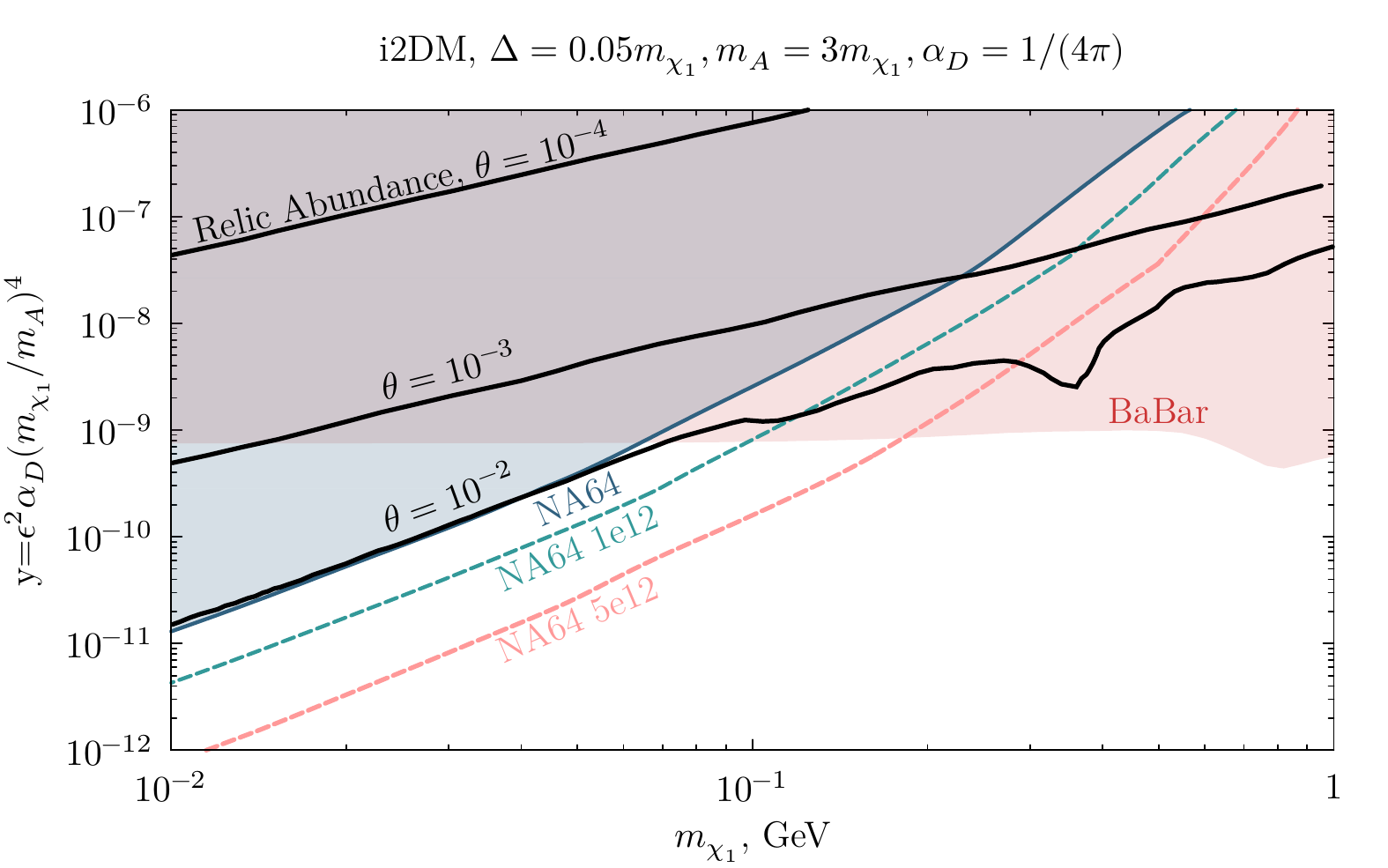} 
    \includegraphics[trim=0.8cm 0cm 0.6cm 0cm,width=\columnwidth]{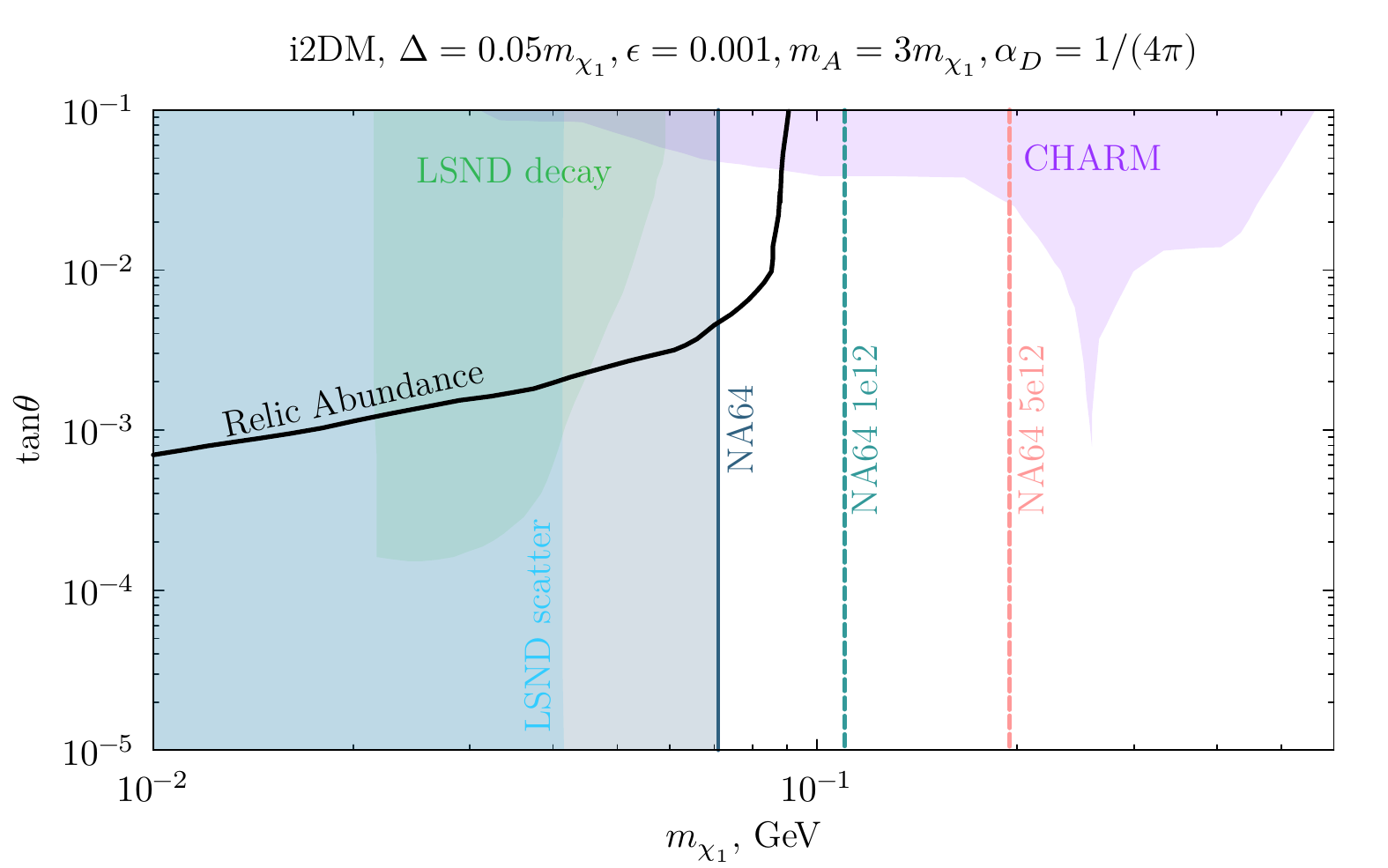}
    \includegraphics[trim=0.8cm 0cm 0.6cm 0cm,width=\columnwidth]{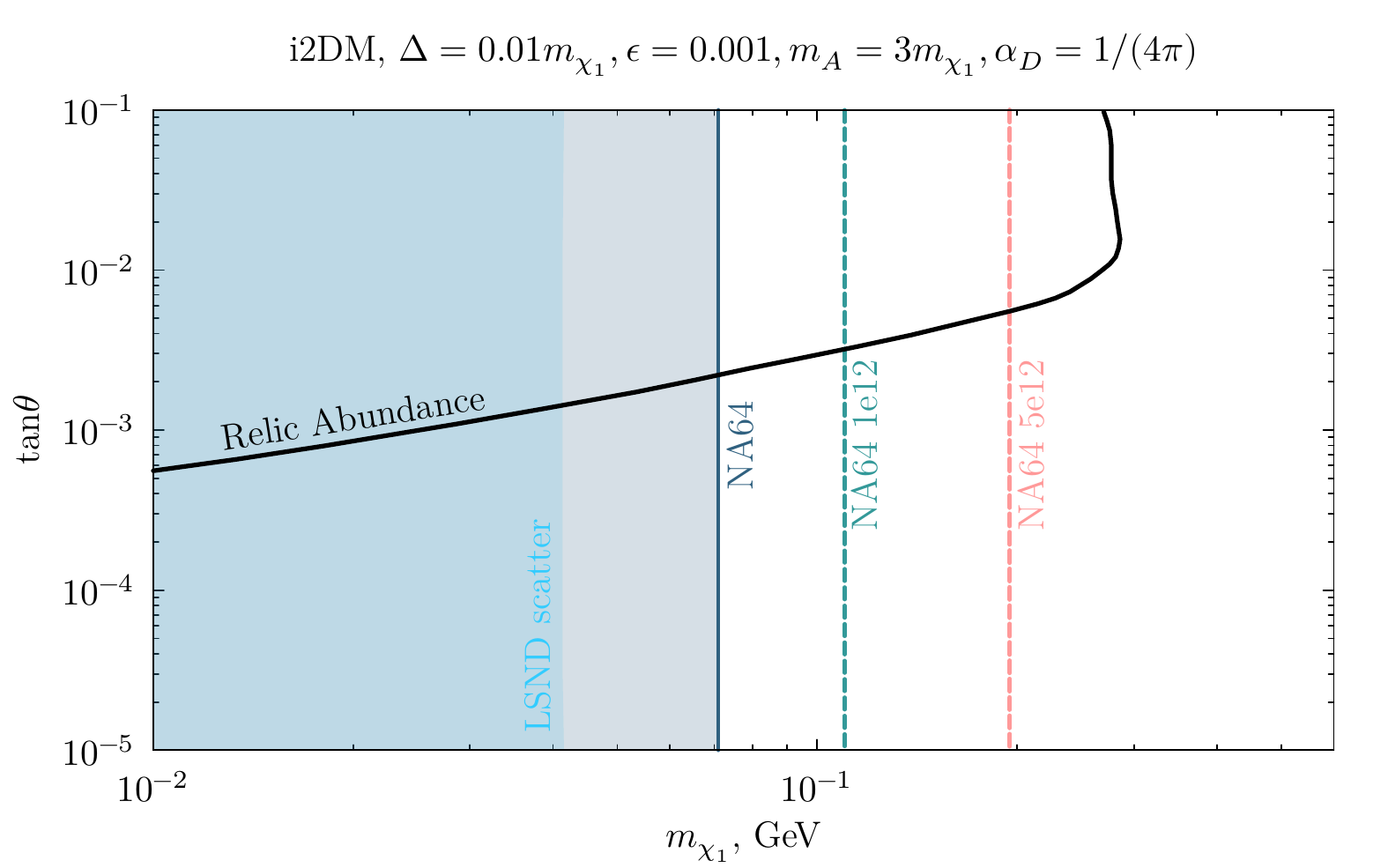}
\caption{\label{fig:i2DM_epsart_relic} i2DM light dark matter parameter space illustrating the available model constraints and the future NA64 sensitivity. First plot: BaBar and NA64 bounds in the DM candidate mass $m_{\chi_1}$ versus $y=\epsilon^2\alpha_D(m_{\chi_1}/m_{A'})^4$ parameter space. The three thick black curves correspond to the relic target for the different mixings $\tan{\theta}=10^{-4},10^{-3},10^{-2}$, taken from~\cite{Filimonova:2022pkj}. Second and third plots: constraints for two mass splittings ($\Delta/m_{\chi_1}=0.05, 0.01$) in the $m_{\chi_1}-\tan{\theta}$ plane obtained by fixing $\epsilon=10^{-3}$ to circumvent the BaBar exclusions. The CHARM and LSND boundaries were re-interpreted from the iDM results in~\refref{Filimonova:2022pkj}. The NA64 projections for the total accumulated statistics at present ($10^{12}$ EOT) and the expected one before LS3 ($5\times 10^{12}$ EOT) are shown with dashed lines.}
\end{figure}

\bigskip
\emph{i2DM --- }The sub-GeV DM parameter space for the i2DM model was assessed in \refref{Filimonova:2022pkj}, the work on which the following discussion is based. 
In this case, the BaBar missing energy searches also apply. 
This is shown in the first plot of~\cref{fig:i2DM_epsart_relic}, where the small mass splitting $\Delta/m_{\chi_1}=0.05$ dictates long, and thus undetectable, $\chi_2$ decays for both the NA64 and BaBar experiments and allows generalizing the bounds to the different $\theta$ values generating the three DM abundance curves depicted. 
The mass splitting dominates the $\chi_2$ lifetime, given that the particle width in the i2DM model scales, similarly to the iDM scenario, as \cite{Filimonova:2022pkj}
\begin{equation} \label{widthi2DM}
    \Gamma_{i2DM}(\chi_2\rightarrow\chi_1 e^+e^-)\simeq\frac{4\epsilon^2\alpha\alpha_D\Delta^5\tan^2{\theta}\cos^4{\theta}}{15\pi m_{A'}^4}.
\end{equation}
Compared to the iDM case, $\chi_2$ decays are furthermore suppressed by a $\tan^2{\theta}\cos^4{\theta}$ factor, which enhances the lifetime of the unstable state. 

The NA64 exclusion line matches closely the DM relic curve for $\theta=10^{-2}$ in the range $m_{\chi_1}\sim0.01-0.05$ GeV, while the projected limits for $10^{12}$ EOT can fully probe the open combinations of parameters resulting in the observed DM yield.
It must be emphasized that the BaBar exclusions can be alleviated through a targeted choice of the mixing strength $\epsilon$; the upper bound on this parameters is set at $\epsilon\sim10^{-3}$ and thereby, below this threshold, BaBar is no longer sensitive to semi-visible dark photons.
To evade these constraints, $\epsilon=10^{-3}$ was imposed to study the limits in the $m_{\chi_1}-\tan{\theta}$ parameter space.

The second and last panel of \cref{fig:i2DM_epsart_relic} illustrate the existing boundaries for two selected splittings obtained by LSND and CHARM \cite{Filimonova:2022pkj}, and the newly derived NA64 exclusions and projections. 
In the two cases shown, the i2DM scenario appears to be poorly covered in the high mass spectrum. 
NA64 sets new limits on the minimum allowed DM candidate mass, leaving just the values above $m_{\chi_1}>71$~MeV allowed. 
With $10^{12}$ EOT, the parameter space below $m_{\chi_1}=110$~MeV, and the whole relic target can be probed, whereas by considering $5\times 10^{12}$ EOT, this value shifts to $m_{\chi_1}=195$~MeV. 
The model point envisioning the larger split spectrum is partly restricted by the two neutrino experiments: the reinterpreted CHARM data rule out part of the large $\tan{\theta}$ mixing region and the sensitivity extends to lower values in the window $m_{\chi_1}\sim0.02-0.03$ GeV, while LSND significantly precludes the lower spectrum end. 
In the last plot, the $\Delta$ parameter is decreased by a factor of 5, and decay-in-flight signatures at CHARM and LSND become much weaker. 
In this context, NA64 can provide the leading limits in the higher mass region with the foreseen research plan. 

Finally, we conclude our discussions of the results with a cautionary statement. 
The beam-dump and collider constraints shown in all of our plots are the result of phenomenological recast analyses, which are often simplified and approximate procedures.
The uncertainty in these curves is, therefore, expected to be much larger than the NA64 limits we derive.
In this sense, a significant fraction of the relic density curves covered by NA64 are being probed by an experimental analysis for the first time.

%%%%%%%%%%%%%%%%%%%%%%%%%%%%%%%%%%%%%
\section{Conclusions}\label{sec7}
%%%%%%%%%%%%%%%%%%%%%%%%%%%%%%%%%%%%

In this article, we evaluate the NA64 potential in the exploration of next-to-minimal vector portal theories predicting a dark sector populated by multiple fermions. This non-minimal scenario has been suggested as it constitutes a natural
extension of the SM, able to relax the existing bounds on both visible and invisible dark photons and allow new discovery prospects.
In particular, the NA64 90\%~C.L. exclusion limits obtained in the invisible search were recasted on two interesting semi-visible dark photon models, the inelastic dark matter (iDM) and inelastic Dirac dark matter (i2DM) models. 
By extending the signal definition w.r.t. the previous iDM publication, we place new limits on regions of parameter space where the semi-visible decay is prompt, thus improving constraints on iDM and i2DM models. 
Nevertheless, decays that take place between the end regions of the ECAL and the end of the setup are often vetoed by the hermeticity requirements imposed in the event selection. 
This loss in coverage, visible in the fluctuations of the NA64 limits (see e.g. \cref{fig:iDMmuon_ma_vs_eps} and \cref{fig:i2DMmuon_ma_vs_eps}), could be recovered with a tailored signal definition or targeted setup optimization.

Rich dark sectors characterized by sizable mass splittings have also been considered in the context of low-energy experimental anomalies, such as that of the muon $(g-2)$. 
This study constrains most of the parameter space where a dark photon can explain the $(g-2)_\mu$ discrepancy. 
Together with BaBar, NA64 plays a central role in ruling out this possibility for both the iDM and i2DM realizations, the latter being already tightly constrained by CMB data. 

In a more general investigation, the NA64 current bounds and projections were compared to the parameter points yielding the observed amount of freeze-out DM. 
With the statistics expected before LS3, NA64 can widely test the parameter space of viable iDM and i2DM models in the MeV to GeV mass range.

We showed that NA64 is a powerful probe of semi-visible dark photons.
Despite the visible products, as the $A'$ decay cascade eventually produces two energetic invisible particles, the missing energy technique is also well-suited to explore these scenarios.
The additional amount of visible energy produced in the semi-visible decays can be directly measured, and NA64 is also sensitive to prompt decays, where $\chi_2$ decays inside the target (this work), and longer lifetimes, corresponding to $\chi_2$ decays inside the hadronic calorimeters~\cite{NA64:2021acr}.
Only intermediate lifetimes, namely $\chi_2$ decays shortly after the target, would not be picked up by current searches.
Beyond iDM and i2DM, this represents an exciting prospect for the exploration of other semi-visible dark photon models with multiple DM particles or heavy neutral leptons~\cite{future}.

In the future, the inclusion of the $A'$ resonant annihilation mechanism, a secondary dark boson production channel investigated in Ref.~\cite{Andreev:2021fzd}, can enhance the limits in the $m_{A'}\sim200-300$~MeV window by almost an order of magnitude. 
In addition, by running with a muon beam, the NA64$\mu$ experiment~\cite{Sieber:2021fue} can have better sensitivity to invisible dark photons in the high-mass region ($m_{A'}\gtrsim100$ MeV), allowing to support the electron mode in the investigation of the entire target mass range with a similar statistics~\cite{Gninenko:2019qiv}.
The NA64 efforts in search of semi-visible dark particles are complementary to the existing DS program at Belle-II~\cite{Belle-II:2018jsg}. 
The next generation of fixed-target experiments, such as the missing momentum experiment LDMX, will push the future exploration of the full parameter space favored by the thermal DM milestones in the MeV–GeV mass range~\cite{Berlin:2018bsc}.
Neutrino experiments are also sensitive to the production, decay, and scattering of inelastic DM.
Data from the existing short-baseline program at Fermilab~\cite{Batell:2021ooj} and the future DUNE long-baseline program~\cite{DeRomeri:2019kic,DeRoeck:2020ntj} can look for the scattering and decay-in-flight signatures in iDM and i2DM. For a broader overview of other sensitive probes and future experimental projections, see e.g. Ref.~\cite{Krnjaic:2022ozp}.

\begin{acknowledgements}

%SP would like to thank the CERN Theory Department for the hospitality during the early stages of this work.
This project has received partial funding from the European Union’s Horizon 2020 research and innovation programme under the Marie Sklodowska-Curie grant agreement No. H2020-MSCA-ITN-2019/860881-HIDDeN (ITN HIDDeN) and from the European Union’s Horizon Europe research and innovation programme under
the Marie Sklodowska-Curie Staff Exchange grant agreement No.~101086085 - ASYMMETRY.
The research at the Perimeter Institute is supported in part by the Government of Canada through NSERC and by the Province of Ontario through the Ministry of Economic Development, Job Creation and Trade, MEDT.
The work of PC, MM and BB is supported by ETH Zürich and SNSF Grants No. 169133, No. 186181, No. 186158, and No. 197346 (Switzerland). The work of LMB is supported by SNSF Grant No. 186158 (Switzerland), RyC-030551-I, and PID2021-123955NA-100 funded by MCIN/AEI /10.13039/501100011033/FEDER, UE(Spain). PC, MM, BBO and LMB would like to acknowledge the NA64 collaboration, and in particular S. Gninenko and E. Depero.

\end{acknowledgements}
\bibliographystyle{apsrev4-1}
\bibliography{bibfile}

\begin{thebibliography}{88}%
\makeatletter
\providecommand \@ifxundefined [1]{%
 \@ifx{#1\undefined}
}%
\providecommand \@ifnum [1]{%
 \ifnum #1\expandafter \@firstoftwo
 \else \expandafter \@secondoftwo
 \fi
}%
\providecommand \@ifx [1]{%
 \ifx #1\expandafter \@firstoftwo
 \else \expandafter \@secondoftwo
 \fi
}%
\providecommand \natexlab [1]{#1}%
\providecommand \enquote  [1]{``#1''}%
\providecommand \bibnamefont  [1]{#1}%
\providecommand \bibfnamefont [1]{#1}%
\providecommand \citenamefont [1]{#1}%
\providecommand \href@noop [0]{\@secondoftwo}%
\providecommand \href [0]{\begingroup \@sanitize@url \@href}%
\providecommand \@href[1]{\@@startlink{#1}\@@href}%
\providecommand \@@href[1]{\endgroup#1\@@endlink}%
\providecommand \@sanitize@url [0]{\catcode `\\12\catcode `\$12\catcode
  `\&12\catcode `\#12\catcode `\^12\catcode `\_12\catcode `\%12\relax}%
\providecommand \@@startlink[1]{}%
\providecommand \@@endlink[0]{}%
\providecommand \url  [0]{\begingroup\@sanitize@url \@url }%
\providecommand \@url [1]{\endgroup\@href {#1}{\urlprefix }}%
\providecommand \urlprefix  [0]{URL }%
\providecommand \Eprint [0]{\href }%
\providecommand \doibase [0]{http://dx.doi.org/}%
\providecommand \selectlanguage [0]{\@gobble}%
\providecommand \bibinfo  [0]{\@secondoftwo}%
\providecommand \bibfield  [0]{\@secondoftwo}%
\providecommand \translation [1]{[#1]}%
\providecommand \BibitemOpen [0]{}%
\providecommand \bibitemStop [0]{}%
\providecommand \bibitemNoStop [0]{.\EOS\space}%
\providecommand \EOS [0]{\spacefactor3000\relax}%
\providecommand \BibitemShut  [1]{\csname bibitem#1\endcsname}%
\let\auto@bib@innerbib\@empty
%</preamble>
\bibitem [{\citenamefont {Feng}(2010)}]{Feng:2010gw}%
  \BibitemOpen
  \bibfield  {author} {\bibinfo {author} {\bibfnamefont {J.~L.}\ \bibnamefont
  {Feng}},\ }\href {\doibase 10.1146/annurev-astro-082708-101659} {\bibfield
  {journal} {\bibinfo  {journal} {Ann. Rev. Astron. Astrophys.}\ }\textbf
  {\bibinfo {volume} {48}},\ \bibinfo {pages} {495} (\bibinfo {year} {2010})},\
  \Eprint {http://arxiv.org/abs/1003.0904} {arXiv:1003.0904 [astro-ph.CO]}
  \BibitemShut {NoStop}%
\bibitem [{\citenamefont {Battaglieri}\ \emph {et~al.}(2017)\citenamefont
  {Battaglieri} \emph {et~al.}}]{Battaglieri:2017aum}%
  \BibitemOpen
  \bibfield  {author} {\bibinfo {author} {\bibfnamefont {M.}~\bibnamefont
  {Battaglieri}} \emph {et~al.},\ }in\ \href@noop {} {\emph {\bibinfo
  {booktitle} {{U.S. Cosmic Visions: New Ideas in Dark Matter}}}}\ (\bibinfo
  {year} {2017})\ \Eprint {http://arxiv.org/abs/1707.04591} {arXiv:1707.04591
  [hep-ph]} \BibitemShut {NoStop}%
\bibitem [{\citenamefont {Alexander}\ \emph {et~al.}(2016)\citenamefont
  {Alexander} \emph {et~al.}}]{Alexander:2016aln}%
  \BibitemOpen
  \bibfield  {author} {\bibinfo {author} {\bibfnamefont {J.}~\bibnamefont
  {Alexander}} \emph {et~al.}\ }(\bibinfo {year} {2016})\ \Eprint
  {http://arxiv.org/abs/1608.08632} {arXiv:1608.08632 [hep-ph]} \BibitemShut
  {NoStop}%
\bibitem [{\citenamefont {Agrawal}\ \emph {et~al.}(2021)\citenamefont {Agrawal}
  \emph {et~al.}}]{Agrawal:2021dbo}%
  \BibitemOpen
  \bibfield  {author} {\bibinfo {author} {\bibfnamefont {P.}~\bibnamefont
  {Agrawal}} \emph {et~al.},\ }\href {\doibase 10.1140/epjc/s10052-021-09703-7}
  {\bibfield  {journal} {\bibinfo  {journal} {Eur. Phys. J. C}\ }\textbf
  {\bibinfo {volume} {81}},\ \bibinfo {pages} {1015} (\bibinfo {year}
  {2021})},\ \Eprint {http://arxiv.org/abs/2102.12143} {arXiv:2102.12143
  [hep-ph]} \BibitemShut {NoStop}%
\bibitem [{\citenamefont {Arkani-Hamed}\ \emph {et~al.}(2009)\citenamefont
  {Arkani-Hamed}, \citenamefont {Finkbeiner}, \citenamefont {Slatyer},\ and\
  \citenamefont {Weiner}}]{ArkaniHamed:2008qn}%
  \BibitemOpen
  \bibfield  {author} {\bibinfo {author} {\bibfnamefont {N.}~\bibnamefont
  {Arkani-Hamed}}, \bibinfo {author} {\bibfnamefont {D.~P.}\ \bibnamefont
  {Finkbeiner}}, \bibinfo {author} {\bibfnamefont {T.~R.}\ \bibnamefont
  {Slatyer}}, \ and\ \bibinfo {author} {\bibfnamefont {N.}~\bibnamefont
  {Weiner}},\ }\href {\doibase 10.1103/PhysRevD.79.015014} {\bibfield
  {journal} {\bibinfo  {journal} {Phys. Rev. D}\ }\textbf {\bibinfo {volume}
  {79}},\ \bibinfo {pages} {015014} (\bibinfo {year} {2009})},\ \Eprint
  {http://arxiv.org/abs/0810.0713} {arXiv:0810.0713 [hep-ph]} \BibitemShut
  {NoStop}%
\bibitem [{\citenamefont {Pospelov}\ and\ \citenamefont
  {Ritz}(2009)}]{Pospelov:2008jd}%
  \BibitemOpen
  \bibfield  {author} {\bibinfo {author} {\bibfnamefont {M.}~\bibnamefont
  {Pospelov}}\ and\ \bibinfo {author} {\bibfnamefont {A.}~\bibnamefont
  {Ritz}},\ }\href {\doibase 10.1016/j.physletb.2008.12.012} {\bibfield
  {journal} {\bibinfo  {journal} {Phys. Lett. B}\ }\textbf {\bibinfo {volume}
  {671}},\ \bibinfo {pages} {391} (\bibinfo {year} {2009})},\ \Eprint
  {http://arxiv.org/abs/0810.1502} {arXiv:0810.1502 [hep-ph]} \BibitemShut
  {NoStop}%
\bibitem [{\citenamefont {Hooper}\ \emph {et~al.}(2012)\citenamefont {Hooper},
  \citenamefont {Weiner},\ and\ \citenamefont {Xue}}]{Hooper:2012cw}%
  \BibitemOpen
  \bibfield  {author} {\bibinfo {author} {\bibfnamefont {D.}~\bibnamefont
  {Hooper}}, \bibinfo {author} {\bibfnamefont {N.}~\bibnamefont {Weiner}}, \
  and\ \bibinfo {author} {\bibfnamefont {W.}~\bibnamefont {Xue}},\ }\href
  {\doibase 10.1103/PhysRevD.86.056009} {\bibfield  {journal} {\bibinfo
  {journal} {Phys. Rev. D}\ }\textbf {\bibinfo {volume} {86}},\ \bibinfo
  {pages} {056009} (\bibinfo {year} {2012})},\ \Eprint
  {http://arxiv.org/abs/1206.2929} {arXiv:1206.2929 [hep-ph]} \BibitemShut
  {NoStop}%
\bibitem [{\citenamefont {Pospelov}\ \emph {et~al.}(2008)\citenamefont
  {Pospelov}, \citenamefont {Ritz},\ and\ \citenamefont
  {Voloshin}}]{Pospelov:2007mp}%
  \BibitemOpen
  \bibfield  {author} {\bibinfo {author} {\bibfnamefont {M.}~\bibnamefont
  {Pospelov}}, \bibinfo {author} {\bibfnamefont {A.}~\bibnamefont {Ritz}}, \
  and\ \bibinfo {author} {\bibfnamefont {M.~B.}\ \bibnamefont {Voloshin}},\
  }\href {\doibase 10.1016/j.physletb.2008.02.052} {\bibfield  {journal}
  {\bibinfo  {journal} {Phys. Lett. B}\ }\textbf {\bibinfo {volume} {662}},\
  \bibinfo {pages} {53} (\bibinfo {year} {2008})},\ \Eprint
  {http://arxiv.org/abs/0711.4866} {arXiv:0711.4866 [hep-ph]} \BibitemShut
  {NoStop}%
\bibitem [{\citenamefont {Pospelov}(2009)}]{Pospelov:2008zw}%
  \BibitemOpen
  \bibfield  {author} {\bibinfo {author} {\bibfnamefont {M.}~\bibnamefont
  {Pospelov}},\ }\href {\doibase 10.1103/PhysRevD.80.095002} {\bibfield
  {journal} {\bibinfo  {journal} {Phys. Rev. D}\ }\textbf {\bibinfo {volume}
  {80}},\ \bibinfo {pages} {095002} (\bibinfo {year} {2009})},\ \Eprint
  {http://arxiv.org/abs/0811.1030} {arXiv:0811.1030 [hep-ph]} \BibitemShut
  {NoStop}%
\bibitem [{\citenamefont {Essig}\ \emph
  {et~al.}(2013{\natexlab{a}})\citenamefont {Essig} \emph
  {et~al.}}]{Essig:2013lka}%
  \BibitemOpen
  \bibfield  {author} {\bibinfo {author} {\bibfnamefont {R.}~\bibnamefont
  {Essig}} \emph {et~al.},\ }in\ \href@noop {} {\emph {\bibinfo {booktitle}
  {{Community Summer Study 2013}: {Snowmass on the Mississippi}}}}\ (\bibinfo
  {year} {2013})\ \Eprint {http://arxiv.org/abs/1311.0029} {arXiv:1311.0029
  [hep-ph]} \BibitemShut {NoStop}%
\bibitem [{\citenamefont {Holdom}(1986)}]{Holdom:1985ag}%
  \BibitemOpen
  \bibfield  {author} {\bibinfo {author} {\bibfnamefont {B.}~\bibnamefont
  {Holdom}},\ }\href {\doibase 10.1016/0370-2693(86)91377-8} {\bibfield
  {journal} {\bibinfo  {journal} {Phys. Lett. B}\ }\textbf {\bibinfo {volume}
  {166}},\ \bibinfo {pages} {196} (\bibinfo {year} {1986})}\BibitemShut
  {NoStop}%
\bibitem [{\citenamefont {Okun}(1982)}]{Okun:1982xi}%
  \BibitemOpen
  \bibfield  {author} {\bibinfo {author} {\bibfnamefont {L.~B.}\ \bibnamefont
  {Okun}},\ }\href@noop {} {\bibfield  {journal} {\bibinfo  {journal} {Sov.
  Phys. JETP}\ }\textbf {\bibinfo {volume} {56}},\ \bibinfo {pages} {502}
  (\bibinfo {year} {1982})}\BibitemShut {NoStop}%
\bibitem [{\citenamefont {Mohlabeng}(2019)}]{Mohlabeng:2019vrz}%
  \BibitemOpen
  \bibfield  {author} {\bibinfo {author} {\bibfnamefont {G.}~\bibnamefont
  {Mohlabeng}},\ }\href {\doibase 10.1103/PhysRevD.99.115001} {\bibfield
  {journal} {\bibinfo  {journal} {Phys. Rev. D}\ }\textbf {\bibinfo {volume}
  {99}},\ \bibinfo {pages} {115001} (\bibinfo {year} {2019})},\ \Eprint
  {http://arxiv.org/abs/1902.05075} {arXiv:1902.05075 [hep-ph]} \BibitemShut
  {NoStop}%
\bibitem [{\citenamefont {Tsai}\ \emph {et~al.}(2021)\citenamefont {Tsai},
  \citenamefont {deNiverville},\ and\ \citenamefont {Liu}}]{Tsai:2019buq}%
  \BibitemOpen
  \bibfield  {author} {\bibinfo {author} {\bibfnamefont {Y.-D.}\ \bibnamefont
  {Tsai}}, \bibinfo {author} {\bibfnamefont {P.}~\bibnamefont {deNiverville}},
  \ and\ \bibinfo {author} {\bibfnamefont {M.~X.}\ \bibnamefont {Liu}},\ }\href
  {\doibase 10.1103/PhysRevLett.126.181801} {\bibfield  {journal} {\bibinfo
  {journal} {Phys. Rev. Lett.}\ }\textbf {\bibinfo {volume} {126}},\ \bibinfo
  {pages} {181801} (\bibinfo {year} {2021})},\ \Eprint
  {http://arxiv.org/abs/1908.07525} {arXiv:1908.07525 [hep-ph]} \BibitemShut
  {NoStop}%
\bibitem [{\citenamefont {Duerr}\ \emph {et~al.}(2020)\citenamefont {Duerr},
  \citenamefont {Ferber}, \citenamefont {Hearty}, \citenamefont {Kahlhoefer},
  \citenamefont {Schmidt-Hoberg},\ and\ \citenamefont
  {Tunney}}]{Duerr:2019dmv}%
  \BibitemOpen
  \bibfield  {author} {\bibinfo {author} {\bibfnamefont {M.}~\bibnamefont
  {Duerr}}, \bibinfo {author} {\bibfnamefont {T.}~\bibnamefont {Ferber}},
  \bibinfo {author} {\bibfnamefont {C.}~\bibnamefont {Hearty}}, \bibinfo
  {author} {\bibfnamefont {F.}~\bibnamefont {Kahlhoefer}}, \bibinfo {author}
  {\bibfnamefont {K.}~\bibnamefont {Schmidt-Hoberg}}, \ and\ \bibinfo {author}
  {\bibfnamefont {P.}~\bibnamefont {Tunney}},\ }\href {\doibase
  10.1007/JHEP02(2020)039} {\bibfield  {journal} {\bibinfo  {journal} {JHEP}\
  }\textbf {\bibinfo {volume} {02}},\ \bibinfo {pages} {039} (\bibinfo {year}
  {2020})},\ \Eprint {http://arxiv.org/abs/1911.03176} {arXiv:1911.03176
  [hep-ph]} \BibitemShut {NoStop}%
\bibitem [{\citenamefont {Abdullahi}\ \emph {et~al.}(2021)\citenamefont
  {Abdullahi}, \citenamefont {Hostert},\ and\ \citenamefont
  {Pascoli}}]{Abdullahi:2020nyr}%
  \BibitemOpen
  \bibfield  {author} {\bibinfo {author} {\bibfnamefont {A.}~\bibnamefont
  {Abdullahi}}, \bibinfo {author} {\bibfnamefont {M.}~\bibnamefont {Hostert}},
  \ and\ \bibinfo {author} {\bibfnamefont {S.}~\bibnamefont {Pascoli}},\ }\href
  {\doibase 10.1016/j.physletb.2021.136531} {\bibfield  {journal} {\bibinfo
  {journal} {Phys. Lett. B}\ }\textbf {\bibinfo {volume} {820}},\ \bibinfo
  {pages} {136531} (\bibinfo {year} {2021})},\ \Eprint
  {http://arxiv.org/abs/2007.11813} {arXiv:2007.11813 [hep-ph]} \BibitemShut
  {NoStop}%
\bibitem [{\citenamefont {Abdullahi}\ \emph {et~al.}(2023)\citenamefont
  {Abdullahi}, \citenamefont {Hostert}, \citenamefont {Massaro},\ and\
  \citenamefont {Pascoli}}]{future}%
  \BibitemOpen
  \bibfield  {author} {\bibinfo {author} {\bibfnamefont {A.}~\bibnamefont
  {Abdullahi}}, \bibinfo {author} {\bibfnamefont {M.}~\bibnamefont {Hostert}},
  \bibinfo {author} {\bibfnamefont {D.}~\bibnamefont {Massaro}}, \ and\
  \bibinfo {author} {\bibfnamefont {S.}~\bibnamefont {Pascoli}},\ }\href@noop
  {} {\enquote {\bibinfo {title} {{Semi-visible Dark Photons below the
  Electroweak Scale}},}\ } (\bibinfo {year} {2023}),\ \bibinfo {note} {{on
  arXiv}}\BibitemShut {NoStop}%
\bibitem [{\citenamefont {Banerjee}\ \emph {et~al.}(2018)\citenamefont
  {Banerjee} \emph {et~al.}}]{NA64:2017vtt}%
  \BibitemOpen
  \bibfield  {author} {\bibinfo {author} {\bibfnamefont {D.}~\bibnamefont
  {Banerjee}} \emph {et~al.} (\bibinfo {collaboration} {NA64}),\ }\href
  {\doibase 10.1103/PhysRevD.97.072002} {\bibfield  {journal} {\bibinfo
  {journal} {Phys. Rev. D}\ }\textbf {\bibinfo {volume} {97}},\ \bibinfo
  {pages} {072002} (\bibinfo {year} {2018})},\ \Eprint
  {http://arxiv.org/abs/1710.00971} {arXiv:1710.00971 [hep-ex]} \BibitemShut
  {NoStop}%
\bibitem [{\citenamefont {Izaguirre}\ \emph
  {et~al.}(2015{\natexlab{a}})\citenamefont {Izaguirre}, \citenamefont
  {Krnjaic}, \citenamefont {Schuster},\ and\ \citenamefont
  {Toro}}]{Izaguirre:2014bca}%
  \BibitemOpen
  \bibfield  {author} {\bibinfo {author} {\bibfnamefont {E.}~\bibnamefont
  {Izaguirre}}, \bibinfo {author} {\bibfnamefont {G.}~\bibnamefont {Krnjaic}},
  \bibinfo {author} {\bibfnamefont {P.}~\bibnamefont {Schuster}}, \ and\
  \bibinfo {author} {\bibfnamefont {N.}~\bibnamefont {Toro}},\ }\href {\doibase
  10.1103/PhysRevD.91.094026} {\bibfield  {journal} {\bibinfo  {journal} {Phys.
  Rev. D}\ }\textbf {\bibinfo {volume} {91}},\ \bibinfo {pages} {094026}
  (\bibinfo {year} {2015}{\natexlab{a}})},\ \Eprint
  {http://arxiv.org/abs/1411.1404} {arXiv:1411.1404 [hep-ph]} \BibitemShut
  {NoStop}%
\bibitem [{\citenamefont {Knapen}\ \emph {et~al.}(2017)\citenamefont {Knapen},
  \citenamefont {Lin},\ and\ \citenamefont {Zurek}}]{Knapen:2017xzo}%
  \BibitemOpen
  \bibfield  {author} {\bibinfo {author} {\bibfnamefont {S.}~\bibnamefont
  {Knapen}}, \bibinfo {author} {\bibfnamefont {T.}~\bibnamefont {Lin}}, \ and\
  \bibinfo {author} {\bibfnamefont {K.~M.}\ \bibnamefont {Zurek}},\ }\href
  {\doibase 10.1103/PhysRevD.96.115021} {\bibfield  {journal} {\bibinfo
  {journal} {Phys. Rev. D}\ }\textbf {\bibinfo {volume} {96}},\ \bibinfo
  {pages} {115021} (\bibinfo {year} {2017})},\ \Eprint
  {http://arxiv.org/abs/1709.07882} {arXiv:1709.07882 [hep-ph]} \BibitemShut
  {NoStop}%
\bibitem [{\citenamefont {Essig}\ \emph {et~al.}(2022)\citenamefont {Essig},
  \citenamefont {Giovanetti}, \citenamefont {Kurinsky}, \citenamefont
  {McKinsey}, \citenamefont {Ramanathan}, \citenamefont {Stifter},\ and\
  \citenamefont {Yu}}]{Essig:2022dfa}%
  \BibitemOpen
  \bibfield  {author} {\bibinfo {author} {\bibfnamefont {R.}~\bibnamefont
  {Essig}}, \bibinfo {author} {\bibfnamefont {G.~K.}\ \bibnamefont
  {Giovanetti}}, \bibinfo {author} {\bibfnamefont {N.}~\bibnamefont
  {Kurinsky}}, \bibinfo {author} {\bibfnamefont {D.}~\bibnamefont {McKinsey}},
  \bibinfo {author} {\bibfnamefont {K.}~\bibnamefont {Ramanathan}}, \bibinfo
  {author} {\bibfnamefont {K.}~\bibnamefont {Stifter}}, \ and\ \bibinfo
  {author} {\bibfnamefont {T.-T.}\ \bibnamefont {Yu}},\ }in\ \href@noop {}
  {\emph {\bibinfo {booktitle} {{2022 Snowmass Summer Study}}}}\ (\bibinfo
  {year} {2022})\ \Eprint {http://arxiv.org/abs/2203.08297} {arXiv:2203.08297
  [hep-ph]} \BibitemShut {NoStop}%
\bibitem [{\citenamefont {Dreiner}\ \emph {et~al.}(2014)\citenamefont
  {Dreiner}, \citenamefont {Fortin}, \citenamefont {Hanhart},\ and\
  \citenamefont {Ubaldi}}]{Dreiner:2013mua}%
  \BibitemOpen
  \bibfield  {author} {\bibinfo {author} {\bibfnamefont {H.~K.}\ \bibnamefont
  {Dreiner}}, \bibinfo {author} {\bibfnamefont {J.-F.}\ \bibnamefont {Fortin}},
  \bibinfo {author} {\bibfnamefont {C.}~\bibnamefont {Hanhart}}, \ and\
  \bibinfo {author} {\bibfnamefont {L.}~\bibnamefont {Ubaldi}},\ }\href
  {\doibase 10.1103/PhysRevD.89.105015} {\bibfield  {journal} {\bibinfo
  {journal} {Phys. Rev. D}\ }\textbf {\bibinfo {volume} {89}},\ \bibinfo
  {pages} {105015} (\bibinfo {year} {2014})},\ \Eprint
  {http://arxiv.org/abs/1310.3826} {arXiv:1310.3826 [hep-ph]} \BibitemShut
  {NoStop}%
\bibitem [{\citenamefont {Chang}\ \emph {et~al.}(2018)\citenamefont {Chang},
  \citenamefont {Essig},\ and\ \citenamefont {McDermott}}]{Chang:2018rso}%
  \BibitemOpen
  \bibfield  {author} {\bibinfo {author} {\bibfnamefont {J.~H.}\ \bibnamefont
  {Chang}}, \bibinfo {author} {\bibfnamefont {R.}~\bibnamefont {Essig}}, \ and\
  \bibinfo {author} {\bibfnamefont {S.~D.}\ \bibnamefont {McDermott}},\ }\href
  {\doibase 10.1007/JHEP09(2018)051} {\bibfield  {journal} {\bibinfo  {journal}
  {JHEP}\ }\textbf {\bibinfo {volume} {09}},\ \bibinfo {pages} {051} (\bibinfo
  {year} {2018})},\ \Eprint {http://arxiv.org/abs/1803.00993} {arXiv:1803.00993
  [hep-ph]} \BibitemShut {NoStop}%
\bibitem [{\citenamefont {Artamonov}\ \emph {et~al.}(2009)\citenamefont
  {Artamonov} \emph {et~al.}}]{BNL-E949:2009dza}%
  \BibitemOpen
  \bibfield  {author} {\bibinfo {author} {\bibfnamefont {A.~V.}\ \bibnamefont
  {Artamonov}} \emph {et~al.} (\bibinfo {collaboration} {BNL-E949}),\ }\href
  {\doibase 10.1103/PhysRevD.79.092004} {\bibfield  {journal} {\bibinfo
  {journal} {Phys. Rev. D}\ }\textbf {\bibinfo {volume} {79}},\ \bibinfo
  {pages} {092004} (\bibinfo {year} {2009})},\ \Eprint
  {http://arxiv.org/abs/0903.0030} {arXiv:0903.0030 [hep-ex]} \BibitemShut
  {NoStop}%
\bibitem [{\citenamefont {Mirra}(2018)}]{Mirra:2018zdd}%
  \BibitemOpen
  \bibfield  {author} {\bibinfo {author} {\bibfnamefont {M.}~\bibnamefont
  {Mirra}} (\bibinfo {collaboration} {NA62}),\ }in\ \href@noop {} {\emph
  {\bibinfo {booktitle} {{53rd Rencontres de Moriond on QCD and High Energy
  Interactions}}}}\ (\bibinfo {year} {2018})\ pp.\ \bibinfo {pages}
  {185--188}\BibitemShut {NoStop}%
\bibitem [{\citenamefont {Cortina~Gil}\ \emph {et~al.}(2021)\citenamefont
  {Cortina~Gil} \emph {et~al.}}]{NA62:2020xlg}%
  \BibitemOpen
  \bibfield  {author} {\bibinfo {author} {\bibfnamefont {E.}~\bibnamefont
  {Cortina~Gil}} \emph {et~al.} (\bibinfo {collaboration} {NA62}),\ }\href
  {\doibase 10.1007/JHEP03(2021)058} {\bibfield  {journal} {\bibinfo  {journal}
  {JHEP}\ }\textbf {\bibinfo {volume} {03}},\ \bibinfo {pages} {058} (\bibinfo
  {year} {2021})},\ \Eprint {http://arxiv.org/abs/2011.11329} {arXiv:2011.11329
  [hep-ex]} \BibitemShut {NoStop}%
\bibitem [{\citenamefont {Batell}\ \emph
  {et~al.}(2009{\natexlab{a}})\citenamefont {Batell}, \citenamefont
  {Pospelov},\ and\ \citenamefont {Ritz}}]{Batell:2009di}%
  \BibitemOpen
  \bibfield  {author} {\bibinfo {author} {\bibfnamefont {B.}~\bibnamefont
  {Batell}}, \bibinfo {author} {\bibfnamefont {M.}~\bibnamefont {Pospelov}}, \
  and\ \bibinfo {author} {\bibfnamefont {A.}~\bibnamefont {Ritz}},\ }\href
  {\doibase 10.1103/PhysRevD.80.095024} {\bibfield  {journal} {\bibinfo
  {journal} {Phys. Rev. D}\ }\textbf {\bibinfo {volume} {80}},\ \bibinfo
  {pages} {095024} (\bibinfo {year} {2009}{\natexlab{a}})},\ \Eprint
  {http://arxiv.org/abs/0906.5614} {arXiv:0906.5614 [hep-ph]} \BibitemShut
  {NoStop}%
\bibitem [{\citenamefont {Batell}\ \emph {et~al.}(2014)\citenamefont {Batell},
  \citenamefont {Essig},\ and\ \citenamefont {Surujon}}]{Batell:2014mga}%
  \BibitemOpen
  \bibfield  {author} {\bibinfo {author} {\bibfnamefont {B.}~\bibnamefont
  {Batell}}, \bibinfo {author} {\bibfnamefont {R.}~\bibnamefont {Essig}}, \
  and\ \bibinfo {author} {\bibfnamefont {Z.}~\bibnamefont {Surujon}},\ }\href
  {\doibase 10.1103/PhysRevLett.113.171802} {\bibfield  {journal} {\bibinfo
  {journal} {Phys. Rev. Lett.}\ }\textbf {\bibinfo {volume} {113}},\ \bibinfo
  {pages} {171802} (\bibinfo {year} {2014})},\ \Eprint
  {http://arxiv.org/abs/1406.2698} {arXiv:1406.2698 [hep-ph]} \BibitemShut
  {NoStop}%
\bibitem [{\citenamefont {Aguilar-Arevalo}\ \emph {et~al.}(2017)\citenamefont
  {Aguilar-Arevalo} \emph {et~al.}}]{MiniBooNE:2017nqe}%
  \BibitemOpen
  \bibfield  {author} {\bibinfo {author} {\bibfnamefont {A.~A.}\ \bibnamefont
  {Aguilar-Arevalo}} \emph {et~al.} (\bibinfo {collaboration} {MiniBooNE}),\
  }\href {\doibase 10.1103/PhysRevLett.118.221803} {\bibfield  {journal}
  {\bibinfo  {journal} {Phys. Rev. Lett.}\ }\textbf {\bibinfo {volume} {118}},\
  \bibinfo {pages} {221803} (\bibinfo {year} {2017})},\ \Eprint
  {http://arxiv.org/abs/1702.02688} {arXiv:1702.02688 [hep-ex]} \BibitemShut
  {NoStop}%
\bibitem [{\citenamefont {Batell}\ \emph
  {et~al.}(2009{\natexlab{b}})\citenamefont {Batell}, \citenamefont
  {Pospelov},\ and\ \citenamefont {Ritz}}]{Batell:2009yf}%
  \BibitemOpen
  \bibfield  {author} {\bibinfo {author} {\bibfnamefont {B.}~\bibnamefont
  {Batell}}, \bibinfo {author} {\bibfnamefont {M.}~\bibnamefont {Pospelov}}, \
  and\ \bibinfo {author} {\bibfnamefont {A.}~\bibnamefont {Ritz}},\ }\href
  {\doibase 10.1103/PhysRevD.79.115008} {\bibfield  {journal} {\bibinfo
  {journal} {Phys. Rev. D}\ }\textbf {\bibinfo {volume} {79}},\ \bibinfo
  {pages} {115008} (\bibinfo {year} {2009}{\natexlab{b}})},\ \Eprint
  {http://arxiv.org/abs/0903.0363} {arXiv:0903.0363 [hep-ph]} \BibitemShut
  {NoStop}%
\bibitem [{\citenamefont {Essig}\ \emph
  {et~al.}(2013{\natexlab{b}})\citenamefont {Essig}, \citenamefont {Mardon},
  \citenamefont {Papucci}, \citenamefont {Volansky},\ and\ \citenamefont
  {Zhong}}]{Essig:2013vha}%
  \BibitemOpen
  \bibfield  {author} {\bibinfo {author} {\bibfnamefont {R.}~\bibnamefont
  {Essig}}, \bibinfo {author} {\bibfnamefont {J.}~\bibnamefont {Mardon}},
  \bibinfo {author} {\bibfnamefont {M.}~\bibnamefont {Papucci}}, \bibinfo
  {author} {\bibfnamefont {T.}~\bibnamefont {Volansky}}, \ and\ \bibinfo
  {author} {\bibfnamefont {Y.-M.}\ \bibnamefont {Zhong}},\ }\href {\doibase
  10.1007/JHEP11(2013)167} {\bibfield  {journal} {\bibinfo  {journal} {JHEP}\
  }\textbf {\bibinfo {volume} {11}},\ \bibinfo {pages} {167} (\bibinfo {year}
  {2013}{\natexlab{b}})},\ \Eprint {http://arxiv.org/abs/1309.5084}
  {arXiv:1309.5084 [hep-ph]} \BibitemShut {NoStop}%
\bibitem [{\citenamefont {Lees}\ \emph {et~al.}(2017)\citenamefont {Lees} \emph
  {et~al.}}]{BaBar:2017tiz}%
  \BibitemOpen
  \bibfield  {author} {\bibinfo {author} {\bibfnamefont {J.~P.}\ \bibnamefont
  {Lees}} \emph {et~al.} (\bibinfo {collaboration} {BaBar}),\ }\href {\doibase
  10.1103/PhysRevLett.119.131804} {\bibfield  {journal} {\bibinfo  {journal}
  {Phys. Rev. Lett.}\ }\textbf {\bibinfo {volume} {119}},\ \bibinfo {pages}
  {131804} (\bibinfo {year} {2017})},\ \Eprint
  {http://arxiv.org/abs/1702.03327} {arXiv:1702.03327 [hep-ex]} \BibitemShut
  {NoStop}%
\bibitem [{\citenamefont {Gninenko}\ and\ \citenamefont
  {Krasnikov}(2001)}]{Gninenko:2001hx}%
  \BibitemOpen
  \bibfield  {author} {\bibinfo {author} {\bibfnamefont {S.~N.}\ \bibnamefont
  {Gninenko}}\ and\ \bibinfo {author} {\bibfnamefont {N.~V.}\ \bibnamefont
  {Krasnikov}},\ }\href {\doibase 10.1016/S0370-2693(01)00693-1} {\bibfield
  {journal} {\bibinfo  {journal} {Phys. Lett. B}\ }\textbf {\bibinfo {volume}
  {513}},\ \bibinfo {pages} {119} (\bibinfo {year} {2001})},\ \Eprint
  {http://arxiv.org/abs/hep-ph/0102222} {arXiv:hep-ph/0102222} \BibitemShut
  {NoStop}%
\bibitem [{\citenamefont {Bennett}\ \emph {et~al.}(2006)\citenamefont {Bennett}
  \emph {et~al.}}]{Muong-2:2006rrc}%
  \BibitemOpen
  \bibfield  {author} {\bibinfo {author} {\bibfnamefont {G.~W.}\ \bibnamefont
  {Bennett}} \emph {et~al.} (\bibinfo {collaboration} {Muon g-2}),\ }\href
  {\doibase 10.1103/PhysRevD.73.072003} {\bibfield  {journal} {\bibinfo
  {journal} {Phys. Rev. D}\ }\textbf {\bibinfo {volume} {73}},\ \bibinfo
  {pages} {072003} (\bibinfo {year} {2006})},\ \Eprint
  {http://arxiv.org/abs/hep-ex/0602035} {arXiv:hep-ex/0602035} \BibitemShut
  {NoStop}%
\bibitem [{\citenamefont {Abi}\ \emph {et~al.}(2021)\citenamefont {Abi} \emph
  {et~al.}}]{Muong-2:2021ojo}%
  \BibitemOpen
  \bibfield  {author} {\bibinfo {author} {\bibfnamefont {B.}~\bibnamefont
  {Abi}} \emph {et~al.} (\bibinfo {collaboration} {Muon g-2}),\ }\href
  {\doibase 10.1103/PhysRevLett.126.141801} {\bibfield  {journal} {\bibinfo
  {journal} {Phys. Rev. Lett.}\ }\textbf {\bibinfo {volume} {126}},\ \bibinfo
  {pages} {141801} (\bibinfo {year} {2021})},\ \Eprint
  {http://arxiv.org/abs/2104.03281} {arXiv:2104.03281 [hep-ex]} \BibitemShut
  {NoStop}%
\bibitem [{\citenamefont {Jegerlehner}\ and\ \citenamefont
  {Nyffeler}(2009)}]{Jegerlehner:2009ry}%
  \BibitemOpen
  \bibfield  {author} {\bibinfo {author} {\bibfnamefont {F.}~\bibnamefont
  {Jegerlehner}}\ and\ \bibinfo {author} {\bibfnamefont {A.}~\bibnamefont
  {Nyffeler}},\ }\href {\doibase 10.1016/j.physrep.2009.04.003} {\bibfield
  {journal} {\bibinfo  {journal} {Phys. Rept.}\ }\textbf {\bibinfo {volume}
  {477}},\ \bibinfo {pages} {1} (\bibinfo {year} {2009})},\ \Eprint
  {http://arxiv.org/abs/0902.3360} {arXiv:0902.3360 [hep-ph]} \BibitemShut
  {NoStop}%
\bibitem [{\citenamefont {Miller}\ \emph {et~al.}(2012)\citenamefont {Miller},
  \citenamefont {de~Rafael}, \citenamefont {Roberts},\ and\ \citenamefont
  {St\"ockinger}}]{Miller:2012opa}%
  \BibitemOpen
  \bibfield  {author} {\bibinfo {author} {\bibfnamefont {J.~P.}\ \bibnamefont
  {Miller}}, \bibinfo {author} {\bibfnamefont {E.}~\bibnamefont {de~Rafael}},
  \bibinfo {author} {\bibfnamefont {B.~L.}\ \bibnamefont {Roberts}}, \ and\
  \bibinfo {author} {\bibfnamefont {D.}~\bibnamefont {St\"ockinger}},\ }\href
  {\doibase 10.1146/annurev-nucl-031312-120340} {\bibfield  {journal} {\bibinfo
   {journal} {Ann. Rev. Nucl. Part. Sci.}\ }\textbf {\bibinfo {volume} {62}},\
  \bibinfo {pages} {237} (\bibinfo {year} {2012})}\BibitemShut {NoStop}%
\bibitem [{\citenamefont {Aoyama}\ \emph {et~al.}(2020)\citenamefont {Aoyama}
  \emph {et~al.}}]{Aoyama:2020ynm}%
  \BibitemOpen
  \bibfield  {author} {\bibinfo {author} {\bibfnamefont {T.}~\bibnamefont
  {Aoyama}} \emph {et~al.},\ }\href {\doibase 10.1016/j.physrep.2020.07.006}
  {\bibfield  {journal} {\bibinfo  {journal} {Phys. Rept.}\ }\textbf {\bibinfo
  {volume} {887}},\ \bibinfo {pages} {1} (\bibinfo {year} {2020})},\ \Eprint
  {http://arxiv.org/abs/2006.04822} {arXiv:2006.04822 [hep-ph]} \BibitemShut
  {NoStop}%
\bibitem [{\citenamefont {Borsanyi}\ \emph {et~al.}(2021)\citenamefont
  {Borsanyi} \emph {et~al.}}]{Borsanyi:2020mff}%
  \BibitemOpen
  \bibfield  {author} {\bibinfo {author} {\bibfnamefont {S.}~\bibnamefont
  {Borsanyi}} \emph {et~al.},\ }\href {\doibase 10.1038/s41586-021-03418-1}
  {\bibfield  {journal} {\bibinfo  {journal} {Nature}\ }\textbf {\bibinfo
  {volume} {593}},\ \bibinfo {pages} {51} (\bibinfo {year} {2021})},\ \Eprint
  {http://arxiv.org/abs/2002.12347} {arXiv:2002.12347 [hep-lat]} \BibitemShut
  {NoStop}%
\bibitem [{\citenamefont {Banerjee}\ \emph {et~al.}(2017)\citenamefont
  {Banerjee} \emph {et~al.}}]{Banerjee:2016tad}%
  \BibitemOpen
  \bibfield  {author} {\bibinfo {author} {\bibfnamefont {D.}~\bibnamefont
  {Banerjee}} \emph {et~al.} (\bibinfo {collaboration} {NA64}),\ }\href
  {\doibase 10.1103/PhysRevLett.118.011802} {\bibfield  {journal} {\bibinfo
  {journal} {Phys. Rev. Lett.}\ }\textbf {\bibinfo {volume} {118}},\ \bibinfo
  {pages} {011802} (\bibinfo {year} {2017})},\ \Eprint
  {http://arxiv.org/abs/1610.02988} {arXiv:1610.02988 [hep-ex]} \BibitemShut
  {NoStop}%
\bibitem [{\citenamefont {Lees}\ \emph {et~al.}(2014)\citenamefont {Lees} \emph
  {et~al.}}]{BaBar:2014zli}%
  \BibitemOpen
  \bibfield  {author} {\bibinfo {author} {\bibfnamefont {J.~P.}\ \bibnamefont
  {Lees}} \emph {et~al.} (\bibinfo {collaboration} {BaBar}),\ }\href {\doibase
  10.1103/PhysRevLett.113.201801} {\bibfield  {journal} {\bibinfo  {journal}
  {Phys. Rev. Lett.}\ }\textbf {\bibinfo {volume} {113}},\ \bibinfo {pages}
  {201801} (\bibinfo {year} {2014})},\ \Eprint {http://arxiv.org/abs/1406.2980}
  {arXiv:1406.2980 [hep-ex]} \BibitemShut {NoStop}%
\bibitem [{\citenamefont {Batley}\ \emph {et~al.}(2015)\citenamefont {Batley}
  \emph {et~al.}}]{NA482:2015wmo}%
  \BibitemOpen
  \bibfield  {author} {\bibinfo {author} {\bibfnamefont {J.~R.}\ \bibnamefont
  {Batley}} \emph {et~al.} (\bibinfo {collaboration} {NA48/2}),\ }\href
  {\doibase 10.1016/j.physletb.2015.04.068} {\bibfield  {journal} {\bibinfo
  {journal} {Phys. Lett. B}\ }\textbf {\bibinfo {volume} {746}},\ \bibinfo
  {pages} {178} (\bibinfo {year} {2015})},\ \Eprint
  {http://arxiv.org/abs/1504.00607} {arXiv:1504.00607 [hep-ex]} \BibitemShut
  {NoStop}%
\bibitem [{\citenamefont {Anastasi}\ \emph {et~al.}(2015)\citenamefont
  {Anastasi} \emph {et~al.}}]{Anastasi:2015qla}%
  \BibitemOpen
  \bibfield  {author} {\bibinfo {author} {\bibfnamefont {A.}~\bibnamefont
  {Anastasi}} \emph {et~al.},\ }\href {\doibase 10.1016/j.physletb.2015.10.003}
  {\bibfield  {journal} {\bibinfo  {journal} {Phys. Lett. B}\ }\textbf
  {\bibinfo {volume} {750}},\ \bibinfo {pages} {633} (\bibinfo {year}
  {2015})},\ \Eprint {http://arxiv.org/abs/1509.00740} {arXiv:1509.00740
  [hep-ex]} \BibitemShut {NoStop}%
\bibitem [{\citenamefont {Tucker-Smith}\ and\ \citenamefont
  {Weiner}(2001)}]{Tucker-Smith:2001myb}%
  \BibitemOpen
  \bibfield  {author} {\bibinfo {author} {\bibfnamefont {D.}~\bibnamefont
  {Tucker-Smith}}\ and\ \bibinfo {author} {\bibfnamefont {N.}~\bibnamefont
  {Weiner}},\ }\href {\doibase 10.1103/PhysRevD.64.043502} {\bibfield
  {journal} {\bibinfo  {journal} {Phys. Rev. D}\ }\textbf {\bibinfo {volume}
  {64}},\ \bibinfo {pages} {043502} (\bibinfo {year} {2001})},\ \Eprint
  {http://arxiv.org/abs/hep-ph/0101138} {arXiv:hep-ph/0101138} \BibitemShut
  {NoStop}%
\bibitem [{\citenamefont {Tucker-Smith}\ and\ \citenamefont
  {Weiner}(2005)}]{Tucker-Smith:2004mxa}%
  \BibitemOpen
  \bibfield  {author} {\bibinfo {author} {\bibfnamefont {D.}~\bibnamefont
  {Tucker-Smith}}\ and\ \bibinfo {author} {\bibfnamefont {N.}~\bibnamefont
  {Weiner}},\ }\href {\doibase 10.1103/PhysRevD.72.063509} {\bibfield
  {journal} {\bibinfo  {journal} {Phys. Rev. D}\ }\textbf {\bibinfo {volume}
  {72}},\ \bibinfo {pages} {063509} (\bibinfo {year} {2005})},\ \Eprint
  {http://arxiv.org/abs/hep-ph/0402065} {arXiv:hep-ph/0402065} \BibitemShut
  {NoStop}%
\bibitem [{\citenamefont {Filimonova}\ \emph {et~al.}(2022)\citenamefont
  {Filimonova}, \citenamefont {Junius}, \citenamefont {Lopez~Honorez},\ and\
  \citenamefont {Westhoff}}]{Filimonova:2022pkj}%
  \BibitemOpen
  \bibfield  {author} {\bibinfo {author} {\bibfnamefont {A.}~\bibnamefont
  {Filimonova}}, \bibinfo {author} {\bibfnamefont {S.}~\bibnamefont {Junius}},
  \bibinfo {author} {\bibfnamefont {L.}~\bibnamefont {Lopez~Honorez}}, \ and\
  \bibinfo {author} {\bibfnamefont {S.}~\bibnamefont {Westhoff}},\ }\href
  {\doibase 10.1007/JHEP06(2022)048} {\bibfield  {journal} {\bibinfo  {journal}
  {JHEP}\ }\textbf {\bibinfo {volume} {06}},\ \bibinfo {pages} {048} (\bibinfo
  {year} {2022})},\ \Eprint {http://arxiv.org/abs/2201.08409} {arXiv:2201.08409
  [hep-ph]} \BibitemShut {NoStop}%
\bibitem [{\citenamefont {Banerjee}\ \emph {et~al.}(2020)\citenamefont
  {Banerjee} \emph {et~al.}}]{NA64:2019auh}%
  \BibitemOpen
  \bibfield  {author} {\bibinfo {author} {\bibfnamefont {D.}~\bibnamefont
  {Banerjee}} \emph {et~al.} (\bibinfo {collaboration} {NA64}),\ }\href
  {\doibase 10.1103/PhysRevD.101.071101} {\bibfield  {journal} {\bibinfo
  {journal} {Phys. Rev. D}\ }\textbf {\bibinfo {volume} {101}},\ \bibinfo
  {pages} {071101} (\bibinfo {year} {2020})},\ \Eprint
  {http://arxiv.org/abs/1912.11389} {arXiv:1912.11389 [hep-ex]} \BibitemShut
  {NoStop}%
\bibitem [{\citenamefont {Banerjee}\ \emph {et~al.}(2019)\citenamefont
  {Banerjee} \emph {et~al.}}]{Banerjee:2019pds}%
  \BibitemOpen
  \bibfield  {author} {\bibinfo {author} {\bibfnamefont {D.}~\bibnamefont
  {Banerjee}} \emph {et~al.},\ }\href {\doibase 10.1103/PhysRevLett.123.121801}
  {\bibfield  {journal} {\bibinfo  {journal} {Phys. Rev. Lett.}\ }\textbf
  {\bibinfo {volume} {123}},\ \bibinfo {pages} {121801} (\bibinfo {year}
  {2019})},\ \Eprint {http://arxiv.org/abs/1906.00176} {arXiv:1906.00176
  [hep-ex]} \BibitemShut {NoStop}%
\bibitem [{\citenamefont {Gninenko}(2014)}]{Gninenko:2013rka}%
  \BibitemOpen
  \bibfield  {author} {\bibinfo {author} {\bibfnamefont {S.~N.}\ \bibnamefont
  {Gninenko}},\ }\href {\doibase 10.1103/PhysRevD.89.075008} {\bibfield
  {journal} {\bibinfo  {journal} {Phys. Rev. D}\ }\textbf {\bibinfo {volume}
  {89}},\ \bibinfo {pages} {075008} (\bibinfo {year} {2014})},\ \Eprint
  {http://arxiv.org/abs/1308.6521} {arXiv:1308.6521 [hep-ph]} \BibitemShut
  {NoStop}%
\bibitem [{\citenamefont {Depero}\ \emph {et~al.}(2017)\citenamefont {Depero}
  \emph {et~al.}}]{Depero:2017mrr}%
  \BibitemOpen
  \bibfield  {author} {\bibinfo {author} {\bibfnamefont {E.}~\bibnamefont
  {Depero}} \emph {et~al.},\ }\href {\doibase 10.1016/j.nima.2017.05.028}
  {\bibfield  {journal} {\bibinfo  {journal} {Nucl. Instrum. Meth. A}\ }\textbf
  {\bibinfo {volume} {866}},\ \bibinfo {pages} {196} (\bibinfo {year}
  {2017})},\ \Eprint {http://arxiv.org/abs/1703.05993} {arXiv:1703.05993
  [physics.ins-det]} \BibitemShut {NoStop}%
\bibitem [{\citenamefont {Bjorken}\ \emph {et~al.}(2009)\citenamefont
  {Bjorken}, \citenamefont {Essig}, \citenamefont {Schuster},\ and\
  \citenamefont {Toro}}]{Bjorken:2009mm}%
  \BibitemOpen
  \bibfield  {author} {\bibinfo {author} {\bibfnamefont {J.~D.}\ \bibnamefont
  {Bjorken}}, \bibinfo {author} {\bibfnamefont {R.}~\bibnamefont {Essig}},
  \bibinfo {author} {\bibfnamefont {P.}~\bibnamefont {Schuster}}, \ and\
  \bibinfo {author} {\bibfnamefont {N.}~\bibnamefont {Toro}},\ }\href {\doibase
  10.1103/PhysRevD.80.075018} {\bibfield  {journal} {\bibinfo  {journal} {Phys.
  Rev. D}\ }\textbf {\bibinfo {volume} {80}},\ \bibinfo {pages} {075018}
  (\bibinfo {year} {2009})},\ \Eprint {http://arxiv.org/abs/0906.0580}
  {arXiv:0906.0580 [hep-ph]} \BibitemShut {NoStop}%
\bibitem [{\citenamefont {Gninenko}\ \emph {et~al.}(2018)\citenamefont
  {Gninenko}, \citenamefont {Kirpichnikov}, \citenamefont {Kirsanov},\ and\
  \citenamefont {Krasnikov}}]{Gninenko:2017yus}%
  \BibitemOpen
  \bibfield  {author} {\bibinfo {author} {\bibfnamefont {S.~N.}\ \bibnamefont
  {Gninenko}}, \bibinfo {author} {\bibfnamefont {D.~V.}\ \bibnamefont
  {Kirpichnikov}}, \bibinfo {author} {\bibfnamefont {M.~M.}\ \bibnamefont
  {Kirsanov}}, \ and\ \bibinfo {author} {\bibfnamefont {N.~V.}\ \bibnamefont
  {Krasnikov}},\ }\href {\doibase 10.1016/j.physletb.2018.05.010} {\bibfield
  {journal} {\bibinfo  {journal} {Phys. Lett. B}\ }\textbf {\bibinfo {volume}
  {782}},\ \bibinfo {pages} {406} (\bibinfo {year} {2018})},\ \Eprint
  {http://arxiv.org/abs/1712.05706} {arXiv:1712.05706 [hep-ph]} \BibitemShut
  {NoStop}%
\bibitem [{\citenamefont {Liu}\ and\ \citenamefont
  {Miller}(2017)}]{Liu:2017htz}%
  \BibitemOpen
  \bibfield  {author} {\bibinfo {author} {\bibfnamefont {Y.-S.}\ \bibnamefont
  {Liu}}\ and\ \bibinfo {author} {\bibfnamefont {G.~A.}\ \bibnamefont
  {Miller}},\ }\href {\doibase 10.1103/PhysRevD.96.016004} {\bibfield
  {journal} {\bibinfo  {journal} {Phys. Rev. D}\ }\textbf {\bibinfo {volume}
  {96}},\ \bibinfo {pages} {016004} (\bibinfo {year} {2017})},\ \Eprint
  {http://arxiv.org/abs/1705.01633} {arXiv:1705.01633 [hep-ph]} \BibitemShut
  {NoStop}%
\bibitem [{\citenamefont {Agostinelli}\ \emph {et~al.}(2003)\citenamefont
  {Agostinelli} \emph {et~al.}}]{GEANT4:2002zbu}%
  \BibitemOpen
  \bibfield  {author} {\bibinfo {author} {\bibfnamefont {S.}~\bibnamefont
  {Agostinelli}} \emph {et~al.} (\bibinfo {collaboration} {GEANT4}),\ }\href
  {\doibase 10.1016/S0168-9002(03)01368-8} {\bibfield  {journal} {\bibinfo
  {journal} {Nucl. Instrum. Meth. A}\ }\textbf {\bibinfo {volume} {506}},\
  \bibinfo {pages} {250} (\bibinfo {year} {2003})}\BibitemShut {NoStop}%
\bibitem [{\citenamefont {Bondi}\ \emph {et~al.}(2021)\citenamefont {Bondi},
  \citenamefont {Celentano}, \citenamefont {Dusaev}, \citenamefont
  {Kirpichnikov}, \citenamefont {Kirsanov}, \citenamefont {Krasnikov},
  \citenamefont {Marsicano},\ and\ \citenamefont {Shchukin}}]{Bondi:2021nfp}%
  \BibitemOpen
  \bibfield  {author} {\bibinfo {author} {\bibfnamefont {M.}~\bibnamefont
  {Bondi}}, \bibinfo {author} {\bibfnamefont {A.}~\bibnamefont {Celentano}},
  \bibinfo {author} {\bibfnamefont {R.~R.}\ \bibnamefont {Dusaev}}, \bibinfo
  {author} {\bibfnamefont {D.~V.}\ \bibnamefont {Kirpichnikov}}, \bibinfo
  {author} {\bibfnamefont {M.~M.}\ \bibnamefont {Kirsanov}}, \bibinfo {author}
  {\bibfnamefont {N.~V.}\ \bibnamefont {Krasnikov}}, \bibinfo {author}
  {\bibfnamefont {L.}~\bibnamefont {Marsicano}}, \ and\ \bibinfo {author}
  {\bibfnamefont {D.}~\bibnamefont {Shchukin}},\ }\href {\doibase
  10.1016/j.cpc.2021.108129} {\bibfield  {journal} {\bibinfo  {journal}
  {Comput. Phys. Commun.}\ }\textbf {\bibinfo {volume} {269}},\ \bibinfo
  {pages} {108129} (\bibinfo {year} {2021})},\ \Eprint
  {http://arxiv.org/abs/2101.12192} {arXiv:2101.12192 [hep-ph]} \BibitemShut
  {NoStop}%
\bibitem [{\citenamefont {Andreev}\ \emph {et~al.}(2021)\citenamefont {Andreev}
  \emph {et~al.}}]{Andreev:2021fzd}%
  \BibitemOpen
  \bibfield  {author} {\bibinfo {author} {\bibfnamefont {Y.~M.}\ \bibnamefont
  {Andreev}} \emph {et~al.},\ }\href {\doibase 10.1103/PhysRevD.104.L091701}
  {\bibfield  {journal} {\bibinfo  {journal} {Phys. Rev. D}\ }\textbf {\bibinfo
  {volume} {104}},\ \bibinfo {pages} {L091701} (\bibinfo {year} {2021})},\
  \Eprint {http://arxiv.org/abs/2108.04195} {arXiv:2108.04195 [hep-ex]}
  \BibitemShut {NoStop}%
\bibitem [{\citenamefont {Cazzaniga}\ \emph {et~al.}(2021)\citenamefont
  {Cazzaniga} \emph {et~al.}}]{NA64:2021acr}%
  \BibitemOpen
  \bibfield  {author} {\bibinfo {author} {\bibfnamefont {C.}~\bibnamefont
  {Cazzaniga}} \emph {et~al.} (\bibinfo {collaboration} {NA64}),\ }\href
  {\doibase 10.1140/epjc/s10052-021-09705-5} {\bibfield  {journal} {\bibinfo
  {journal} {Eur. Phys. J. C}\ }\textbf {\bibinfo {volume} {81}},\ \bibinfo
  {pages} {959} (\bibinfo {year} {2021})},\ \Eprint
  {http://arxiv.org/abs/2107.02021} {arXiv:2107.02021 [hep-ex]} \BibitemShut
  {NoStop}%
\bibitem [{\citenamefont {Bergsma}\ \emph {et~al.}(1983)\citenamefont {Bergsma}
  \emph {et~al.}}]{CHARM:1983ayi}%
  \BibitemOpen
  \bibfield  {author} {\bibinfo {author} {\bibfnamefont {F.}~\bibnamefont
  {Bergsma}} \emph {et~al.} (\bibinfo {collaboration} {CHARM}),\ }\href
  {\doibase 10.1016/0370-2693(83)90275-7} {\bibfield  {journal} {\bibinfo
  {journal} {Phys. Lett. B}\ }\textbf {\bibinfo {volume} {128}},\ \bibinfo
  {pages} {361} (\bibinfo {year} {1983})}\BibitemShut {NoStop}%
\bibitem [{\citenamefont {Blumlein}\ \emph {et~al.}(1991)\citenamefont
  {Blumlein} \emph {et~al.}}]{Blumlein:1990ay}%
  \BibitemOpen
  \bibfield  {author} {\bibinfo {author} {\bibfnamefont {J.}~\bibnamefont
  {Blumlein}} \emph {et~al.},\ }\href {\doibase 10.1007/BF01548556} {\bibfield
  {journal} {\bibinfo  {journal} {Z. Phys. C}\ }\textbf {\bibinfo {volume}
  {51}},\ \bibinfo {pages} {341} (\bibinfo {year} {1991})}\BibitemShut
  {NoStop}%
\bibitem [{\citenamefont {Blumlein}\ \emph {et~al.}(1992)\citenamefont
  {Blumlein} \emph {et~al.}}]{Blumlein:1991xh}%
  \BibitemOpen
  \bibfield  {author} {\bibinfo {author} {\bibfnamefont {J.}~\bibnamefont
  {Blumlein}} \emph {et~al.},\ }\href {\doibase 10.1142/S0217751X9200171X}
  {\bibfield  {journal} {\bibinfo  {journal} {Int. J. Mod. Phys. A}\ }\textbf
  {\bibinfo {volume} {7}},\ \bibinfo {pages} {3835} (\bibinfo {year}
  {1992})}\BibitemShut {NoStop}%
\bibitem [{\citenamefont {Blumlein}\ and\ \citenamefont
  {Brunner}(2011)}]{Blumlein:2011mv}%
  \BibitemOpen
  \bibfield  {author} {\bibinfo {author} {\bibfnamefont {J.}~\bibnamefont
  {Blumlein}}\ and\ \bibinfo {author} {\bibfnamefont {J.}~\bibnamefont
  {Brunner}},\ }\href {\doibase 10.1016/j.physletb.2011.05.046} {\bibfield
  {journal} {\bibinfo  {journal} {Phys. Lett. B}\ }\textbf {\bibinfo {volume}
  {701}},\ \bibinfo {pages} {155} (\bibinfo {year} {2011})},\ \Eprint
  {http://arxiv.org/abs/1104.2747} {arXiv:1104.2747 [hep-ex]} \BibitemShut
  {NoStop}%
\bibitem [{\citenamefont {Gninenko}(2012)}]{Gninenko:2012eq}%
  \BibitemOpen
  \bibfield  {author} {\bibinfo {author} {\bibfnamefont {S.~N.}\ \bibnamefont
  {Gninenko}},\ }\href {\doibase 10.1016/j.physletb.2012.06.002} {\bibfield
  {journal} {\bibinfo  {journal} {Phys. Lett. B}\ }\textbf {\bibinfo {volume}
  {713}},\ \bibinfo {pages} {244} (\bibinfo {year} {2012})},\ \Eprint
  {http://arxiv.org/abs/1204.3583} {arXiv:1204.3583 [hep-ph]} \BibitemShut
  {NoStop}%
\bibitem [{\citenamefont {deNiverville}\ \emph {et~al.}(2018)\citenamefont
  {deNiverville}, \citenamefont {Lee},\ and\ \citenamefont
  {Seo}}]{deNiverville:2018hrc}%
  \BibitemOpen
  \bibfield  {author} {\bibinfo {author} {\bibfnamefont {P.}~\bibnamefont
  {deNiverville}}, \bibinfo {author} {\bibfnamefont {H.-S.}\ \bibnamefont
  {Lee}}, \ and\ \bibinfo {author} {\bibfnamefont {M.-S.}\ \bibnamefont
  {Seo}},\ }\href {\doibase 10.1103/PhysRevD.98.115011} {\bibfield  {journal}
  {\bibinfo  {journal} {Phys. Rev. D}\ }\textbf {\bibinfo {volume} {98}},\
  \bibinfo {pages} {115011} (\bibinfo {year} {2018})},\ \Eprint
  {http://arxiv.org/abs/1806.00757} {arXiv:1806.00757 [hep-ph]} \BibitemShut
  {NoStop}%
\bibitem [{\citenamefont {Auerbach}\ \emph {et~al.}(2001)\citenamefont
  {Auerbach} \emph {et~al.}}]{LSND:2001akn}%
  \BibitemOpen
  \bibfield  {author} {\bibinfo {author} {\bibfnamefont {L.~B.}\ \bibnamefont
  {Auerbach}} \emph {et~al.} (\bibinfo {collaboration} {LSND}),\ }\href
  {\doibase 10.1103/PhysRevD.63.112001} {\bibfield  {journal} {\bibinfo
  {journal} {Phys. Rev. D}\ }\textbf {\bibinfo {volume} {63}},\ \bibinfo
  {pages} {112001} (\bibinfo {year} {2001})},\ \Eprint
  {http://arxiv.org/abs/hep-ex/0101039} {arXiv:hep-ex/0101039} \BibitemShut
  {NoStop}%
\bibitem [{\citenamefont {deNiverville}\ \emph {et~al.}(2011)\citenamefont
  {deNiverville}, \citenamefont {Pospelov},\ and\ \citenamefont
  {Ritz}}]{deNiverville:2011it}%
  \BibitemOpen
  \bibfield  {author} {\bibinfo {author} {\bibfnamefont {P.}~\bibnamefont
  {deNiverville}}, \bibinfo {author} {\bibfnamefont {M.}~\bibnamefont
  {Pospelov}}, \ and\ \bibinfo {author} {\bibfnamefont {A.}~\bibnamefont
  {Ritz}},\ }\href {\doibase 10.1103/PhysRevD.84.075020} {\bibfield  {journal}
  {\bibinfo  {journal} {Phys. Rev. D}\ }\textbf {\bibinfo {volume} {84}},\
  \bibinfo {pages} {075020} (\bibinfo {year} {2011})},\ \Eprint
  {http://arxiv.org/abs/1107.4580} {arXiv:1107.4580 [hep-ph]} \BibitemShut
  {NoStop}%
\bibitem [{\citenamefont {Izaguirre}\ \emph {et~al.}(2017)\citenamefont
  {Izaguirre}, \citenamefont {Kahn}, \citenamefont {Krnjaic},\ and\
  \citenamefont {Moschella}}]{Izaguirre:2017bqb}%
  \BibitemOpen
  \bibfield  {author} {\bibinfo {author} {\bibfnamefont {E.}~\bibnamefont
  {Izaguirre}}, \bibinfo {author} {\bibfnamefont {Y.}~\bibnamefont {Kahn}},
  \bibinfo {author} {\bibfnamefont {G.}~\bibnamefont {Krnjaic}}, \ and\
  \bibinfo {author} {\bibfnamefont {M.}~\bibnamefont {Moschella}},\ }\href
  {\doibase 10.1103/PhysRevD.96.055007} {\bibfield  {journal} {\bibinfo
  {journal} {Phys. Rev. D}\ }\textbf {\bibinfo {volume} {96}},\ \bibinfo
  {pages} {055007} (\bibinfo {year} {2017})},\ \Eprint
  {http://arxiv.org/abs/1703.06881} {arXiv:1703.06881 [hep-ph]} \BibitemShut
  {NoStop}%
\bibitem [{\citenamefont {Bjorken}\ \emph {et~al.}(1988)\citenamefont
  {Bjorken}, \citenamefont {Ecklund}, \citenamefont {Nelson}, \citenamefont
  {Abashian}, \citenamefont {Church}, \citenamefont {Lu}, \citenamefont {Mo},
  \citenamefont {Nunamaker},\ and\ \citenamefont {Rassmann}}]{Bjorken:1988as}%
  \BibitemOpen
  \bibfield  {author} {\bibinfo {author} {\bibfnamefont {J.~D.}\ \bibnamefont
  {Bjorken}}, \bibinfo {author} {\bibfnamefont {S.}~\bibnamefont {Ecklund}},
  \bibinfo {author} {\bibfnamefont {W.~R.}\ \bibnamefont {Nelson}}, \bibinfo
  {author} {\bibfnamefont {A.}~\bibnamefont {Abashian}}, \bibinfo {author}
  {\bibfnamefont {C.}~\bibnamefont {Church}}, \bibinfo {author} {\bibfnamefont
  {B.}~\bibnamefont {Lu}}, \bibinfo {author} {\bibfnamefont {L.~W.}\
  \bibnamefont {Mo}}, \bibinfo {author} {\bibfnamefont {T.~A.}\ \bibnamefont
  {Nunamaker}}, \ and\ \bibinfo {author} {\bibfnamefont {P.}~\bibnamefont
  {Rassmann}},\ }\href {\doibase 10.1103/PhysRevD.38.3375} {\bibfield
  {journal} {\bibinfo  {journal} {Phys. Rev. D}\ }\textbf {\bibinfo {volume}
  {38}},\ \bibinfo {pages} {3375} (\bibinfo {year} {1988})}\BibitemShut
  {NoStop}%
\bibitem [{\citenamefont {Cortina~Gil}\ \emph {et~al.}(2019)\citenamefont
  {Cortina~Gil} \emph {et~al.}}]{NA62:2019meo}%
  \BibitemOpen
  \bibfield  {author} {\bibinfo {author} {\bibfnamefont {E.}~\bibnamefont
  {Cortina~Gil}} \emph {et~al.} (\bibinfo {collaboration} {NA62}),\ }\href
  {\doibase 10.1007/JHEP05(2019)182} {\bibfield  {journal} {\bibinfo  {journal}
  {JHEP}\ }\textbf {\bibinfo {volume} {05}},\ \bibinfo {pages} {182} (\bibinfo
  {year} {2019})},\ \Eprint {http://arxiv.org/abs/1903.08767} {arXiv:1903.08767
  [hep-ex]} \BibitemShut {NoStop}%
\bibitem [{\citenamefont {Kribs}\ \emph {et~al.}(2021)\citenamefont {Kribs},
  \citenamefont {McKeen},\ and\ \citenamefont {Raj}}]{Kribs:2020vyk}%
  \BibitemOpen
  \bibfield  {author} {\bibinfo {author} {\bibfnamefont {G.~D.}\ \bibnamefont
  {Kribs}}, \bibinfo {author} {\bibfnamefont {D.}~\bibnamefont {McKeen}}, \
  and\ \bibinfo {author} {\bibfnamefont {N.}~\bibnamefont {Raj}},\ }\href
  {\doibase 10.1103/PhysRevLett.126.011801} {\bibfield  {journal} {\bibinfo
  {journal} {Phys. Rev. Lett.}\ }\textbf {\bibinfo {volume} {126}},\ \bibinfo
  {pages} {011801} (\bibinfo {year} {2021})},\ \Eprint
  {http://arxiv.org/abs/2007.15655} {arXiv:2007.15655 [hep-ph]} \BibitemShut
  {NoStop}%
\bibitem [{\citenamefont {Carrazza}\ \emph {et~al.}(2019)\citenamefont
  {Carrazza}, \citenamefont {Degrande}, \citenamefont {Iranipour},
  \citenamefont {Rojo},\ and\ \citenamefont {Ubiali}}]{Carrazza:2019sec}%
  \BibitemOpen
  \bibfield  {author} {\bibinfo {author} {\bibfnamefont {S.}~\bibnamefont
  {Carrazza}}, \bibinfo {author} {\bibfnamefont {C.}~\bibnamefont {Degrande}},
  \bibinfo {author} {\bibfnamefont {S.}~\bibnamefont {Iranipour}}, \bibinfo
  {author} {\bibfnamefont {J.}~\bibnamefont {Rojo}}, \ and\ \bibinfo {author}
  {\bibfnamefont {M.}~\bibnamefont {Ubiali}},\ }\href {\doibase
  10.1103/PhysRevLett.123.132001} {\bibfield  {journal} {\bibinfo  {journal}
  {Phys. Rev. Lett.}\ }\textbf {\bibinfo {volume} {123}},\ \bibinfo {pages}
  {132001} (\bibinfo {year} {2019})},\ \Eprint
  {http://arxiv.org/abs/1905.05215} {arXiv:1905.05215 [hep-ph]} \BibitemShut
  {NoStop}%
\bibitem [{\citenamefont {Thomas}\ \emph {et~al.}(2022)\citenamefont {Thomas},
  \citenamefont {Wang},\ and\ \citenamefont {Williams}}]{Thomas:2021lub}%
  \BibitemOpen
  \bibfield  {author} {\bibinfo {author} {\bibfnamefont {A.~W.}\ \bibnamefont
  {Thomas}}, \bibinfo {author} {\bibfnamefont {X.~G.}\ \bibnamefont {Wang}}, \
  and\ \bibinfo {author} {\bibfnamefont {A.~G.}\ \bibnamefont {Williams}},\
  }\href {\doibase 10.1103/PhysRevD.105.L031901} {\bibfield  {journal}
  {\bibinfo  {journal} {Phys. Rev. D}\ }\textbf {\bibinfo {volume} {105}},\
  \bibinfo {pages} {L031901} (\bibinfo {year} {2022})},\ \Eprint
  {http://arxiv.org/abs/2111.05664} {arXiv:2111.05664 [hep-ph]} \BibitemShut
  {NoStop}%
\bibitem [{\citenamefont {Abramowicz}\ \emph {et~al.}(2015)\citenamefont
  {Abramowicz} \emph {et~al.}}]{H1:2015ubc}%
  \BibitemOpen
  \bibfield  {author} {\bibinfo {author} {\bibfnamefont {H.}~\bibnamefont
  {Abramowicz}} \emph {et~al.} (\bibinfo {collaboration} {H1, ZEUS}),\ }\href
  {\doibase 10.1140/epjc/s10052-015-3710-4} {\bibfield  {journal} {\bibinfo
  {journal} {Eur. Phys. J. C}\ }\textbf {\bibinfo {volume} {75}},\ \bibinfo
  {pages} {580} (\bibinfo {year} {2015})},\ \Eprint
  {http://arxiv.org/abs/1506.06042} {arXiv:1506.06042 [hep-ex]} \BibitemShut
  {NoStop}%
\bibitem [{\citenamefont {Hook}\ \emph {et~al.}(2011)\citenamefont {Hook},
  \citenamefont {Izaguirre},\ and\ \citenamefont {Wacker}}]{Hook:2010tw}%
  \BibitemOpen
  \bibfield  {author} {\bibinfo {author} {\bibfnamefont {A.}~\bibnamefont
  {Hook}}, \bibinfo {author} {\bibfnamefont {E.}~\bibnamefont {Izaguirre}}, \
  and\ \bibinfo {author} {\bibfnamefont {J.~G.}\ \bibnamefont {Wacker}},\
  }\href {\doibase 10.1155/2011/859762} {\bibfield  {journal} {\bibinfo
  {journal} {Adv. High Energy Phys.}\ }\textbf {\bibinfo {volume} {2011}},\
  \bibinfo {pages} {859762} (\bibinfo {year} {2011})},\ \Eprint
  {http://arxiv.org/abs/1006.0973} {arXiv:1006.0973 [hep-ph]} \BibitemShut
  {NoStop}%
\bibitem [{\citenamefont {Curtin}\ \emph {et~al.}(2015)\citenamefont {Curtin},
  \citenamefont {Essig}, \citenamefont {Gori},\ and\ \citenamefont
  {Shelton}}]{Curtin:2014cca}%
  \BibitemOpen
  \bibfield  {author} {\bibinfo {author} {\bibfnamefont {D.}~\bibnamefont
  {Curtin}}, \bibinfo {author} {\bibfnamefont {R.}~\bibnamefont {Essig}},
  \bibinfo {author} {\bibfnamefont {S.}~\bibnamefont {Gori}}, \ and\ \bibinfo
  {author} {\bibfnamefont {J.}~\bibnamefont {Shelton}},\ }\href {\doibase
  10.1007/JHEP02(2015)157} {\bibfield  {journal} {\bibinfo  {journal} {JHEP}\
  }\textbf {\bibinfo {volume} {02}},\ \bibinfo {pages} {157} (\bibinfo {year}
  {2015})},\ \Eprint {http://arxiv.org/abs/1412.0018} {arXiv:1412.0018
  [hep-ph]} \BibitemShut {NoStop}%
\bibitem [{\citenamefont {Slatyer}(2016)}]{Slatyer:2015jla}%
  \BibitemOpen
  \bibfield  {author} {\bibinfo {author} {\bibfnamefont {T.~R.}\ \bibnamefont
  {Slatyer}},\ }\href {\doibase 10.1103/PhysRevD.93.023527} {\bibfield
  {journal} {\bibinfo  {journal} {Phys. Rev. D}\ }\textbf {\bibinfo {volume}
  {93}},\ \bibinfo {pages} {023527} (\bibinfo {year} {2016})},\ \Eprint
  {http://arxiv.org/abs/1506.03811} {arXiv:1506.03811 [hep-ph]} \BibitemShut
  {NoStop}%
\bibitem [{\citenamefont {Izaguirre}\ \emph {et~al.}(2016)\citenamefont
  {Izaguirre}, \citenamefont {Krnjaic},\ and\ \citenamefont
  {Shuve}}]{Izaguirre:2015zva}%
  \BibitemOpen
  \bibfield  {author} {\bibinfo {author} {\bibfnamefont {E.}~\bibnamefont
  {Izaguirre}}, \bibinfo {author} {\bibfnamefont {G.}~\bibnamefont {Krnjaic}},
  \ and\ \bibinfo {author} {\bibfnamefont {B.}~\bibnamefont {Shuve}},\ }\href
  {\doibase 10.1103/PhysRevD.93.063523} {\bibfield  {journal} {\bibinfo
  {journal} {Phys. Rev. D}\ }\textbf {\bibinfo {volume} {93}},\ \bibinfo
  {pages} {063523} (\bibinfo {year} {2016})},\ \Eprint
  {http://arxiv.org/abs/1508.03050} {arXiv:1508.03050 [hep-ph]} \BibitemShut
  {NoStop}%
\bibitem [{\citenamefont {Berlin}\ \emph {et~al.}(2019)\citenamefont {Berlin},
  \citenamefont {Blinov}, \citenamefont {Krnjaic}, \citenamefont {Schuster},\
  and\ \citenamefont {Toro}}]{Berlin:2018bsc}%
  \BibitemOpen
  \bibfield  {author} {\bibinfo {author} {\bibfnamefont {A.}~\bibnamefont
  {Berlin}}, \bibinfo {author} {\bibfnamefont {N.}~\bibnamefont {Blinov}},
  \bibinfo {author} {\bibfnamefont {G.}~\bibnamefont {Krnjaic}}, \bibinfo
  {author} {\bibfnamefont {P.}~\bibnamefont {Schuster}}, \ and\ \bibinfo
  {author} {\bibfnamefont {N.}~\bibnamefont {Toro}},\ }\href {\doibase
  10.1103/PhysRevD.99.075001} {\bibfield  {journal} {\bibinfo  {journal} {Phys.
  Rev. D}\ }\textbf {\bibinfo {volume} {99}},\ \bibinfo {pages} {075001}
  (\bibinfo {year} {2019})},\ \Eprint {http://arxiv.org/abs/1807.01730}
  {arXiv:1807.01730 [hep-ph]} \BibitemShut {NoStop}%
\bibitem [{\citenamefont {Ilten}\ \emph {et~al.}(2018)\citenamefont {Ilten},
  \citenamefont {Soreq}, \citenamefont {Williams},\ and\ \citenamefont
  {Xue}}]{Ilten:2018crw}%
  \BibitemOpen
  \bibfield  {author} {\bibinfo {author} {\bibfnamefont {P.}~\bibnamefont
  {Ilten}}, \bibinfo {author} {\bibfnamefont {Y.}~\bibnamefont {Soreq}},
  \bibinfo {author} {\bibfnamefont {M.}~\bibnamefont {Williams}}, \ and\
  \bibinfo {author} {\bibfnamefont {W.}~\bibnamefont {Xue}},\ }\href {\doibase
  10.1007/JHEP06(2018)004} {\bibfield  {journal} {\bibinfo  {journal} {JHEP}\
  }\textbf {\bibinfo {volume} {06}},\ \bibinfo {pages} {004} (\bibinfo {year}
  {2018})},\ \Eprint {http://arxiv.org/abs/1801.04847} {arXiv:1801.04847
  [hep-ph]} \BibitemShut {NoStop}%
\bibitem [{\citenamefont {Izaguirre}\ \emph
  {et~al.}(2015{\natexlab{b}})\citenamefont {Izaguirre}, \citenamefont
  {Krnjaic}, \citenamefont {Schuster},\ and\ \citenamefont
  {Toro}}]{Izaguirre:2015yja}%
  \BibitemOpen
  \bibfield  {author} {\bibinfo {author} {\bibfnamefont {E.}~\bibnamefont
  {Izaguirre}}, \bibinfo {author} {\bibfnamefont {G.}~\bibnamefont {Krnjaic}},
  \bibinfo {author} {\bibfnamefont {P.}~\bibnamefont {Schuster}}, \ and\
  \bibinfo {author} {\bibfnamefont {N.}~\bibnamefont {Toro}},\ }\href {\doibase
  10.1103/PhysRevLett.115.251301} {\bibfield  {journal} {\bibinfo  {journal}
  {Phys. Rev. Lett.}\ }\textbf {\bibinfo {volume} {115}},\ \bibinfo {pages}
  {251301} (\bibinfo {year} {2015}{\natexlab{b}})},\ \Eprint
  {http://arxiv.org/abs/1505.00011} {arXiv:1505.00011 [hep-ph]} \BibitemShut
  {NoStop}%
\bibitem [{\citenamefont {Feng}\ and\ \citenamefont
  {Smolinsky}(2017)}]{Feng:2017drg}%
  \BibitemOpen
  \bibfield  {author} {\bibinfo {author} {\bibfnamefont {J.~L.}\ \bibnamefont
  {Feng}}\ and\ \bibinfo {author} {\bibfnamefont {J.}~\bibnamefont
  {Smolinsky}},\ }\href {\doibase 10.1103/PhysRevD.96.095022} {\bibfield
  {journal} {\bibinfo  {journal} {Phys. Rev. D}\ }\textbf {\bibinfo {volume}
  {96}},\ \bibinfo {pages} {095022} (\bibinfo {year} {2017})},\ \Eprint
  {http://arxiv.org/abs/1707.03835} {arXiv:1707.03835 [hep-ph]} \BibitemShut
  {NoStop}%
\bibitem [{\citenamefont {Berlin}\ \emph {et~al.}(2020)\citenamefont {Berlin},
  \citenamefont {deNiverville}, \citenamefont {Ritz}, \citenamefont
  {Schuster},\ and\ \citenamefont {Toro}}]{Berlin:2020uwy}%
  \BibitemOpen
  \bibfield  {author} {\bibinfo {author} {\bibfnamefont {A.}~\bibnamefont
  {Berlin}}, \bibinfo {author} {\bibfnamefont {P.}~\bibnamefont
  {deNiverville}}, \bibinfo {author} {\bibfnamefont {A.}~\bibnamefont {Ritz}},
  \bibinfo {author} {\bibfnamefont {P.}~\bibnamefont {Schuster}}, \ and\
  \bibinfo {author} {\bibfnamefont {N.}~\bibnamefont {Toro}},\ }\href {\doibase
  10.1103/PhysRevD.102.095011} {\bibfield  {journal} {\bibinfo  {journal}
  {Phys. Rev. D}\ }\textbf {\bibinfo {volume} {102}},\ \bibinfo {pages}
  {095011} (\bibinfo {year} {2020})},\ \Eprint
  {http://arxiv.org/abs/2003.03379} {arXiv:2003.03379 [hep-ph]} \BibitemShut
  {NoStop}%
\bibitem [{\citenamefont {Sieber}\ \emph {et~al.}(2022)\citenamefont {Sieber},
  \citenamefont {Banerjee}, \citenamefont {Crivelli}, \citenamefont {Depero},
  \citenamefont {Gninenko}, \citenamefont {Kirpichnikov}, \citenamefont
  {Kirsanov}, \citenamefont {Poliakov},\ and\ \citenamefont
  {Molina~Bueno}}]{Sieber:2021fue}%
  \BibitemOpen
  \bibfield  {author} {\bibinfo {author} {\bibfnamefont {H.}~\bibnamefont
  {Sieber}}, \bibinfo {author} {\bibfnamefont {D.}~\bibnamefont {Banerjee}},
  \bibinfo {author} {\bibfnamefont {P.}~\bibnamefont {Crivelli}}, \bibinfo
  {author} {\bibfnamefont {E.}~\bibnamefont {Depero}}, \bibinfo {author}
  {\bibfnamefont {S.~N.}\ \bibnamefont {Gninenko}}, \bibinfo {author}
  {\bibfnamefont {D.~V.}\ \bibnamefont {Kirpichnikov}}, \bibinfo {author}
  {\bibfnamefont {M.~M.}\ \bibnamefont {Kirsanov}}, \bibinfo {author}
  {\bibfnamefont {V.}~\bibnamefont {Poliakov}}, \ and\ \bibinfo {author}
  {\bibfnamefont {L.}~\bibnamefont {Molina~Bueno}},\ }\href {\doibase
  10.1103/PhysRevD.105.052006} {\bibfield  {journal} {\bibinfo  {journal}
  {Phys. Rev. D}\ }\textbf {\bibinfo {volume} {105}},\ \bibinfo {pages}
  {052006} (\bibinfo {year} {2022})},\ \Eprint
  {http://arxiv.org/abs/2110.15111} {arXiv:2110.15111 [hep-ex]} \BibitemShut
  {NoStop}%
\bibitem [{\citenamefont {Gninenko}\ \emph {et~al.}(2019)\citenamefont
  {Gninenko}, \citenamefont {Kirpichnikov}, \citenamefont {Kirsanov},\ and\
  \citenamefont {Krasnikov}}]{Gninenko:2019qiv}%
  \BibitemOpen
  \bibfield  {author} {\bibinfo {author} {\bibfnamefont {S.~N.}\ \bibnamefont
  {Gninenko}}, \bibinfo {author} {\bibfnamefont {D.~V.}\ \bibnamefont
  {Kirpichnikov}}, \bibinfo {author} {\bibfnamefont {M.~M.}\ \bibnamefont
  {Kirsanov}}, \ and\ \bibinfo {author} {\bibfnamefont {N.~V.}\ \bibnamefont
  {Krasnikov}},\ }\href {\doibase 10.1016/j.physletb.2019.07.015} {\bibfield
  {journal} {\bibinfo  {journal} {Phys. Lett. B}\ }\textbf {\bibinfo {volume}
  {796}},\ \bibinfo {pages} {117} (\bibinfo {year} {2019})},\ \Eprint
  {http://arxiv.org/abs/1903.07899} {arXiv:1903.07899 [hep-ph]} \BibitemShut
  {NoStop}%
\bibitem [{\citenamefont {Altmannshofer}\ \emph {et~al.}(2019)\citenamefont
  {Altmannshofer} \emph {et~al.}}]{Belle-II:2018jsg}%
  \BibitemOpen
  \bibfield  {author} {\bibinfo {author} {\bibfnamefont {W.}~\bibnamefont
  {Altmannshofer}} \emph {et~al.} (\bibinfo {collaboration} {Belle-II}),\
  }\href {\doibase 10.1093/ptep/ptz106} {\bibfield  {journal} {\bibinfo
  {journal} {PTEP}\ }\textbf {\bibinfo {volume} {2019}},\ \bibinfo {pages}
  {123C01} (\bibinfo {year} {2019})},\ \bibinfo {note} {[Erratum: PTEP 2020,
  029201 (2020)]},\ \Eprint {http://arxiv.org/abs/1808.10567} {arXiv:1808.10567
  [hep-ex]} \BibitemShut {NoStop}%
\bibitem [{\citenamefont {Batell}\ \emph {et~al.}(2021)\citenamefont {Batell},
  \citenamefont {Berger}, \citenamefont {Darm\'e},\ and\ \citenamefont
  {Frugiuele}}]{Batell:2021ooj}%
  \BibitemOpen
  \bibfield  {author} {\bibinfo {author} {\bibfnamefont {B.}~\bibnamefont
  {Batell}}, \bibinfo {author} {\bibfnamefont {J.}~\bibnamefont {Berger}},
  \bibinfo {author} {\bibfnamefont {L.}~\bibnamefont {Darm\'e}}, \ and\
  \bibinfo {author} {\bibfnamefont {C.}~\bibnamefont {Frugiuele}},\ }\href
  {\doibase 10.1103/PhysRevD.104.075026} {\bibfield  {journal} {\bibinfo
  {journal} {Phys. Rev. D}\ }\textbf {\bibinfo {volume} {104}},\ \bibinfo
  {pages} {075026} (\bibinfo {year} {2021})},\ \Eprint
  {http://arxiv.org/abs/2106.04584} {arXiv:2106.04584 [hep-ph]} \BibitemShut
  {NoStop}%
\bibitem [{\citenamefont {De~Romeri}\ \emph {et~al.}(2019)\citenamefont
  {De~Romeri}, \citenamefont {Kelly},\ and\ \citenamefont
  {Machado}}]{DeRomeri:2019kic}%
  \BibitemOpen
  \bibfield  {author} {\bibinfo {author} {\bibfnamefont {V.}~\bibnamefont
  {De~Romeri}}, \bibinfo {author} {\bibfnamefont {K.~J.}\ \bibnamefont
  {Kelly}}, \ and\ \bibinfo {author} {\bibfnamefont {P.~A.~N.}\ \bibnamefont
  {Machado}},\ }\href {\doibase 10.1103/PhysRevD.100.095010} {\bibfield
  {journal} {\bibinfo  {journal} {Phys. Rev. D}\ }\textbf {\bibinfo {volume}
  {100}},\ \bibinfo {pages} {095010} (\bibinfo {year} {2019})},\ \Eprint
  {http://arxiv.org/abs/1903.10505} {arXiv:1903.10505 [hep-ph]} \BibitemShut
  {NoStop}%
\bibitem [{\citenamefont {De~Roeck}\ \emph {et~al.}(2020)\citenamefont
  {De~Roeck}, \citenamefont {Kim}, \citenamefont {Moghaddam}, \citenamefont
  {Park}, \citenamefont {Shin},\ and\ \citenamefont
  {Whitehead}}]{DeRoeck:2020ntj}%
  \BibitemOpen
  \bibfield  {author} {\bibinfo {author} {\bibfnamefont {A.}~\bibnamefont
  {De~Roeck}}, \bibinfo {author} {\bibfnamefont {D.}~\bibnamefont {Kim}},
  \bibinfo {author} {\bibfnamefont {Z.~G.}\ \bibnamefont {Moghaddam}}, \bibinfo
  {author} {\bibfnamefont {J.-C.}\ \bibnamefont {Park}}, \bibinfo {author}
  {\bibfnamefont {S.}~\bibnamefont {Shin}}, \ and\ \bibinfo {author}
  {\bibfnamefont {L.~H.}\ \bibnamefont {Whitehead}},\ }\href {\doibase
  10.1007/JHEP11(2020)043} {\bibfield  {journal} {\bibinfo  {journal} {JHEP}\
  }\textbf {\bibinfo {volume} {11}},\ \bibinfo {pages} {043} (\bibinfo {year}
  {2020})},\ \Eprint {http://arxiv.org/abs/2005.08979} {arXiv:2005.08979
  [hep-ph]} \BibitemShut {NoStop}%
\bibitem [{\citenamefont {Krnjaic}\ \emph {et~al.}(2022)\citenamefont {Krnjaic}
  \emph {et~al.}}]{Krnjaic:2022ozp}%
  \BibitemOpen
  \bibfield  {author} {\bibinfo {author} {\bibfnamefont {G.}~\bibnamefont
  {Krnjaic}} \emph {et~al.},\ }\href@noop {} {\  (\bibinfo {year} {2022})},\
  \Eprint {http://arxiv.org/abs/2207.00597} {arXiv:2207.00597 [hep-ph]}
  \BibitemShut {NoStop}%
\end{thebibliography}%
%%%%%%%%%%%%%%%%%%%%%%%%%%%%%%%%%%%%%%%%
% supplemental materials
\pagebreak
\appendix
%%%%%%%%%%%%%%%%%%%%%%%%%%%%%%%%%%%%%%%%

\end{document}